\documentclass[10pt,onecolumn]{IEEEtran}
\usepackage{epsfig}
\usepackage{amssymb}
\usepackage[usenames]{color}
\usepackage{umoline}
\usepackage{setspace}
\usepackage{algorithm}
\usepackage{algorithmic}
\usepackage[tbtags]{amsmath}
\usepackage{mathtools}
\usepackage{algorithm}
\usepackage{algorithmic}
\usepackage{url}
\usepackage{enumerate}

\newcommand{\INPUT}{\item[\textbf{Input:}]}
\newcommand{\OUTPUT}{\item[\textbf{Output:}]}

\newcommand{\norm}[1]{\Vert#1\Vert}

\newcommand{\rank}{\mathrm{rank}}
\newcommand{\krank}{\mathrm{krank}}

\newcommand{\spark}{\mathrm{spark}}

\newtheorem{theorem}{Theorem}[section]
\newtheorem{lemma}[theorem]{Lemma}
\newtheorem{proposition}[theorem]{Proposition}
\newtheorem{corollary}[theorem]{Corollary}

\numberwithin{equation}{section}
\renewcommand{\theequation}{\arabic{section}.\arabic{equation}}

\newtheorem{definition}[theorem]{Definition}
\newtheorem{remark}[theorem]{Remark}

\newcommand{\qed}{\nobreak \ifvmode \relax \else
      \ifdim\lastskip<1.5em \hskip-\lastskip
      \hskip1.5em plus0em minus0.5em \fi \nobreak
      \vrule height0.75em width0.5em depth0.25em\fi}

\title{
Subspace Methods for Joint Sparse Recovery
}
\author{Kiryung Lee, Yoram Bresler~\IEEEmembership{Fellow,~IEEE}, and Marius Junge
\thanks{This work was supported in part by NSF grant No. CCF 06-35234, NSF grant No. CCF 10-18660, and NSF grant No. DMS 09-01457.}
\thanks{K. Lee and Y. Bresler are with Coordinated Science Laboratory and Department of ECE, University of Illinois at Urbana-Champaign, IL 61801, USA, e-mail: \{klee81,ybresler\}@illinois.edu}
\thanks{M. Junge is with Department of Mathematics, University of Illinois at Urbana-Champaign, IL 61801, USA, e-mail: junge@math.uiuc.edu}
\thanks{The results in this paper have been partially presented at the 6th IEEE Sensor Array and Multichannel Signal Processing Workshop \cite{LeeBre10sam}.
}}

\begin{document}
\doublespacing

%%%% Table of Contents %%%%
%\clearpage
%\tableofcontents
%\clearpage

\vspace{0.5in}

\maketitle

\begin{abstract}
We propose robust and efficient algorithms for the joint sparse recovery problem in compressed sensing,
which simultaneously recover the supports of jointly sparse signals
from their multiple measurement vectors obtained through a common sensing matrix.
In a favorable situation, the unknown matrix, which consists of the jointly sparse signals, has linearly independent nonzero rows.
In this case, the MUSIC (MUltiple SIgnal Classification) algorithm, originally proposed by Schmidt for the direction of arrival problem in sensor array processing and later proposed and analyzed for joint sparse recovery by Feng and Bresler,
provides a guarantee with the minimum number of measurements.
We focus instead on the unfavorable but practically significant case of rank-defect or ill-conditioning.
This situation arises with limited number of measurement vectors, or with highly correlated signal components.
In this case MUSIC fails, and in practice none of the existing methods can consistently approach the fundamental limit.
We propose subspace-augmented MUSIC (SA-MUSIC), which improves on MUSIC so that
the support is reliably recovered under such unfavorable conditions.
Combined with subspace-based greedy algorithms also proposed and analyzed in this paper,
SA-MUSIC provides a computationally efficient algorithm with a performance guarantee.
The performance guarantees are given in terms of a version of restricted isometry property.
In particular, we also present a non-asymptotic perturbation analysis of the signal subspace estimation
that has been missing in the previous study of MUSIC.
\end{abstract}

\begin{IEEEkeywords}
Compressed sensing, joint sparsity, multiple measurement vectors (MMV), subspace estimation,
restricted isometry property (RIP), sensor array processing, spectrum-blind sampling.
\end{IEEEkeywords}

%%%%%%%%%%%%
\section{Introduction}\label{sec:intro}
%%%%%%%%%%%%

The problem of computing a sparse approximate solution to a linear system has been
studied as the subset selection problem in matrix computations \cite{GolVan96}
with applications in statistical regression \cite{Mil02} and signal processing \cite{FenBre96}, \cite{MalCW05}.
The matrix representing the linear system and its columns are called a dictionary and atoms, respectively.
The sparse recovery problem addresses the identification of the support,
which denotes the indices of the atoms that contribute to the sparse solution,
or equivalently, the indices of the nonzero rows of the unknown matrix.
Once the support is determined, the recovery of the sparse signals reduces to standard overdetermined linear inverse problems.

The study of sparse solutions to underdetermined systems dates back to 70's \cite{ClaMui73}, \cite{TayBM79}.
Relevant theories and algorithms have been further developed in the 80's \cite{LevFul81}, \cite{SanSym86}
and in the 90's \cite{Nat95}, \cite{RaoKre98}, \cite{CouBre00}.
Recently, this subject became more popular in the signal processing community with the name of \textit{compressed sensing} \cite{Don06}.
In particular, the elegant analysis derived with modern probability theory \cite{Don06}, \cite{CanRT06}
provided performance guarantees for polynomial-time algorithms in terms of properties of random dictionaries.
These might be the most important contributions in recent years.

In some applications, there are multiple measurement vectors (righthand sides of the linear system of equations),
each corresponding to a different unknown vector, with the special property that all unknown vectors share a common support.
The sparse recovery problem with this joint structure in the sparsity pattern is called
\textit{joint sparse recovery} or the \textit{multiple measurement vector (MMV)} problem
and is often an easier problem with better performance.

In the mid 1990's, Bresler and Feng introduced ``spectrum-blind sampling'' \cite{FenBre96}, \cite{Feng97}, \cite{BreFen96}.
Their scheme enables sub-Nyquist minimum-rate sampling and perfect reconstruction of multi-band signals
(analog or discrete, in one or more dimensions) with unknown but sparse spectral support.
They reduced the spectrum-blind reconstruction problem to a finite-dimensional joint sparse recovery problem.
Mishali and Eldar elaborated the spectrum-blind sampling approach in \cite{MisEld09} (\textit{cf.} \cite{Bre08} for a more detailed discussion of the relationship of \cite{MisEld09} to the earlier work from the 1990's \cite{FenBre96}, \cite{Feng97}, \cite{BreFen96}), 
and they also proposed ``modulated wideband converter'' in \cite{MisEld10mwc},
which improves on spectrum-blind sampling by adding robustness against jitter.
The reconstruction in the modulated wideband converter too is reduced to a finite dimensional joint sparse recovery problem.
Rao \textit{et al.} (\textit{cf.} \cite{RaoKre98} and the references therein) introduced
a joint sparse recovery formulation and methods for the recovery of sparse brain excitations.
Obozinski \textit{et. al.} \cite{OboTJ10} formulated variable selection in multivariate regression as a joint sparse recovery problem.
The design matrix in regression corresponds to the linear system matrix of the joint sparse recovery
and the indicator function of the variables that mostly contribute to the given data is assumed to be sparse.
Malioutov \textit{et al.} posed the direction of arrival (DOA) estimation problem as a joint sparse recovery problem \cite{MalCW05}.
For the typically small number of sources in this problem, the indicator function of the quantized angles is modeled to be sparse.

Algorithms that exploit the structure in the sparsity pattern have been developed for the joint sparse recovery problem.
Bresler and Feng proposed to use a version of the MUSIC algorithm \cite{Sch86} from sensor array processing for the \textit{full row rank case} where the nonzero rows of the unknown matrix have full row rank \cite{FenBre96}, \cite{Feng97}, \cite{BreFen96}.
They also proposed methods based on a greedy search
inspired by the alternating projections algorithm \cite{ZisWax88} in DOA estimation, later dubbed orthogonal least squares (OLS) \cite{CheBL89}.
Existing solutions to the sparse recovery problem for the single measurement vector (SMV) case have been extended to the MMV case.
Greedy algorithms \cite{CotREK05}, \cite{TroGS06}, \cite{GRSV08} extend orthogonal matching pursuit (OMP) \cite{PatiRK93}
and convex optimization formulations with the mixed norm \cite{MalCW05}, \cite{Tro06}, \cite{CheHuo06}, \cite{OboWJ11}
extend the corresponding SMV solution such as basis pursuit (BP) \cite{CheDS01} and LASSO \cite{Tib96}.
Sparse Bayesian learning (SBL) \cite{WipRao04} has been also extended to the MMV case \cite{WipRao07}, \cite{ZhaRao11}.

Owing to the similarity between the joint sparse recovery problem and DOA estimation,
theories developed for DOA estimation affected those for joint sparse recovery.
For example, the fundamental limit on the performance of DOA estimation \cite{WaxZis89} also applies to joint sparse recovery.
Wax and Ziskind \cite{WaxZis89} showed the condition for the unique identification of DOA in the sensor array processing,
which has been applied to joint sparse recovery by Feng and Bresler \cite{FenBre96}
to determine the condition for the unique identification of the support.
The condition has been further studied in more general settings \cite{CotREK05}, \cite{CheHuo06}.
MUSIC applied to joint sparse recovery \cite{FenBre96} was the first method
that was guaranteed with the tightest sufficient condition,
which also coincides with a necessary condition required for the support identification by any method.
However, the guarantee only applies to the case
where the nonzero rows of the unknown matrix have full row rank and there is no noise in the measurement vectors.

Performance guarantees of greedy algorithms and of convex optimization formulations for joint sparse recovery
have been also studied extensively in the literature \cite{Tro04}, \cite{TroGS06}, %greedy
\cite{CheHuo06}, \cite{Tro06}, % convex
\cite{GRSV08}, \cite{FleRan09NIPS}, \cite{DavWak10}, % greedy
\cite{VanFri10}. % convex
The guarantees of such methods have not been proved to be strictly better than the guarantees for the SMV case.
Moreover, unlike the guarantee of MUSIC, such methods are not guaranteed with the minimal requirement
for the full row rank case in the absence of noise.

Performance guarantees aside,
the empirical performance and computational cost of any method are of key importance
and usually determines its adoption in practice.
Empirically, the optimization schemes with diversity measures (\textit{e.g.}, the mixed norm) perform better than greedy algorithms.
In particular, the rate of exact support recovery in existing algorithms 
does not improve with increasing rank of the unknown matrix beyond a certain level.
Furthermore, under unfavorable settings such as rank-defect or ill-conditioning,
existing algorithms for joint sparse recovery, while not failing, are far from achieving the guarantee of MUSIC for the full row rank case.

While the optimization scheme with diversity measures perform better empirically than the greedy algorithms,
this improved performance comes at a much higher computational cost.
Convex optimization formulations \cite{MalCW05}, \cite{Tro06}, \cite{CheHuo06}, \cite{OboWJ11}
are usually cast as second order cone programming (SOCP),
which is more difficult to solve
compared to its analogues in the SMV case, which are cast as linear programming (LP) or quadratic programming (QP). 
In contrast, greedy algorithms and MUSIC are computationally efficient.
As a summary, none of the listed methods enjoys both good empirical performance and computational speed at the same time.

In view of the various drawbacks of the existing algorithms for joint sparse recovery, MUSIC, when it works, is extremely attractive.
In a favorable setting, where the matrix composed of the nonzero rows of the unknown signal matrix has full row rank,
MUSIC is guaranteed to recover the support and hence provides a guarantee with minimal requirement.
Moreover, MUSIC is highly efficient computationally.
However, the full row rank condition is often violated in practice.
For example, if the number of measurement vectors $N$ is smaller than the sparsity level $s$,
then no more than $N$ rows can be linearly independent, and the nonzero rows do not have full row rank.
In other applications, such as spectrum-blind sampling or the DOA problem, $N$ is large or even infinite.
Even in this case though, the rank might be smaller than $s$, or the submatrix of nonzero rows can be ill-conditioned.
For example, in the DOA problem, correlation between sources or multi-path propagation can cause a large condition number.
It is well-known that MUSIC fails in this practically important ``rank-defective'' case
and this has motivated numerous attempts to overcome this problem, without resorting to infeasible multi-dimensional search.
However, all these previous methods use special structure of the linear system -- such as shift invariance that enables to apply so-called spatial smoothing \cite{ShaWK85}.
Previous extension of MUSIC are therefore not applicable to the general joint sparse recovery problem.

The main contributions of this paper are summarized as follows.
First, we propose a new class of algorithms, \textit{subspace-augmented MUSIC} (SA-MUSIC)
that overcome the limitations of existing algorithms and provide the best of both worlds:
good empirical performance at all rank conditions; and efficient computation.
In particular, SA-MUSIC algorithms improve on MUSIC so that the support is recovered even in the case
that the unknown matrix has rank-defect and/or ill-conditioning.
Compared to MUSIC \cite{FenBre96}, in the presence of a rank-defect,
SA-MUSIC has additional steps of partial support recovery and subspace augmentation.
Combined with partial support recovery by the subspace-based greedy algorithms also introduced in this paper,
SA-MUSIC provides a computationally efficient solution to joint sparse recovery with a performance guarantee.
In fact, the computational requirements of SA-MUSIC algorithms are similar to those of greedy algorithms and of MUSIC.
Secondly, we derive explicit conditions that guarantee each step of SA-MUSIC for the noisy and/or rank-defective case.
The performance is analyzed in terms of a property \cite{EldRau09},
which is a local version of the restricted isometry property (RIP) \cite{CanTao05}.
We call this property the weak-1 restricted isometry property (weak-1 RIP).
Most importantly, compared to the relevant work \cite{KimLY10} with similar but independently developed ideas,
the analysis in this paper is non-asymptotic
and applies to wider class of matrices including Gaussian, random Fourier, and incoherent unit-norm tight frames.

Contributions of independent interest include
two new subspace-based greedy algorithms for joint sparse recovery with performance guarantees,
extension of the analysis of MUSIC for joint sparse recovery to the noisy case with imperfect subspace estimation,
and non-asymptotic analysis of subspace estimation from finitely many snapshots.
The latter analysis is different from previous analysis of subspace methods, 
which were based on the law of large numbers, asymptotic normality, or low order expansion.

The remainder of this paper is organized as follows.
After introducing notations in Section~\ref{sec:notation},
the joint sparse recovery problem is stated in Section~\ref{sec:problemstatement}.
MUSIC for joint sparse recovery is reviewed in Section~\ref{sec:music} with discussion that motivates the current work.
We also propose an algorithm for signal subspace estimation in Section~\ref{sec:music}.
In Section~\ref{sec:samusic}, we propose SA-MUSIC and subspace-based greedy algorithms.
We review the notion of the weak-1 RIP in Section~\ref{sec:w1rip}.
The weak-1 RIP analysis of various matrices that arise commonly in compressed sensing is given without any ambiguous constant,
which might be of independent interest.
In Section~\ref{sec:analysisww1rip},
we provide non-asymptotic analysis of the algorithms of Sections~\ref{sec:music} and \ref{sec:samusic} by using the weak-1 RIP.
In Section~\ref{sec:estss}, we analyze subspace estimation using a random signal model.
The empirical performance of SA-MUSIC is compared to other methods in Section~\ref{sec:numerical_results}
and the relation to relevant works is discussed in Section~\ref{sec:discussion}.

%%%%%%%%%%%%
\section{Notation}\label{sec:notation}
%%%%%%%%%%%%

Symbol $\mathbb{K}$ denotes a scalar field, which is either the real field $\mathbb{R}$ or the complex field $\mathbb{C}$.
The vector space of $d$-tuples over $\mathbb{K}$ is denoted by $\mathbb{K}^d$ for $d \in \mathbb{N}$ where $\mathbb{N}$ is the set of natural numbers (excluding zero).
Similarly, for $m,n \in \mathbb{N}$, the vector space of $m \times n$ matrices over $\mathbb{K}$ is denoted by $\mathbb{K}^{m \times n}$.

We will use various notations on a matrix $A \in \mathbb{K}^{m \times n}$.
The range space spanned by the columns of $A$ will be denoted by $\mathcal{R}(A)$.
The Hermitian transpose of $A$ will be denoted by $A^*$.
The $j$-th column of $A$ is denoted by $a_j$ and the submatrix of $A$ with columns indexed by $J \subset [n]$ is denoted by $A_J$, 
where $[\ell]$ denotes the set $\{1,\ldots,\ell\}$ for $\ell \in \mathbb{N}$.
The $k$-th row of $A$ is denoted by $a^k$, 
and the submatrix of $A$ with rows indexed by $K \subset [m]$ is denoted by $A^K$.
Symbol $e_k$ will denote the $k$-th standard basis vector of $\mathbb{K}^d$,
where $d$ is implicitly determined for compatibility.
The $k$-th largest singular value of $A$ will be denoted by $\sigma_k(A)$.
For Hermitian symmetric $A$, $\lambda_k(A)$ will denote the $k$-the largest eigenvalue of $A$.
The Frobenius norm and the spectral norm of $A$ are denoted by $\norm{A}_F$ and $\norm{A}$, respectively.
For $p,q \in [1,\infty]$, the mixed $\ell_{p,q}$ norm of $A$ is defined by
\begin{equation*}
\norm{A}_{p,q} \triangleq
\left\{
\begin{array}{ll}
\displaystyle \left( \sum_{k=1}^m \norm{a^k}_p^q \right)^{\frac{1}{q}} & \text{if}~ q < \infty, \\
\displaystyle \max_{k \in [m]} \norm{a^k}_p & \text{else}.
\end{array}
\right.
\end{equation*}

The inner product is denoted by $\langle \cdot, \cdot \rangle$.
The embedding Hilbert space, where the inner product is defined, is not explicitly mentioned when it is obvious from the context.

For a subspace $S$ of $\mathbb{K}^d$, matrices $P_S \in \mathbb{K}^{d \times d}$ and $P_S^\perp \in \mathbb{K}^{d \times d}$ denote the orthogonal projectors onto $S$ and its orthogonal complement $S^\perp$, respectively.

For two matrices $A$ and $B$ of the same dimension, $A \geq B$ if and only if $A - B$ is positive semidefinite.

Symbols $\mathbb{P}$ and $\mathbb{E}$ will denote the probability and the expectation with respect to a certain distribution.
Unless otherwise mentioned, the distribution shall be obvious from the context.

%%%%%%%%%%%%
\section{Problem Statement}\label{sec:problemstatement}
%%%%%%%%%%%%

A vector $x \in \mathbb{K}^n$ is \textit{$s$-sparse} if it has at most $s$ nonzero components.
The \textit{support} of $x$ is defined as the set of the indices of nonzero components.
The \textit{sparsity level} of $x$ is defined as the number of nonzero components of $x$.
The \textit{sparse recovery problem} is to reconstruct an $s$-sparse vector $x_0 \in \mathbb{K}^n$
from its linear measurement vector $y \in \mathbb{K}^m$ through sensing matrix $A \in \mathbb{K}^{m \times n}$.
In particular, if there is no noise in $y$, then $x_0$ is a solution to the linear system $A x = y$.
In this case, under certain conditions on $A$, unknown vector $x_0$ is recovered as the unique $s$-sparse solution.
For example, any $s$-sparse $x_0$ is recovered if and only if any $2s$ columns of $A$ are linearly independent \cite{DonEla03}.
For $x_0$ to be the unique $s$-sparse solution to $A x = y$,
it is necessary that submatrix $A_{J_0}$ have full column rank, where $J_0$ is the support of $x_0$.
Otherwise, there exists another $s$-sparse solution and this contradicts uniqueness.
Note that once the support of $x_0$ is determined from $(y,A)$, then $x_0$ is easily computed as $x_0 = A_{J_0}^\dagger y$
where $A_{J_0}^\dagger \in \mathbb{K}^{s \times m}$ denotes the Moore-Penrose pseudo inverse of $A_{J_0}$.
Therefore, the key step in solving the sparse recovery problem is the identification of the support.

In practice, the measurement vector $y$ is perturbed by noise.
Usually, we assume that $y$ is given by
\begin{equation*}
y = A x_0 + w
\end{equation*}
with additive noise $w \in \mathbb{K}^m$.
In this case, $x_0$ is no longer a solution to the linear system $A x = y$.
Instead, minimizing $\norm{y - A x}_2$ with the sparsity constraint that $x$ is $s$-sparse provides a solution to the sparse recovery problem.
Alternatively, various convex optimization methods with sparsity inducing metrics such as the $\ell_1$ norm have been proposed.
However, the solution provided by such methods is not exactly $s$-sparse in the presence of noise.
In some applications, the support has important physical meaning and hence the identification of the support is explicitly required.
For example, in imaging applications of compressed sensing the support corresponds to the location of the target object, and in sparse linear regression the most contributing variables are identified by the support (\textit{cf.} \cite{TanNeh10}).
In such applications, unless the solution obtained to the sparse recovery problem is exactly $s$-sparse,
a step of thresholding the obtained solution to the nearest $s$-sparse vector is necessary.
For this reason, in this paper, the success of the sparse recovery problem is defined as the exact identification of the support of $x_0$.

Let us now turn to the main problem of this paper, where there exist multiple sparse signal vectors
$\{x_i\}_{i=1}^N \subset \mathbb{K}^n$ that share the same (or similar) sparsity pattern(s)
and the measurement vectors $\{y_i\}_{i=1}^N \subset \mathbb{K}^m$ are obtained through a common sensing matrix $A \in \mathbb{K}^{m \times n}$.
We assume that the union of the supports of the $x_i$ for $i=1,\ldots,N$ has at most $s$ elements.
Then, $X_0 = [x_1,\ldots,x_N] \in \mathbb{K}^{n \times N}$ has at most $s$ nonzero rows and is called \textit{row $s$-sparse}.
The \textit{row support} of $X_0$ is defined as the set of indices of nonzero rows.
The \textit{joint sparse recovery problem} is to find the row support of the unknown signal matrix $X_0$
from the matrix $Y \in \mathbb{K}^{m \times N}$ with multiple measurement vectors (MMV) given by
\begin{equation*}
Y = A X_0 + W
\end{equation*}
with common sensing matrix $A \in \mathbb{K}^{m \times n}$ and with perturbation $W \in \mathbb{K}^{m \times N}$.
Let $J_0$ denote the row support of $X_0$.
Then, $A X_0$ is compactly rewritten as $A_{J_0} X_0^{J_0}$
where $X_0^{J_0}$ is the matrix composed of the nonzero rows of $X_0$, and
$A_{J_0}$ is the submatrix of $A$ with the corresponding columns.
The prior knowledge that $X_0$ is row $s$-sparse will be assumed
\footnote{
This is only for convenience of the analysis but is not a limitation of the proposed algorithms.
See~\ref{subsec:ssomp} for more detailed discussion.
}.
An important parameter in the problem will be the rank of the unknown signal matrix $X_0$,
$\rank(X_0) = \rank(X_0^{J_0})$, which will be assumed unknown as well.
When matrix $X_0^{J_0}$ has full row rank,
$\rank(X_0^{J_0})$ assumes its maximum value, $\rank(X_0^{J_0}) = s$, and we will refer to this as the \textit{full row rank case}.
This is the case preferred by the algorithms in this paper.
Otherwise, $\rank(X_0^{J_0}) < s$, considered as violation of the full row rank case, will be called the \textit{rank-defective case}.

%%%%%%%%%%%%
\section{MUSIC Revisited}\label{sec:music}
%%%%%%%%%%%%

The similarity between the joint sparse recovery (or MMV) problem and
the direction of arrival (DOA) estimation problem in sensor array processing
has been well studied before (\textit{e.g.} \cite{FenBre96}).
In particular, it has been shown that
the joint sparse recovery problem can be regarded as a special case of the DOA problem with discretized angles \cite{FenBre96}.
Through this analogy between the two problems,
the algorithms and their analysis developed for the DOA problem have been applied to the joint sparse recovery problem \cite{FenBre96}.
In this section, we review a subspace-based algorithm proposed by Feng and Bresler \cite{FenBre96}, on which our new algorithm in Section~\ref{sec:samusic} improves.
We also elaborate the subspace-based algorithm in \cite{FenBre96} to work without ideal assumptions.

%%%%%%
\subsection{MUSIC for the Joint Sparse Recovery Problem Revisited}\label{subsec:music}
%%%%%%
Inspired by the success of the MUSIC algorithm \cite{Sch86} in sensor array processing,
Bresler and Feng \cite{FenBre96}, \cite{BreFen96}, \cite{Feng97}
proposed to use a version of MUSIC for joint sparse recovery.
As in the original MUSIC algorithm in sensor array processing \cite{Sch86},
the first step is to estimate the so-called \textit{signal subspace} $S$ defined by
\begin{equation*}
S \triangleq \mathcal{R}(A X_0) = \mathcal{R}(A_{J_0} X_0^{J_0})
\end{equation*}
from the \textit{snapshot matrix}\footnote{
We adopt the terminology from the sensor array processing literature.
To emphasize the analogy between the joint sparse recovery problem and DOA estimation, we also call each of the $N$ columns of $Y$ a snapshot.
Then, $N$ will denote the number of snapshots.
}
$Y = A_{J_0} X_0^{J_0} + W \in \mathbb{K}^{m \times N}$.
In sensor array processing, MUSIC \cite{Sch86} estimates $S$ using the eigenvalue decomposition (EVD) of $\frac{YY^*}{N}$.
The same method is applied to the joint sparse recovery problem under the assumption that
\begin{equation}
\frac{YY^*}{N} = \frac{A_{J_0} X_0^{J_0} (X_0^{J_0})^* A_{J_0}^*}{N} + \sigma_w^2 I,
\label{eq:asympN}
\end{equation}
which is achieved with statistical assumptions on $A_{J_0} X_0^{J_0}$ and $W$
under the asymptotic in $N$ (with infinitely many snapshots) \cite{Feng97}.
It is also assumed that $A_{J_0} \in \mathbb{K}^{m \times s}$ has full column rank and $X_0^{J_0} \in \mathbb{K}^{s \times N}$ has full row rank.
Then, the dimension of $S$ coincides with $\rank(X_0^{J_0}) = s$.
The assumed relation (\ref{eq:asympN}) implies that the smallest eigenvalue of $\frac{YY^*}{N}$ has multiplicity $m - s$,
whence the dimension of $S$ is exactly determined.
The signal subspace $S$ is then exactly computed as the invariant subspace spanned by the $s$ dominant eigenvectors of $\frac{YY^*}{N}$
since the noise part $\sigma_w^2 I$ in (\ref{eq:asympN}) only shifts the eigenvalues of $\frac{A_{J_0} X_0^{J_0} (X_0^{J_0})^* A_{J_0}^*}{N}$.
The joint sparse recovery problem then reduces to the case where the subspace estimation is error-free
and hence the subsequent analysis of the support recovery in \cite{FenBre96}, \cite{BreFen96}, \cite{Feng97} considered this error-free case
\footnote{
Obviously, for the noiseless case, the subspace estimation is error-free without relying on the asymptotic $N$.
}.

Given the signal subspace $S$, MUSIC for joint sparse recovery
identifies the row-support $J_0$ as the set of the indices $k$ of columns of $A$ such that $P_S^\perp a_k = 0$.
In other words, MUSIC accepts the index $k$ as an element of the support if $a_k \in S$ and rejects it otherwise.
In the remainder of this paper, the acronym MUSIC will denote the version for joint sparse recovery \cite{FenBre96}
rather than the original MUSIC algorithm for sensor array processing \cite{Sch86}.

A sufficient condition that guarantees the success of MUSIC for the special case where $X_0^{J_0}$ has full row rank,
is given in terms of the Kruskal rank \cite{Kru77} defined below.
\begin{definition}
The Kruskal rank of a matrix $A$, denoted by $\krank(A)$, is the maximal number $k$ such that
any $k$ columns of $A$ are linearly independent.
\end{definition}
\begin{proposition}[\cite{FenBre96}, \cite{Feng97}]
Let $J_0$ be an arbitrary subset of $[n]$ with $s$ elements.
Suppose that $X_0 \in \mathbb{K}^{n \times N}$ is row $s$-sparse with support $J_0$ and $\rank(X_0^{J_0}) = s$.
If $A$ satisfies
\begin{equation}
\krank(A) = s,
\label{eq:prop:musicFenBre:cond}
\end{equation}
then
\begin{equation}
P_S^\perp a_k = 0
\label{eq:prop:musicFenBre:res}
\end{equation}
for $S \triangleq \mathcal{R}(A_{J_0} X_0^{J_0})$ if and only if $k \in J_0$.
\label{prop:musicFenBre}
\end{proposition}
The following result directly follows from Proposition~\ref{prop:musicFenBre}.
\begin{corollary}
Under the conditions on $A$ and $X_0^{J_0}$ in Proposition~\ref{prop:musicFenBre},
given the exact signal subspace $S$, MUSIC is guaranteed to recover $J_0$.
\end{corollary}
\begin{remark}
In \cite{Feng97}, Condition (\ref{eq:prop:musicFenBre:cond}) was stated in terms of the ``universality level'' of $A$,
which is in fact identical to $\krank(A)$.
A similar notion called the ``spark'' of $A$ was later introduced in \cite{DonEla03},
which is related to the Kruskal rank by $\spark(A) = \krank(A) + 1$.
\end{remark}
\begin{remark}
Condition (\ref{eq:prop:musicFenBre:cond}) is satisfied by certain $A \in \mathbb{K}^{m \times n}$ with $m > s$.
For example, it has been shown that the matrix composed of 
any consecutive $m$ rows of the $n \times n$ DFT matrix,
which corresponds to the ``bunched sampling pattern'' in spectrum-blind sampling,
satisfies (\ref{eq:prop:musicFenBre:cond}) when $m > s$ \cite{Feng97}.
\end{remark}

MUSIC with its performance guarantee in Proposition~\ref{prop:musicFenBre}
is remarkable in the context of compressive sensing for the following reasons.
First, MUSIC is a polynomial-time algorithm with a performance guarantee
under a condition that coincides with the necessary condition $m > s$ for unique recovery (by \textit{any} algorithm, no matter how complex)
\footnote{
Since the mid 1990's, when these results (for what became known later as compressive sampling) \cite{FenBre96}, \cite{BreFen96}, \cite{Feng97} were published, until recently, MUSIC was the only polynomial-time algorithm with such a guarantee.
Recent results \cite{DavEld10} showed another (greedy) algorithm with the same guarantee.
}.
Second, MUSIC is simple and cheap, involving little more than a single EVD of the data covariance matrix.
In fact, efficient methods for partial EVD, or other rank-revealing decompositions can further reduce the cost
\footnote{
The rank revealing decompositions can be computed by the Lanczos method \cite{GolVan96} for the truncated singular value decomposition (SVD).
If the matrix is large, randomized algorithms \cite{Har06}, \cite{WLRT08} can be also used to compute the truncated SVD.
}.

Unfortunately, as is well-known in the sensor array processing literature \cite{Sch86}, \cite{StoNeh89},
and also demonstrated by numerical experiments later in Section~\ref{sec:numerical_results},
MUSIC is prone to failure when $X_0^{J_0}$ does not have full row rank, or when it is ill-conditioned in the presence of noise.
In sensor array processing, this is known as the ``source coherence problem'' \cite{KriVib96},
and (with the exception of the case of a Vandermonde matrix $A$) no general solution to this problem are known.
This motivates our work to propose a new subspace-based method for joint sparse recovery that improves on MUSIC.

For the noisy case, the analysis of signal subspace estimation based on the asymptotic in $N$ is not practical.
In particular, from a perspective of compressed sensing (with joint sparsity),
recovery of the support from a finite (often small) number of snapshots is desired.
In the next subsection, we propose a subspace estimation scheme that works with finitely many snapshots,
which will be applied to both MUSIC and the new subspace-based methods in this paper.
In Section~\ref{subsec:musicwn}, we formalize the MUSIC algorithm for support recovery in the presence of a perturbation in the estimate of $S$.
This will lay the ground for the subsequent analysis of MUSIC in the noisy scenario, and for its extension in the same scenario to SA-MUSIC.

%%%%%%
\subsection{Signal Subspace Estimation from Finitely Many Snapshots}\label{subsec:estss}
%%%%%%
We study the problem of signal subspace estimation from finitely many snapshots in this subsection.
For later use in other subspace-based algorithms in Section~\ref{sec:samusic},
we weaken the assumptions in the previous section.
In particular, we assume finite $N$ and no longer assume that $X_0^{J_0}$ has full row rank.
We also assume that the columns of the noise matrix $W$ are i.i.d. random vectors with white spectrum,
\textit{i.e.}, $\mathbb{E} \frac{WW^*}{N} = \sigma_w^2 I_m$.
(Otherwise, the standard pre-whitening schemes developed in sensor array processing may be applied prior to subspace estimation.)

When $A_{J_0} X_0^{J_0}$ is ill-conditioned,
the last few singular values of $A_{J_0} X_0^{J_0}$ are small, making the estimation of $S \triangleq \mathcal{R}(A_{J_0} X_0^{J_0})$
highly sensitive to the noise in the snapshots.
To improve the robustness against noise, instead of estimating the whole subspace $S$, we only estimate an $r$-dimensional subspace of $S$ for $r < s$.
The dimension $r$ is determined so that the gap between $\sigma_r(A_{J_0} X_0^{J_0})$ and $\sigma_{r+1}(A_{J_0} X_0^{J_0})$ is significant.

We propose and analyze a simple scheme that determines the dimension $r$ by thresholding the eigenvalues of $\frac{YY^*}{N}$
so that the estimated signal subspace $\widehat{S}$
spanned by the $r$ dominant eigenvectors of $\frac{YY^*}{N}$ is close to an $r$-dimensional subspace of $S$.
The procedure for estimating the signal subspace and its dimension is described next.

Given the snapshot matrix $Y = A_{J_0} X_0^{J_0} + W$, we compute its sample covariance matrix $\Gamma_Y$ defined by
\begin{equation*}
\Gamma_Y \triangleq \frac{Y Y^*}{N}.
\end{equation*}
Then, we compute a biased matrix $\widehat{\Gamma}$ by
\begin{equation*}
\widehat{\Gamma} \triangleq \Gamma_Y - \lambda_m(\Gamma_Y) I_m.
\end{equation*}
Note that $\widehat{\Gamma}$ and $\Gamma_Y$ have the same eigenvectors.
Recall that our goal is to find an $r$-dimensional subspace $\widehat{S}$ for some $r \leq s$ from $\widehat{\Gamma}$
such that there exists an $r$-dimensional subspace $\bar{S}$ of the signal subspace $S \triangleq \mathcal{R}(A_{J_0} X_0^{J_0})$
satisfying $\norm{P_{\widehat{S}} - P_{\bar{S}}} \leq \eta$ for small $\eta$.
For better performance of support recovery by algorithms in Section~\ref{sec:samusic}, larger $r$ is preferred.

For an ideal case where (\ref{eq:asympN}) holds,
$\widehat{\Gamma}$ reduces to $\Gamma_S$ defined by
\begin{equation*}
\Gamma_S \triangleq \frac{A_{J_0} X_0^{J_0} (X_0^{J_0})^* A_{J_0}^*}{N}.
\end{equation*}
Since $S = \mathcal{R}(\Gamma_S)$, by setting $r$ to $\rank(\Gamma_S)$, we compute $S$ itself rather than a proper subspace of $S$.

For finite $N$, usually, the cross correlation terms in the sample covariance matrix $\Gamma_Y$
between the noise term $W$ and the signal term $A_{J_0} X_0^{J_0}$ are smaller than the autocorrelation terms.
Since the autocorrelation term of $W$ is nearly removed in $\widehat{\Gamma}$,
we will show that it is likely that $D \triangleq \widehat{\Gamma} - \Gamma_S$ is small in the spectral norm.
In particular, let $\widehat{S}$ be the subspace spanned by the $r$ dominant eigenvectors of $\widehat{\Gamma}$; 
if $\norm{D}$ is small compared to the gap between $\lambda_r(\widehat{\Gamma})$ and $\lambda_{r+1}(\widehat{\Gamma})$, 
then there exists an $r$-dimensional subspace $\bar{S}$ of $S$ with small $\norm{P_{\widehat{S}} - P_{\bar{S}}}$. 
Since $\norm{D}$ is not available, we determine the dimension $r$ by simple thresholding.
More specifically, given a threshold $\tau > 0$,
the dimension of $\widehat{S}$ is determined as the maximal number $r$ satisfying
\begin{equation}
\frac{\lambda_r(\widehat{\Gamma}) - \lambda_{r+1}(\widehat{\Gamma})}{\lambda_1(\widehat{\Gamma})}
\geq \tau
> \frac{\lambda_k(\widehat{\Gamma}) - \lambda_{k+1}(\widehat{\Gamma})}{\lambda_1(\widehat{\Gamma})},
\quad \forall k > r.
\label{eq:estss:thresrank}
\end{equation}

When $A_{J_0} X_0^{J_0}$ (and hence $\Gamma_S$) is ill-conditioned,
it is likely that the gap between the consecutive eigenvalues $\lambda_{r_0}(\widehat{\Gamma})$ and $\lambda_{r_0+1}(\widehat{\Gamma})$
where $r_0 = \rank(\Gamma_S)$ is small compared to $\norm{D}$,
which is roughly proportional to $\sigma_w^2 / \lambda_1(\widehat{\Gamma})$.
The aforementioned increased robustness to noise is provided by choosing $r < r_0$ so
that the gap $\lambda_r(\widehat{\Gamma}) - \lambda_{r+1}(\widehat{\Gamma})$ is large.,
which will in turn reduce the estimated subspace dimension $r$.
More sophisticated methods for determining $r$ are possible, but this simple method suffices for our purposes
and is amenable to analysis (see Section~\ref{sec:estss}).
This algorithm for estimating the signal subspace and its dimension is summarized as Algorithm~\ref{alg:estss}.
\begin{algorithm}
\caption{Signal Subspace Estimation}
\begin{algorithmic}[1]
\INPUT $Y \in \mathbb{K}^{m \times N}$ and $\tau > 0$.
\OUTPUT $r \in \mathbb{N}$ and $P_{\widehat{S}} \in \mathbb{K}^{m \times m}$
\STATE $\Gamma_Y \leftarrow \frac{Y Y^*}{N}$;
\STATE $\widehat{\Gamma} \leftarrow \Gamma_Y - \lambda_m(\Gamma_Y) I_m$;
\STATE $r \leftarrow m-1$;
\WHILE{$\lambda_r(\widehat{\Gamma}) - \lambda_{r+1}(\widehat{\Gamma}) < \tau \lambda_1(\widehat{\Gamma})$}
\STATE $r \leftarrow r-1$;
\ENDWHILE
\STATE $U \leftarrow$ $r$ dominant eigenvectors of $\widehat{\Gamma}$;
\STATE $P_{\widehat{S}} \leftarrow U U^*$;
\RETURN $r$, $P_{\widehat{S}}$;
\end{algorithmic}
\label{alg:estss}
\end{algorithm}

%%%%%%
\subsection{MUSIC Applied to an Inaccurate Estimate of the Signal Subspace}\label{subsec:musicwn}
%%%%%%
In the presence of a perturbation in the estimated signal subspace $\widehat{S}$,
MUSIC finds the set $\widehat{J}$ of $s$ indices that satisfy
\begin{equation}
\min_{k \in \widehat{J}} \frac{\norm{P_{\widehat{S}} a_k}_2}{\norm{a_k}_2}
> \max_{k \in [n] \setminus \widehat{J}} \frac{\norm{P_{\widehat{S}} a_k}_2}{\norm{a_k}_2}.
\label{eq:musicthres2}
\end{equation}
The corresponding algorithm is summarized in Algorithm~\ref{alg:music}.
To provide an intuition for the selection criterion in (\ref{eq:musicthres2}), we use the notion of the ``angle function'' \cite{Wed83}.
\begin{definition}[\cite{Wed83}]
The angle function between two subspaces $S_1$ and $S_2$ is defined by
\begin{equation}
\sphericalangle_2(S_1,S_2)
\triangleq \sin^{-1}\left( \min\{\norm{P_{S_1}^\perp P_{S_2}} ,~  \norm{P_{S_1}^\perp P_{S_2}}\} \right).
\label{eq:def_dwed}
\end{equation}
\end{definition}
\begin{remark}
The angle function $\sphericalangle_2(S_1,S_2)$ is different from the largest principal angle between $S_1$ and $S_2$ \cite{GolVan96}.
Unlike the largest principal angle, the angle function satisfies the metric properties even when $S_1$ and $S_2$ have different dimensions.
In particular, when $\dim(S_1) \geq \dim(S_2)$, the expression on the right hand side of (\ref{eq:def_dwed}) reduces to
\begin{equation}
\sphericalangle_2(S_1,S_2)
= \sin^{-1} \left( \norm{P_{S_1}^\perp P_{S_2}} \right).
\label{eq:def_dwed2}
\end{equation}
\end{remark}

By (\ref{eq:def_dwed2}), the criterion in (\ref{eq:musicthres2}) is equivalent to
\begin{equation}
\max_{k \in \widehat{J}} \sphericalangle_2(\widehat{S},\mathcal{R}(a_k))
< \min_{k \in [n] \setminus \widehat{J}} \sphericalangle_2(\widehat{S},\mathcal{R}(a_k)).
\label{eq:musicthres1}
\end{equation}
That is, MUSIC finds, among all subspaces spanned by a single column of $A$, $s$ subspaces nearest to $S$ (in the angle function metric).

\begin{algorithm}
\caption{MUSIC}
\begin{algorithmic}[1]
\INPUT $Y \in \mathbb{K}^{m \times N}$, $A \in \mathbb{K}^{m \times n}$, $s \in \mathbb{N}$.
\OUTPUT $J \subset [n]$
\STATE $P_{\widehat{S}} \in \mathbb{R}^{m \times m}, r \in \mathbb{N} \leftarrow$ \textit{estimate signal subspace from $Y$};
\STATE $J \leftarrow \emptyset$;
\FOR{$\ell = 1,\ldots,n$}
\STATE $\zeta_\ell \leftarrow \norm{P_{\widehat{S}} a_\ell}_2/\norm{a_\ell}_2$
\ENDFOR
\STATE $J \leftarrow$ indices of the $s$-largest $\zeta_\ell$'s;
\RETURN $J$
\end{algorithmic}
\label{alg:music}
\end{algorithm}

%%%%%%%%%%%%
\section{Subspace-Augmented MUSIC}\label{sec:samusic}
%%%%%%%%%%%%

%%%%%%
\subsection{MUSIC Applied to an Augmented Signal Subspace}\label{subsec:samusic}
%%%%%%
When $X_0^{J_0}$ has full row rank, the signal subspace $S \triangleq \mathcal{R}(A_{J_0} X_0^{J_0})$ coincides with $\mathcal{R}(A_{J_0})$.
In this case, given the exact $S$, MUSIC is guaranteed to recover the support $J_0$
because \textit{(i)} $a_k \in S$ for all $k \in J_0$; and \textit{(ii)} $a_k \not\in S$ for all $k \in [n] \setminus J_0$,
which is implied by $\krank(A) = s+1$ (Proposition~\ref{prop:musicFenBre}).
However, in the rank-defective case, when $X_0^{J_0}$ does not have full row rank,
\textit{i.e.}, when $\rank(X_0^{J_0})$ is strictly smaller than the sparsity level $s$,
we have $\dim(S) \leq \rank(X_0^{J_0}) < s = \dim(\mathcal{R}(A_{J_0})$.
Therefore, $S$ is a proper subspace of $\mathcal{R}(A_{J_0})$ and it may happen that $a_k \not\in S$ for some (or all) $k \in J_0$.
This will cause MUSIC to miss valid components of $J_0$.
Because, in the presence of noise (imperfect subspace estimate), MUSIC selects, by (\ref{eq:musicthres1}),
the $s$ indices $k \in [n]$ for which $\mathcal{R}(a_k)$ is closets (in the sense of the angle function) to $S$,
this may result in the selection of spurious indices into the estimate of $J_0$.
This explains the well-known fact that in the rank-defective case MUSIC is prone to failure.

Subspace-augmented MUSIC overcomes the limitation of MUSIC to the full row rank case by capitalizing on the following observation:
MUSIC fails in the rank-defective case because $S$ is a proper subspace of $\mathcal{R}(A_{J_0})$;
however, if another subspace $T$ that complements $S$ is given so that $S + T = \mathcal{R}(A_{J_0})$,
then MUSIC applied to the augmented subspace $S + T$ will be successful.

Unfortunately, in general, finding such an oracle subspace is not feasible.
The search procedure cannot even be enumerated.
However, if $X_0^{J_0}$, the matrix of nonzero rows of $X_0$, or more generally,
the subspace $\mathcal{R}(X_0^{J_0})$, satisfies a mild condition,
then the search may be restricted without loss of generality to subspaces spanned by $s-r$ columns of $A$.
The following proposition states this result.
\begin{definition}
Matrix $X$ is \textit{row-nondegenerate} if
\begin{equation}
\krank(X^*) = \rank(X).
\label{eq:def:nondeg}
\end{equation}
\end{definition}
\begin{remark}
Condition (\ref{eq:def:nondeg}) says that every $k$ rows of $X$ are linearly independent for $k \leq \rank(X)$.
This is satisfied by $X$ that is generic in the set of full rank matrices of the same size.
In fact, an even weaker requirement on $X$ suffices, as shown by the next argument that reduces the requirement to $\mathcal{R}(X)$.
\label{rem:x0spark}
\end{remark}
\begin{remark}
Condition (\ref{eq:def:nondeg}) is invariant to multiplication of $X$ by any full row rank matrix of compatible size on the right.
In particular, Condition (\ref{eq:def:nondeg}) holds
if and only if
\begin{equation}
\krank(Q^*) = \rank(X)
\label{eq:def:nondegU0}
\end{equation}
for any orthonormal basis $Q$ of $\mathcal{R}(X)$.
It follows that (\ref{eq:def:nondeg}) is a property of the subspace $\mathcal{R}(X)$.
Furthermore, (\ref{eq:def:nondeg}) also implies that any $\widetilde{Q}$
such that $\mathcal{R}(\widetilde{Q}) \subset \mathcal{R}(X)$ is also row-nondegenerate (Lemma~\ref{lemma:fullkrank}).
\label{rem:u0sparkSP}
\end{remark}
\begin{remark}
The condition on $Q$ in (\ref{eq:def:nondegU0}) that any subset of rows of $Q$ up to size $\rank(X)$
are linearly independent is purely algebraic
and can be paraphrased to say that the rows of $Q$ are in general position. This is a mild condition satisfied by generic $Q$.
For example, if the rows of $Q$ are independently distributed with respect to any absolutely continuous probability measure,
then (\ref{eq:def:nondegU0}) is satisfied with probability 1.
\label{rem:u0spark}
\end{remark}
\begin{remark}
Remarks~\ref{rem:x0spark}--\ref{rem:u0spark} validate the definition of Condition~\ref{eq:def:nondeg} as a row-nondegeneracy condition.
\label{rem:rownondeg}
\end{remark}

\begin{proposition}[Subspace Augmentation]
Suppose that $X_0 \in \mathbb{K}^{n \times N}$ is row $s$-sparse with support $J_0 \subset [n]$ and $X_0^{J_0}$ is row-nondegenerate.
Let $\bar{S}$ be an arbitrary $r$-dimensional subspace of $\mathcal{R}(A_{J_0} X_0^{J_0})$ where $r < s$.
Let $J_1$ be an arbitrary subset of $J_0$ with $s-r$ elements.
If $A_{J_0}$ has full column rank, then
\begin{equation}
\bar{S} + \mathcal{R}(A_{J_1}) = \mathcal{R}(A_{J_0}).
\label{eq:prop:augssp:res}
\end{equation}
\label{prop:augssp}
\end{proposition}
\begin{IEEEproof}
See Appendix~\ref{subsec:prop:augssp}.
\end{IEEEproof}
\begin{remark}
Note that $\dim(\mathcal{R}(A_{J_1})) \geq \dim(\mathcal{R}(A_{J_0})) - \dim(\bar{S}) = s - r$ is a necessary condition for (\ref{eq:prop:augssp:res}).
Therefore, $J_1$ should be a subset of $J_0$ with at least $s-r$ elements for the success of the subspace augmentation.
\end{remark}
\begin{remark}
The row-nondegeneracy condition on $X_0^{J_0}$ is a necessary condition
to guarantee (\ref{eq:prop:augssp:res}) for an arbitrary subset $J_1$ of $J_0$ with $s-r$ elements.
Suppose that $X_0^{J_0}$ fails to satisfy the row-nondegeneracy condition, \textit{i.e.}, $\krank((X_0^{J_0})^*) < \rank(X_0^{J_0})$.
By the assumption on $\bar{S}$,
there exists a row $s$-sparse matrix $U \in \mathbb{K}^{n \times r}$ with support $J_0$ such that $\bar{S} = \mathcal{R}(A_{J_0} U^{J_0})$.
Since $\bar{S}$ was an arbitrary $r$-dimensional subspace of $S$, without loss of generality, we may assume that $\krank((U^{J_0})^*) < r$.
By the projection update formula, it follows that
$P_{\mathcal{R}(A_{J_1}) + \bar{S}} = P_{\mathcal{R}(A_{J_1})} + P_{P_{\mathcal{R}(A_{J_1})}^\perp \bar{S}}$
and hence it suffices to show $\dim(P_{\mathcal{R}(A_{J_1})}^\perp \bar{S}) < r$ for the failure of (\ref{eq:prop:augssp:res}).
Since $\krank((U^{J_0})^*) < r$, there exists $J_1 \subset J_0$ of size $s-r$ such that $\rank(U^{J_0 \setminus J_1}) < r$.
Then, $\dim(P_{\mathcal{R}(A_{J_1})}^\perp \bar{S}) = \rank(P_{\mathcal{R}(A_{J_1})}^\perp A_{J_0} U^{J_0}) = \rank(P_{\mathcal{R}(A_{J_1})}^\perp A_{J_0 \setminus J_1} U^{J_0 \setminus J_1}) \leq \rank(U^{J_0 \setminus J_1}) < r$.
It follows that (\ref{eq:prop:augssp:res}) fails for this specific $J_1$.
Therefore, the row-nondegeneracy condition on $X_0^{J_0}$ is essential.
Furthermore, by Remarks \ref{rem:x0spark}--\ref{rem:rownondeg}, the row-nondegeneracy is a mild condition; it will be assumed to hold henceforth.
\end{remark}

Let $X_0^{J_0}$ be row-nondegenerate and suppose that an error-free estimate of $S$ is available.
In this case, Proposition~\ref{prop:augssp} implies that, given a correct partial support $J_1$ of size $s-r$,
MUSIC applied to the augmented subspace $S + \mathcal{R}(A_{J_1})$ enjoys the same guarantee
as MUSIC for the full row rank case (Proposition~\ref{prop:musicFenBre}).
We will see in Section~\ref{sec:analysisww1rip} that a similar statement applies even with an imperfect estimate $\widehat{S}$.

Based on the above result, we propose a class of methods for joint sparse recovery
called \textit{subspace-augmented MUSIC (SA-MUSIC)} consisting of the following steps:
\begin{enumerate}
\item Signal subspace estimation: compute an estimate $\widehat{S}$ of the signal subspace $S \triangleq \mathcal{R}(A_{J_0} X_0^{J_0}) = \mathcal{R}(A X_0)$.
\item Partial support recovery: compute a partial support $J_1$ of size $s-r$ from $\widehat{S}$ and $A$, where $r = \dim(\widehat{S})$ and $s$ is the sparsity level known \textit{a priori}.
\item Augment signal subspace: compute the \textit{augmented subspace} $\widetilde{S}$
    \begin{equation*}
    \widetilde{S} = \widehat{S} + \mathcal{R}(A_{J_1}).
    \end{equation*}
\item Support completion: complete $J_1$ to produce $J_0 \supset J_1$, by adding $r$ more support elements obtained by applying ``MUSIC'' to $\widetilde{S}$, that is, finding $\widehat{J} \subset [n]$ satisfying
    \begin{equation*}
    \min_{k \in \widehat{J} \setminus J_1} \frac{\norm{P_{\widetilde{S}} a_k}_2}{\norm{a_k}_2}
    > \max_{k \in [n] \setminus \widehat{J}} \frac{\norm{P_{\widetilde{S}} a_k}_2}{\norm{a_k}_2}.
    \end{equation*}
\end{enumerate}

The general SA-MUSIC algorithm is summarized as Algorithm~\ref{alg:samusic}.
An actual implementation might use orthonormal bases $\widehat{Q}$ and $\widetilde{Q}$ for the subspaces $\widehat{S}$
and $\widetilde{S}$ instead of constructing the projection operators $P_{\widehat{S}}$ and $P_{\widetilde{S}}$.
Step~\ref{alg:samusic:step:augss} could then be performed by a QR decomposition of matrix $[\widehat{Q}, A_{J_1}]$.

\begin{algorithm}
\caption{Subspace Augmented MUSIC}
\begin{algorithmic}[1]
\INPUT $Y \in \mathbb{K}^{m \times N}$, $A \in \mathbb{K}^{m \times n}$, $s \in \mathbb{N}$.
\OUTPUT $J \subset [n]$
\STATE $P_{\widehat{S}} \in \mathbb{R}^{m \times m}, r \in \mathbb{N} \leftarrow$ \textit{estimate signal subspace from $Y$};
\STATE $J_1 \leftarrow$ \textit{partial support recovery of size $s-r$};
\STATE $P_{\widetilde{S}} \leftarrow P_{\widehat{S}} + (P_{\widehat{S}}^\perp A_{J_1}) (P_{\widehat{S}}^\perp A_{J_1})^\dagger$; \label{alg:samusic:step:augss}
\FOR{$\ell \in [n] \setminus J_1$}
\STATE $\zeta_\ell \leftarrow \norm{P_{\widetilde{S}} a_\ell}_2/\norm{a_\ell}_2$
\ENDFOR
\STATE $J \leftarrow J_1 \cup \{\text{indices of the $r$-largest $\zeta_\ell$'s}\}$;
\RETURN $J$
\end{algorithmic}
\label{alg:samusic}
\end{algorithm}

A particular instance of the SA-MUSIC algorithm is specified by
the particular methods used for the steps of signal subspace estimation and partial support recovery.
For subspace estimation, in the analysis in Sections~\ref{sec:analysisww1rip} and \ref{sec:estss},
we will consider the EVD-based scheme in Section~\ref{subsec:estss}.
However, as mentioned earlier, the subspace estimation scheme is not restricted to the given method.
For example, if the noise $W$ is also sparse,
then robust principal component analysis (RPCA) \cite{CLMW10} will provide a better estimate of $S$ than the usual SVD.

The choice of method for partial support recovery is discussed in the next subsection.
Here we note some special cases.
As $r$ increases, the size of the partial support required in SA-MUSIC decreases.
For small $s-r$, we can use an exhaustive search over $J_1$,
the computational cost of which also decreases in $r$.
In particular, for the special case where $r = s$, the step of partial support recovery is eliminated, 
and SA-MUSIC reduces to MUSIC \cite{FenBre96}.

%%%%%%
\subsection{Partial Support Recovery with Practical Algorithms}\label{subsec:ssomp}
%%%%%%
When there is a ``rank-defect'' in $X_0^{J_0}$, \textit{i.e.}, $\rank(X_0^{J_0}) < s$,
SA-MUSIC requires partial support recovery of size $s-r$.
In addition to computational efficiency, a key desirable property of an algorithm to accomplish this is that
it solves the partial support recovery problem more easily than solving the full support recovery problem.

From this perspective, greedy algorithms for the joint sparse recovery problem are attractive candidates.
Both empirical observations and the performance guarantees in the sequel suggest that
the first few steps of greedy algorithms are more likely to succeed than the entire greedy algorithms.
In other words, greedy algorithms take advantage of the reduction to partial support recovery when they are combined with SA-MUSIC.

Any of the known greedy algorithms for joint sparse recovery may be used in SA-MUSIC, producing a different version of SA-MUSIC.
In particular, we may consider variations on orthogonal matching pursuit (OMP) \cite{PatiRK93},
such as MMV orthogonal matching pursuit (M-OMP) \cite{CotREK05}, simultaneous orthogonal matching pursuit (S-OMP) \cite{TroGS06},
or their generalization to $p$-SOMP \cite{GRSV08}. 
The $p$-SOMP algorithm incrementally updates the support by the following selection rule:
given an index set $J$ from the previous steps, the algorithm adds to $J$ the index $k$ that satisfies
\begin{equation}
k = \arg\max_{\ell \in [n] \setminus J} \norm{Y^* P_{\mathcal{R}(A_J)}^\perp a_\ell}_p
\label{eq:selpomp}
\end{equation}
where $p \in [1,\infty]$ is a parameter of the algorithm.
M-OMP and S-OMP correspond to 2-SOMP and 1-SOMP, respectively.

In particular, we propose and analyze two different algorithms for the partial support recovery step in SA-MUSIC.
The first is the \textit{signal subspace orthogonal matching pursuit} (SS-OMP) algorithm.
SS-OMP is a subspace-based variation of M-OMP (2-SOMP) that replaces the snapshot matrix $Y$ in (\ref{eq:selpomp})
by the orthogonal projector $P_{\widehat{S}}$ onto the estimated signal subspace. 
(Equivalently, $Y$ is replaced by an orthogonal basis matrix for the estimated signal subspace).
Hence, given the estimated support $J$ from the previous steps, SS-OMP updates $J$ by adding $k$ selected by
\begin{equation}
k = \arg\max_{\ell \in [n] \setminus J} \norm{P_{\widehat{S}} P_{\mathcal{R}(A_J)}^\perp a_\ell}_2.
\label{eq:updatessomp}
\end{equation}
The complete SS-OMP is summarized as Algorithm~\ref{alg:ssomp}.

\begin{algorithm}
\caption{Signal Subspace Orthogonal Matching Pursuit (SS-OMP)}
\begin{algorithmic}[1]
\INPUT $Y \in \mathbb{K}^{m \times N}$, $A \in \mathbb{K}^{m \times n}$, $s \in \mathbb{N}$.
\OUTPUT $J \subset [n]$
\STATE $P_{\widehat{S}} \in \mathbb{R}^{m \times m}, r \in \mathbb{N} \leftarrow$ \textit{estimate signal subspace from $Y$};
\STATE $J \leftarrow \emptyset$;
\WHILE{$|J| < s$}
\STATE $\displaystyle k \leftarrow \arg\max_{\ell \in [n] \setminus J} \norm{P_{\widehat{S}} P_{\mathcal{R}(A_J)}^\perp a_\ell}_2$;
\STATE $J \leftarrow J \cup \{k\}$;
\ENDWHILE
\RETURN $J$
\end{algorithmic}
\label{alg:ssomp}
\end{algorithm}

The second algorithm we propose for the partial support recovery step in SA-MUSIC is
\textit{signal subspace orthogonal matching subspace pursuit} (SS-OMSP).
SS-OMSP is a modification of another greedy algorithm,
\textit{rank-aware order recursive matching pursuit} (RA-ORMP) \cite{DavEld10}
\footnote{
The name ``Rank-Aware ORMP'' proposed for this algorithm appears to be a misnomer.
RA-ORMP, as originally proposed \cite{DavEld10}, does not have any feature to determine rank.
It computes an orthonormal basis for $Y$.
However, whereas in the ideal, noiseless case this basis will have dimension $r$ equal to the rank of $X_0^{J_0}$,
with any noise present $Y$ will have full rank equal to $\min\{m,N\}$, 
and this will also be the dimension of the computed orthonormal basis.
Hence, the algorithm does not seem to have any built-in rank-awareness.}.
SS-OMSP replaces the snapshot matrix $Y$ in RA-ORMP by the orthogonal projector $P_S$ onto the estimated signal subspace,
or equivalently, by an orthogonal basis matrix for the estimated signal subspace.
Given the estimated support $J$ from the previous steps, RA-ORMP updates $J$ by adding $k$ selected by
\begin{equation}
k = \arg\max_{\ell \in [n] \setminus J} \norm{(P_{\mathcal{R}(P_{\mathcal{R}(A_J)}^\perp Y)}) a_\ell}_2 / \norm{P_{\mathcal{R}(A_J)}^\perp a_\ell}_2.
\label{eq:updateraormp}
\end{equation}
Similarly, SS-OMSP updates the support by adding $k$ selected by
\begin{equation}
k = \arg\max_{\ell \in [n] \setminus J} \norm{(P_{P_{\mathcal{R}(A_J)}^\perp \widehat{S}}) a_\ell}_2 / \norm{P_{\mathcal{R}(A_J)}^\perp a_\ell}_2.
\label{eq:updatessomsp1}
\end{equation}
In general, $P_{\mathcal{R}(A_J)}^\perp Y$ and $P_{\mathcal{R}(A_J)}^\perp P_{\widehat{S}}$ span different subspaces,
and hence the two projection operators used in (\ref{eq:updateraormp}) and (\ref{eq:updatessomsp1}) are different.
The two projection operators coincide only for the special case when there is no noise.

We interpret (\ref{eq:updatessomsp1}) using the angle function between two subspaces,
\textit{i.e.}, (\ref{eq:updatessomsp1}) is equivalent to
\begin{equation}
k = \arg\min_{\ell \in [n] \setminus J} \sphericalangle_2(P_{\mathcal{R}(A_J)}^\perp \widehat{S},P_{\mathcal{R}(A_J)}^\perp \mathcal{R}(a_\ell)).
\label{eq:updatessomsp2}
\end{equation}
Given the subspace $\mathcal{R}(A_J)$, which is spanned by the columns of $A$ corresponding to support elements $J$
determined in the preceding steps of the algorithm, the selection rule finds the nearest subspace to $P_{\mathcal{R}(A_J)}^\perp \widehat{S}$ among
all subspaces, each of which is spanned by $P_{\mathcal{R}(A_J)}^\perp a_\ell$ for $\ell \in [n] \setminus J$.
The name ``orthogonal matching subspace pursuit'' is intended to distinguish the matching using a subspace metric in SS-OMSP
from that of OMP and its variations.

Again, we use a partial run ($s-r$ steps) of SS-OMSP for partial support recovery and switch to MUSIC applied to the augmented subspace.
The complete SS-OMSP is summarized as Algorithm~\ref{alg:ssomsp}.
\begin{algorithm}
\caption{Signal Subspace Orthogonal Matching Subspace Pursuit (SS-OMSP)}
\begin{algorithmic}[1]
\INPUT $Y \in \mathbb{K}^{m \times N}$, $A \in \mathbb{K}^{m \times n}$, $s \in \mathbb{N}$.
\OUTPUT $J \subset [n]$
\STATE $P_{\widehat{S}} \in \mathbb{R}^{m \times m}, r \in \mathbb{N} \leftarrow$ \textit{estimate signal subspace from $Y$};
\STATE $J \leftarrow \emptyset$;
\WHILE{$|J| < s$}
\STATE $\displaystyle k \leftarrow \arg\max_{\ell \in [n] \setminus J} \norm{(P_{P_{\mathcal{R}(A_J)}^\perp \widehat{S}}) a_\ell}_2 / \norm{P_{\mathcal{R}(A_J)}^\perp a_\ell}_2$;
\STATE $J \leftarrow J \cup \{k\}$;
\ENDWHILE
\RETURN $J$
\end{algorithmic}
\label{alg:ssomsp}
\end{algorithm}

\subsection{Stopping Conditions for Unknown Sparsity Level}\label{subsec:stopcond}
In most analyses of greedy algorithms, the sparsity level $s$ is assumed to be known \textit{a priori}.
In fact, this assumption is only for simplicity of the analysis and not a limitation of the greedy algorithms.
For example, M-OMP recovers a support of unknown size
by running the steps until the residual $\norm{P_{A_J}^\perp Y}_F$ falls below a certain threshold that depends on the noise level
(or vanishes, in the noiseless case).
SS-OMP recovers a support of unknown size similarly
but a different criterion may be used for the stopping condition.
Assume that the estimated signal subspace $\widehat{S}$ satisfies $\dim(\widehat{S}) \leq \dim(S)$.
After $k$ steps, SS-OMP returns $\widehat{J}$ of size $k$.
Whether to continue to the next step is determined by the stopping condition given by
\begin{equation*}
\sphericalangle_2(\widehat{S},\mathcal{R}(A_J)) = \norm{P_{\mathcal{R}(A_{\widehat{J}})}^\perp P_{\widehat{S}}} \leq \eta
\end{equation*}
for a threshold $\eta$.
For the noiseless case, $\eta$ is set to 0 and, for the noisy case, $\eta$ is set to an estimate of $\sphericalangle_2(\widehat{S},\mathcal{R}(A_{J_0}))$.
Let us consider the solution $\widehat{J}_\text{NS}$ given by the enumeration of all possible supports:
\begin{equation*}
\begin{array}{clll}
\widehat{J}_\text{NS} & \triangleq & \displaystyle \min_{J \subset [n], |J| \geq r} & |J| \\
& & \text{subject~to} & \sphericalangle_2(\widehat{S},\mathcal{R}(A_J)) \leq \eta.
\end{array}
\end{equation*}
We assume that $|\widehat{J}_\text{NS}| = s$.
Otherwise, in the noiseless case, this implies that the support is not uniquely determined.
If $|\widehat{J}_\text{NS}| = s$, then SS-OMP stops after $s$ steps whenever it is guaranteed to recover the support with known $s$.

Similarly, SA-MUSIC with SS-OMP (SA-MUSIC+SS-OMP henceforth) can recover the support without knowledge of $s$ by applying the same stopping criterion for SS-OMP, which is summarized in Algorithm~\ref{alg:samusicunknowns}.
The update criterion in Step~\ref{alg:samusicunknowns:update} of Algorithm~\ref{alg:samusicunknowns} determines the SA-MUSIC algorithm.
For example, SA-MUSIC+SS-OMP uses the condition in (\ref{eq:updatessomp})
whereas SA-MUSIC with SS-OMSP (SA-MUSIC+SS-OMSP henceforth) uses the condition in (\ref{eq:updatessomsp1}).
If $|\widehat{J}_\text{NS}| = s$, then SA-MUSIC algorithms return support of size $s$ whenever they are guaranteed to recover the support with known $s$.
For simplicity, the analyses in Section~\ref{sec:analysisww1rip} will assume known sparsity level.
\begin{algorithm}
\caption{SA-MUSIC for unknown $s$}
\begin{algorithmic}[1]
\INPUT $Y \in \mathbb{K}^{m \times N}$, $A \in \mathbb{K}^{m \times n}$, $\eta > 0$.
\OUTPUT $J \subset [n]$
\STATE $P_{\widehat{S}} \in \mathbb{R}^{m \times m}, r \in \mathbb{N} \leftarrow$ \textit{estimate signal subspace from $Y$};
\STATE $J \leftarrow \emptyset$;
\FOR{$\ell = 1,\ldots,n$}
\STATE $\zeta_\ell \leftarrow \norm{P_{\widehat{S}} a_\ell}_2/\norm{a_\ell}_2$
\ENDFOR
\STATE $J \leftarrow$ indices of the $r$-largest $\zeta_k$'s;
\STATE $J_1 \leftarrow \emptyset$;
\WHILE{$\norm{P_{\mathcal{R}(A_J)}^\perp P_{\widehat{S}}} > \eta$}
%\STATE $\displaystyle k \leftarrow \arg\max_{\ell \in [n] \setminus J_1} \norm{P_{\widehat{S}} P_{\mathcal{R}(A_{J_1})}^\perp a_\ell}_2$;
\STATE Select $k$ by an update criterion; \label{alg:samusicunknowns:update}
\STATE $J_1 \leftarrow J_1 \cup \{k\}$;
\STATE $P_{\widetilde{S}} \leftarrow P_{\widehat{S}} + (P_{\widehat{S}}^\perp A_{J_1}) (P_{\widehat{S}}^\perp A_{J_1})^\dagger$;
\FOR{$\ell \in [n] \setminus J_1$}
\STATE $\zeta_\ell \leftarrow \norm{P_{\widetilde{S}} a_\ell}_2/\norm{a_\ell}_2$
\ENDFOR
\STATE $J \leftarrow J_1 \cup \{\text{indices of the $r$-largest $\zeta_\ell$'s}\}$;
\ENDWHILE
\RETURN $J$
\end{algorithmic}
\label{alg:samusicunknowns}
\end{algorithm}

%%%%%%%%%%%%
\section{Weak Restricted Isometry Property}\label{sec:w1rip}
%%%%%%%%%%%%

%%%%%%
\subsection{Uniform Restricted Isometry Property}\label{subsec:urip}
%%%%%%
The \textit{restricted isometry property (RIP)} has been proposed
in the study of the reconstruction of sparse vectors by $\ell_1$ norm minimization \cite{CanTao05}.
Matrix $A \in \mathbb{K}^{m \times n}$ satisfies the RIP of order $s$ if there exists a constant $\delta \in (0,1)$ such that
\begin{equation}
(1 - \delta) \norm{x}_2^2 \leq \norm{A x}_2^2 \leq (1 + \delta) \norm{x}_2^2, \quad \forall x,~ \norm{x}_0 \leq s.
\label{eq:rip}
\end{equation}
The smallest $\delta$ that satisfies (\ref{eq:rip}) is called the \textit{restricted isometry constant (RIC)} of order $s$
and is denoted by $\delta_s(A)$.
Note that (\ref{eq:rip}) is equivalent to
\begin{equation}
(1 - \delta) I_s \leq A_J^* A_J \leq (1 + \delta) I_s, \quad \forall J \subset [n], ~ |J| = s
\label{eq:rip2}
\end{equation}
and hence $\delta_s(A)$ satisfies
\begin{equation*}
\delta_s(A) = \max_{|J| = s} \norm{A_J^* A_J - I_s}.
\end{equation*}
The RIP of order $s$ implies that all submatrices of $A$ with $s$ columns are uniformly well conditioned.

%%%%%%
\subsection{Weak Restricted Isometry Property}\label{subsec:lwrip}
%%%%%%
In many analyses of sparse signal recovery, the uniform RIP is unnecessarily strong
and requires a demanding condition on $\delta_s(A)$ that is not satisfied by the matrices that arise in applications.
Therefore, weaker versions of RIP have been proposed, tailored to specific analyses \cite{EldRau09}, \cite{CanPla10ripless}.

Matrix $A \in \mathbb{K}^{m \times n}$ satisfies the \textit{weak restricted isometry property (weak RIP)} \cite{CanPla10ripless}
with parameter $(J,s,t,\delta)$, where $s,t \in \mathbb{N}$, $J \subset [n]$ with $|J| = s$, and $\delta \in (0,1)$, if
\begin{equation}
(1 - \delta) I_{s+t} \leq A_K^* A_K \leq (1 + \delta) I_{s+t}, \quad \forall K \supset J, ~ |K| = s+t.
\label{eq:wrip}
\end{equation}
The corresponding weak restricted isometry constant is given by
\begin{equation*}
\delta_{s+t}^\text{weak}(A;J) = \max_{\begin{subarray}{c} K \supset J \\ |K| = s+t \end{subarray}} \norm{A_K^* A_K - I_{s+t}}.
\end{equation*}

The special case of the weak RIP with $t = 1$ has been previously proposed \cite{EldRau09}
to derive an average case analysis of the solution of the MMV problem by the mixed $\ell_{2,1}$ norm minimization,
also known as MMV basis pursuit (M-BP) \cite{CheHuo06}.
This specific case of the weak RIP with $t = 1$, which we call the \textit{weak-1 RIP},
is useful for the analysis in this paper.
Obviously, the weak-1 RIP is satisfied by a less stringent condition on $A$.
In the following, we list some matrices that satisfy the weak-1 RIP along with the required conditions
\footnote{
We show that, compared to the uniform RIP, the requirement on the number of measurements to satisfy the weak-1 RIP is reduced by large factors, ranging between 200 to thousands fold.}.
Importantly, in addition to Gaussian matrices, which lend themselves to relatively easy analysis,
this list includes deterministic matrices that arise in applications, and provides reasonable constants for them.

%%%%%%
\subsection{Gaussian Matrix}\label{subsec:iidGaussian}
%%%%%%
Eldar and Rauhut derived a condition for the weak-1 RIP of an i.i.d. Gaussian matrix \cite[Proposition~5.3]{EldRau09}.
Their proof starts with the concentration of the quadratic form $\norm{Gx}_2^2$ around its expectation
where $G$ is an i.i.d. Gaussian matrix, to bound the singular values of $G$, which has been originally proposed in \cite{BDDVW08}.
We provide an alternative and much tighter condition for the weak-1 RIP of $A$
directly using the concentration of the singular values of an i.i.d. Gaussian matrix \cite{DavSza01}.
\begin{proposition}
Given $m,n \in \mathbb{N}$ with $m \leq n$, let $A \in \mathbb{R}^{m \times n}$ be an i.i.d. Gaussian matrix
whose entries follow $\mathcal{N}(0,\frac{1}{n})$.
Suppose that $J$ is a subset of $[n]$ with $s$ elements.
For any $\epsilon,\delta \in (0,1)$, if
\begin{equation}
\sqrt{m} \geq \frac{\sqrt{s+1} + \sqrt{2 \ln\left(\frac{2(n-s)}{\epsilon}\right)}}{\sqrt{1+\delta}-1},
\label{eq:prop:weak1iidg2:cond1}
\end{equation}
then,
\begin{equation}
\mathbb{P}(\delta_{s+1}^\text{weak}(A;J) \geq \delta) \leq \epsilon,
\label{eq:prop:weak1iidg2:res}
\end{equation}
where the probability is with respect to $A$.
\label{prop:weak1iidg2}
\end{proposition}
\begin{IEEEproof}
See Appendix~\ref{subsec:prop:weak1iidg2}.
\end{IEEEproof}
\begin{remark}
For large $s$ such that the log term is negligible,
Condition (\ref{eq:prop:weak1iidg2:cond1}) reduces to $m/s \geq (\sqrt{1+\delta}-1)^{-2}$.
For $\delta \approx 1$, the oversampling factor is approximately 6.
\end{remark}
\begin{remark}
Using the concavity of the square root function, it follows that
\begin{equation}
m \geq \frac{2}{(\sqrt{1+\delta}-1)^2} \left\{ s + 2 \ln\left( \frac{2(n-s)}{\epsilon} \right) + 1 \right\},
\label{eq:prop:weak1iidg2:cond2}
\end{equation}
is a sufficient condition for (\ref{eq:prop:weak1iidg2:cond1}) and hence also guarantees (\ref{eq:prop:weak1iidg2:res}).
\end{remark}
\begin{remark}
By slightly modifying the proof of Proposition~\ref{prop:weak1iidg2}, we obtain the uniform RIP of $A$: if
\begin{equation}
m \geq \frac{2}{(\sqrt{1+\delta}-1)^2} \left\{ \left[3 + \ln\left(\frac{n}{s}\right)\right] s + 2\ln\left( \frac{2}{\epsilon} \right) + 1 \right\},
\label{eq:unifripiidg}
\end{equation}
then
\begin{equation*}
\mathbb{P}(\delta_{s}(A) \geq \delta) \leq \epsilon,
\end{equation*}
where the probability is with respect to $A$.
Compared to Condition (\ref{eq:prop:weak1iidg2:cond2}) required for the weak-1 RIP,
in Condition (\ref{eq:unifripiidg}) for the uniform RIP,
the required oversampling factor $\frac{m}{s}$ has been increased roughly by the factor $3 + 2\ln\left(\frac{n}{s}\right)$.
\end{remark}

We also consider variations of the \textit{asymmetric RIP} \cite{FouLai09}, \cite{BlaCT11}.
Similarly to the weak-1 RIP, $A$ satisfies the weak-1 asymmetric RIP if there exist $\alpha,\beta > 0$ such that
\begin{equation*}
\alpha \leq \sigma_{s+1}(A_{J \cup \{j\}})
\leq \sigma_1 (A_{J \cup \{j\}}) \leq \beta, \quad \forall j \in [n] \setminus J.
\end{equation*}
The corresponding weak-1 asymmetric RICs are defined as follows:
\begin{align*}
\alpha_{s+1}^\text{weak}(A;J) {} & \triangleq \min_{j \in [n] \setminus J} \sigma_{s+1}(A_{J \cup \{j\}}), \\
\beta_{s+1}^\text{weak}(A;J) {} & \triangleq \max_{j \in [n] \setminus J} \sigma_1(A_{J \cup \{j\}}).
\end{align*}

\begin{proposition}
Given $m,n \in \mathbb{N}$ with $m \leq n$, let $A \in \mathbb{R}^{m \times n}$ be an i.i.d. Gaussian matrix
whose entries follow $\mathcal{N}(0,\frac{1}{n})$.
Suppose that $J$ is a subset of $[n]$ with $s$ elements.
For any $\epsilon,\gamma \in (0,1)$, if
\begin{equation}
\sqrt{m} \geq \frac{\sqrt{s+1} + \sqrt{2 \ln\left(\frac{n-s}{\epsilon}\right)}}{\gamma},
\label{eq:prop:weak1aiidg2:cond1}
\end{equation}
then,
\begin{equation}
\mathbb{P}\Big([\alpha_{s+1}^\text{weak}(A;J) \leq 1-\gamma] \vee [\beta_{s+1}^\text{weak}(A;J) \geq 1+\gamma]\Big) \leq \epsilon,
\label{eq:prop:weak1aiidg2:res}
\end{equation}
where the probability is with respect to $A$.
\label{prop:weak1aiidg2}
\end{proposition}
\begin{IEEEproof}
See Appendix~\ref{subsec:prop:weak1aiidg2}.
\end{IEEEproof}
\begin{remark}
For large $s$ such that the log term is negligible,
Condition (\ref{eq:prop:weak1aiidg2:cond1}) reduces to $\frac{m}{s} \geq \frac{1}{\gamma^2}$.
For $\gamma \approx 1$, the oversampling factor is approximately 1.
\end{remark}
\begin{remark}
Using the concavity of the square root function, it follows that
\begin{equation}
m \geq \frac{2}{\gamma^2} \left\{ s + 2 \ln\left( \frac{2(n-s)}{\epsilon} \right) + 1 \right\},
\label{eq:prop:weak1aiidg2:cond2}
\end{equation}
is a sufficient condition for (\ref{eq:prop:weak1aiidg2:cond1}) and hence also guarantees (\ref{eq:prop:weak1aiidg2:res}).
\end{remark}
\begin{remark}
Proposition~\ref{prop:weak1aiidg2} provides a sufficient condition,
which is not necessarily tightest, in particular, in the limit when $\gamma \rightarrow 1$.
Indeed, an i.i.d. Gaussian $A$ satisfies $\alpha_{s+1}^\text{weak}(A;J) > 0$ with probability 1 if $m > s$,
but Condition (\ref{eq:prop:weak1aiidg2:cond1}) does not converge to $m > s$ as $\delta$ approaches 1.
Nevertheless, this gap vanishes if $s$ goes to infinity with $n = o(e^s)$, that is, $s$ grows faster than $\ln n$. 
\end{remark}

%%%%%%
\subsection{Uniformly Random Partial Fourier Matrix}\label{subsec:randFourier}
%%%%%%
Candes and Plan \cite{CanPla10ripless} showed that a matrix $A$ composed of randomly selected rows of a DFT matrix
satisfies the following local restricted isometry property under a certain mild condition:
$\norm{A_J^* A_J - I_s} \leq \delta$ with high probability for a fixed $J \subset [n]$ of size $s$ (\cite[Lemma~2.1]{CanPla10ripless})
\footnote{
In fact, the result in \cite{CanPla10ripless} and hence our argument derived from their result apply to a wider class of matrices.
}.
It is not difficult to derive the weak-1 RIP of such a matrix from its local RIP.
We only need to consider the union of the events corresponding to all subset of $[n]$ that include the support $J$ and one more element outside $J$.

\begin{proposition}[{Corollary to \cite[Lemma~2.1]{CanPla10ripless}}]
Suppose that $A$ is obtained by randomly selecting $m$ rows of the $n \times n$ DFT matrix independently,
each with probability $\frac{m}{n}$, followed by normalization of each column in $\ell_2$ norm.
Also suppose that $J$ is a fixed subset of $[n]$ of cardinality $s$.
For any $\epsilon,\delta \in (0,1)$, if
\begin{equation}
m \geq \left(\frac{2(3+\delta)}{3\delta^2}\right) \left\{ \ln\left(\frac{2(n-s)}{\epsilon}\right) + \ln(s+1) \right\} (s+1),
\label{eq:prop:weak1randfourier:cond}
\end{equation}
then
\begin{equation}
\mathbb{P}(\delta_{s+1}^\text{weak}(A;J) \geq \delta) \leq \epsilon.
\label{eq:prop:weak1randfourier:res}
\end{equation}
where the probability is with respect to $A$.
\label{prop:weak1randfourier}
\end{proposition}
\begin{IEEEproof}
See Appendix~\ref{subsec:prop:weak1randfourier}.
\end{IEEEproof}

\begin{remark}
The uniform RIP of a random partial Fourier matrix has been studied before \cite{CanTao06}, \cite{RudVer08}, \cite{Rau10}.
In particular, Rauhut \cite[Theorem~8.4]{Rau10} showed a sufficient condition with explicit constants given by
\begin{align*}
\frac{m}{\ln(10m)} {} & \geq C \delta^{-2} \ln^2(100s) \ln(4n) s, \\
m {} & \geq D \delta^{-2} \ln(\epsilon^{-1}) s,
\end{align*}
where $C \leq 17,190$ and $D \leq 456$,
which is a much more demanding condition than (\ref{eq:prop:weak1randfourier:cond}).
\end{remark}

\begin{remark}
As discussed in the previous subsection, the algebraic analogue of the weak-1 RIP condition is usually much easier to satisfy.
Feng and Bresler \cite{Feng97} showed that a matrix $A$ composed of $m$ consecutive rows of the $n \times n$ DFT matrix with $m \geq s+1$
\footnote{The selection of the rows in this pattern was called the ``bunched sampling pattern'' in \cite{Feng97}} satisfies
\begin{equation}
\min_{J \subset [n], |J| = s} \alpha_{s+1}^\text{weak}(A;J) > 0
\label{eq:fullspark}
\end{equation}
Candes \textit{et al.} \cite{CanRT06} showed that if $n$ is a prime number, then the above result holds for any $m$ rows.
Note that $m \geq s+1$ is a much milder requirement than $(\ref{eq:prop:weak1randfourier:cond})$
and the property is deterministic, \textit{i.e.}, holds always.
However, the properties mentioned in this remark are purely algebraic and cannot be used in the analysis with noise.
\end{remark}

%%%%%%
\subsection{Incoherent Unit-Norm Tight Frame with Random Support}\label{subsec:untf}
%%%%%%
Incoherent unit-norm tight frames \cite{Chr03frame}
also satisfy the weak-1 RIP with high probability under a mild condition if the support $J$ is uniformly random.
Let $A \in \mathbb{K}^{m \times n}$ be a unit-norm tight frame.
Then, each column of $A$ has unit $\ell_2$ norm and the rows of $A$ are orthogonal \cite{Chr03frame}.
A unit-norm tight frame also satisfies $\norm{A} = \sqrt{\frac{n}{m}}$.
The coherence $\mu$ of $A$ is defined by
\begin{equation*}
\mu \triangleq \max_{k \neq \ell} \frac{|\langle a_k, a_\ell \rangle|}{\norm{a_k}_2 \norm{a_\ell}_2}
\end{equation*}
and is always bounded from below by the Welch bound \cite{Wel74}
\begin{equation*}
\mu \geq \sqrt{\frac{n-m}{m(n-1)}}.
\end{equation*}
For example, the rows of DFT matrix selected by the difference set method \cite{XiaZG05}
form a unit-norm tight frame that achieves the Welch bound.

Based on the statistical RIP analysis in \cite{Tro08}, we derive the following result.
\begin{proposition}
Suppose that $A \in \mathbb{K}^{m \times n}$ is a fixed unit-form tight frame with coherence $\mu$ satisfying
\begin{equation*}
\mu \leq \frac{K}{\sqrt{m}}
\end{equation*}
for some constant $K > 0$.
Also suppose that $J$ is uniformly random among all subsets of $[n]$ with $s$ elements.
For any $\epsilon,\delta \in (0,1)$, if
\begin{equation}
\begin{aligned}
m \geq \frac{4\sqrt{e}}{\delta^2} \left\{ s + 288 K^2 \ln\left( \frac{n-s}{\epsilon} \right) + 1 \right\},
\end{aligned}
\label{eq:prop:weak1untf:cond}
\end{equation}
then
\begin{equation*}
\mathbb{P}(\delta_{s+1}^\text{weak}(A;J) \geq \delta) \leq \epsilon.
\end{equation*}
where the probability is with respect to $J$.
\label{prop:weak1untf}
\end{proposition}
\begin{IEEEproof}
See Appendix~\ref{subsec:prop:weak1untf}.
\end{IEEEproof}
\begin{remark}
If $A$ achieves the Welch bound, then the constant $K$ in Proposition~\ref{prop:weak1untf} is no greater than 1.
\end{remark}
\begin{remark}
By slightly modifying the proof of Proposition~\ref{prop:weak1untf}, we obtain the uniform RIP of $A$
\begin{equation*}
\mathbb{P}(\delta_s(A) \geq \delta) \leq \epsilon
\end{equation*}
if
\begin{equation}
\begin{aligned}
m \geq \frac{4\sqrt{e}}{\delta^2} \left[ \left\{ 1 + 576 K^2 \ln\left(\frac{en}{s}\right) \right\} s + 288 K^2 \ln\left( \frac{1}{\epsilon}\right) + 1 \right].
\end{aligned}
\end{equation}
Compared to the weak-1 RIP, for the uniform RIP,
the oversampling factor $\frac{m}{s}$ has increased roughly by the factor $1 + 576 K^2 \ln\left(\frac{en}{s}\right)$.
\end{remark}

%%%%%%%%%%%%
\section{Performance Guarantees}\label{sec:analysisww1rip}
%%%%%%%%%%%%

%%%%%%
\subsection{MUSIC for the Full Row Rank Case}\label{subsec:musicw1a}
%%%%%%
With an imperfect estimate of the signal subspace,
the support recovery by MUSIC is no longer guaranteed by an algebraic property of $A$.
Instead, in the following proposition, we provide a new guarantee.
\begin{theorem}[MUSIC, noisy, full row rank case]
Assume that $X_0 \in \mathbb{K}^{n \times N}$ is row $s$-sparse with support $J_0 \subset [n]$ and $X_0^{J_0}$ has full row rank.
Let $\widehat{S}$ be an $s$-dimensional estimate of $S \triangleq \mathcal{R}(A_{J_0} X_0^{J_0})$
such that $\norm{P_{\widehat{S}} - P_S} \leq \eta$ for some $\eta < 0.5$.
If $A$ satisfies one direction of the weak-1 asymmetric RIP
\begin{equation}
\alpha_{s+1}^\text{weak}(A;J_0) > \alpha
\label{eq:thm:music:cond1}
\end{equation}
for
\begin{equation}
\alpha \geq 2 \sqrt{\eta(1-\eta)} \norm{A^*}_{2,\infty},
\label{eq:thm:music:cond2}
\end{equation}
then MUSIC applied to $P_{\widehat{S}}$ will identify $J_0$.
\label{thm:music}
\end{theorem}
\begin{IEEEproof}
See Appendix~\ref{subsec:thm:music}.
\end{IEEEproof}
\begin{remark}
The columns of the sensing matrix $A$ are often normalized in the $\ell_2$ norm (\textit{e.g.}, for a partial Fourier matrix)
or their $\ell_2$ norms are highly concentrated around 1 (\textit{e.g.}, i.i.d. for Gaussian matrix).
We therefore consider the quantity $\norm{A^*}_{2,\infty}$ to be 1 or close to 1.
In particular, we assume that all columns of $A$ are normalized in the $\ell_2$ norm (``$A$ is normalized,'' in short) in all the figures and the numerical experiments in this paper.
\end{remark}
\begin{remark}
With normalized $A$, we have $\norm{A^*}_{2,\infty} = 1$,
and the weak-1 RIP $\delta_{s+1}^\text{weak}(A;J_0) < 1 - \alpha^2$ is a sufficient condition for (\ref{eq:thm:music:cond1}).
\label{rem:aw1rip2w1rip}
\end{remark}
\begin{remark}
When the signal subspace estimation is perfect (\textit{i.e.}, in the noiseless case),
Conditions (\ref{eq:thm:music:cond1}) and (\ref{eq:thm:music:cond2}) reduce to $\alpha_{s+1}^\text{weak}(A;J_0) > 0$,
which is an algebraic condition implying $\rank[A_{J_0} ,~ a_j] = s+1$ for all $j \in [n] \setminus J_0$.
This algebraic condition is implied by a much milder condition than the weak-1 asymmetric RIP, which is an analytic condition.
For example, an i.i.d. Gaussian $A$ with $m \geq s+1$ satisfies this with probability 1.
\label{rem:music}
\end{remark}
\begin{remark}
Theorem~\ref{thm:music} guarantees that MUSIC recovers a fixed support $J_0$.
Replacing Condition (\ref{eq:thm:music:cond1}) by its uniform analog, $\sigma_{s+1}(A_J) > \alpha$ for all $J \subset [n]$ with $|J| = s+1$,
provides a uniform guarantee that MUSIC recovers an arbitrary support of size $s$.
With perfect subspace estimation, the uniform guarantee reduces to Proposition~\ref{prop:musicFenBre}.
\end{remark}

%%%%%%
\subsection{SA-MUSIC with Given Partial Support}\label{subsec:samusicoraclew1a}
%%%%%%
SA-MUSIC finds the support by using the augmented subspace $\widetilde{S}$ constructed as $\widetilde{S} = \mathcal{R}(A_{J_1}) + \widehat{S}$,
where $J_1$ is a subset of $J_0$ of size $s-r$ and $\widehat{S}$ is the estimated signal subspace of dimension $r$.
We assume that there exists an $r$-dimensional subspace $\bar{S}$ of the signal subspace $S \triangleq \mathcal{R}(A_{J_0} X_0^{J_0})$ satisfying $\norm{P_{\widehat{S}} - P_{\bar{S}}} \leq \eta$.
It will be shown in Section~\ref{sec:estss} that, if the number of snapshots is large enough relative to the noise variance,
then Algorithm~\ref{alg:estss} computes $\widehat{S}$ with the property that such an $\bar{S}$ exists.
Recall (by Proposition~\ref{prop:augssp}) that assuming row-nondegenerate $X_0^{J_0}$
implies $\mathcal{R}(A_{J_1}) + \bar{S} = R(A_{J_0})$, which is desired by the MUSIC step in SA-MUSIC.
However, because $\widetilde{S}$ is constructed using $\widehat{S}$ rather $\bar{S}$, which is not available,
to show noise robustness of support recovery, we need to bound $\norm{P_{\widetilde{S}} - P_{\mathcal{R}(A_{J_0})}}$.
By the projection update formula, it follows that
\begin{equation*}
P_{\widetilde{S}} - P_{\mathcal{R}(A_{J_0})} = P_{P_{\mathcal{R}(A_{J_1})}^\perp \widehat{S}} - P_{P_{\mathcal{R}(A_{J_1})}^\perp \bar{S}}.
\end{equation*}
Therefore, we need to consider the distance between the two subspaces $\widehat{S}$ and $\bar{S}$ as projected onto $\mathcal{R}(A_{J_1})^\perp$.
However, in general, projecting onto another subspace can either increase or decrease the distance between subspaces arbitrarily.
In our specific case, the distance is bounded depending on the condition number of $A_{J_0}$.
We state the result in a formal way in the following proposition.

\begin{proposition}
Assume that $X_0 \in \mathbb{K}^{n \times N}$ is row $s$-sparse with support $J_0 \subset [n]$ and $X_0^{J_0}$ is row-nondegenerate.
Let $\widehat{S}$ be the estimated signal subspace of dimension $r$ where $r < s$.
Suppose that there exists an $r$-dimensional subspace $\bar{S}$ of the signal subspace $S \triangleq \mathcal{R}(A_{J_0} X_0^{J_0})$ satisfying $\norm{P_{\widehat{S}} - P_{\bar{S}}} \leq \eta$. Let $J$ be a proper subset of $J_0$.
If $A$ satisfies
\begin{equation}
\frac{\sigma_s(A_{J_0})}{\sigma_1(A_{J_0})} > \eta,
\label{eq:prop:errinaugssp:cond2}
\end{equation}
then
\begin{equation}
\norm{P_{P_{\mathcal{R}(A_J)}^\perp \widehat{S}} - P_{P_{\mathcal{R}(A_J)}^\perp \bar{S}}}
\leq \frac{\eta\sigma_1(A_{J_0})}{\sigma_s(A_{J_0}) - \eta\sigma_1(A_{J_0})}.
\label{eq:prop:errinaugssp:res}
\end{equation}
\label{prop:errinaugssp}
\end{proposition}
\begin{IEEEproof}
See Appendix~\ref{subsec:prop:errinaugssp}.
\end{IEEEproof}

\begin{figure}[htb]
\begin{center}
\begin{minipage}[htb]{0.45\linewidth}
\centerline{\includegraphics[height=55mm]{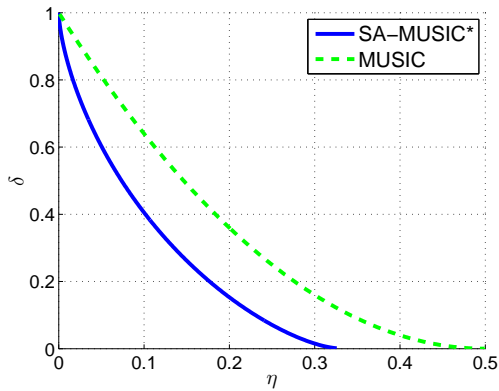}}
\end{minipage}
\end{center}
\caption{Comparison of MUSIC for full rank case vs. SA-MUSIC with known partial support in the rank defective case:
trade-off between parameter $\delta$ (for the weak-1 RIP) and $\eta$ (for subspace estimate perturbation).
The region below the curve provides a guarantee.}
\label{fig:tradeoff_samusico}
\end{figure}

We are now ready to state one of the main results of this paper. 
\begin{theorem}[SA-MUSIC with Correct Partial Support]
Assume that $X_0 \in \mathbb{K}^{n \times N}$ is row $s$-sparse with support $J_0 \subset [n]$ and $X_0^{J_0}$ is row-nondegenerate.
Let $\widehat{S}$ be the estimated signal subspace of dimension $r$, where $r < s$.
Suppose that there exists an $r$-dimensional subspace $\bar{S}$ of the signal subspace $S \triangleq \mathcal{R}(A_{J_0} X_0^{J_0})$ satisfying $\norm{P_{\widehat{S}} - P_{\bar{S}}} \leq \eta$.
Let $J_1$ be an arbitrary subset of $J_0$ with $s-r$ elements.
If $A$ satisfies the weak-1 asymmetric RIP
\begin{equation}
\alpha < \alpha_{s+1}^\text{weak}(A;J_0) \leq \beta_{s+1}^\text{weak}(A;J_0) < \beta
\label{eq:thm:samusicoracle:cond2}
\end{equation}
for $\alpha$ and $\beta$ satisfying
\begin{equation}
1 - \sqrt{1 - \frac{\alpha^2}{\norm{A^*}_{2,\infty}^2}}
\geq \frac{2\eta\beta}{\alpha - \eta\beta},
\label{eq:thm:samusicoracle:cond3}
\end{equation}
then MUSIC applied to $\widetilde{S} = \widehat{S} + \mathcal{R}(A_{J_1})$ using the criterion
\begin{equation}
\min_{k \in J_0 \setminus J_1} \frac{\norm{P_{\widetilde{S}} a_k}_2}{\norm{a_k}_2}
> \max_{k \in [n] \setminus J_0} \frac{\norm{P_{\widetilde{S}} a_k}_2}{\norm{a_k}_2}.
\end{equation}
will identify $J_0 \setminus J_1$.
\label{thm:samusicoracle}
\end{theorem}
\begin{IEEEproof}
See Appendix~\ref{subsec:thm:samusicoracle}.
\end{IEEEproof}
\begin{remark}
With normalized $A$, Condition (\ref{eq:thm:samusicoracle:cond2})+(\ref{eq:thm:samusicoracle:cond3}) is implied by the weak-1 RIP of $A$ given by $\delta_{s+1}^\text{weak}(A;J_0) < \delta$ for
\begin{equation}
\eta \leq \sqrt{\frac{1-\delta}{1+\delta}} \cdot \frac{1-\sqrt{\delta}}{3-\sqrt{\delta}}.
\label{eq:samusicoracle:eta}
\end{equation}
\end{remark}
\begin{remark}
Compared to the guarantee on MUSIC (full row rank case, Theorem~\ref{thm:music}),
the guarantee for SA-MUSIC (for the rank defective case, Theorem~\ref{thm:samusicoracle})
additionally requires $X_0^{J_0}$ to be row-nondegenerate.
For the noiseless case, with normalized $A$, both Theorem~\ref{thm:music} and Theorem~\ref{thm:samusicoracle} require only a mild algebraic condition $\alpha_{s+1}^\text{weak}(A;J_0) > 0$.
However, as shown in Fig.~\ref{fig:tradeoff_samusico}, even with \textit{known partial support SA-MUSIC for the rank defective case}
suffers more from the perturbation in the subspace estimate than does MUSIC in the full row rank case.
This difference in the sensitivity to the subspace estimation error is due to the reduced dimension $r < s$ of the estimated signal subspace,
which in turn is due to the rank defect of $X_0^{J_0}$.
\end{remark}
\begin{remark}
Theorem~\ref{thm:samusicoracle} provides a performance guarantee for SA-MUSIC with a correct partial support estimate, but noisy subspace estimate.
In this scenario, SA-MUSIC provides its best performance.
How realistic is this assumption?
In the next subsections, we will show that if the error in the subspace estimate is small enough,
suboptimal greedy algorithms are indeed guaranteed to recover the partial support exactly.
In particular, when $r/s$ is large, SA-MUSIC combined with partial support recovery by greedy algorithms provides guarantees comparable to 
that given in Theorem~\ref{thm:samusicoracle} for SA-MUSIC with correct partial support estimate.
\end{remark}

%%%%%%
\subsection{SA-MUSIC with Partial Support Recovery by SS-OMP}\label{subsec:samusicw1a}
%%%%%%
We first propose a new sufficient condition for a guarantee of M-OMP, which is of independent interest in its own right.
\begin{proposition}[M-OMP]
Suppose that $X_0 \in \mathbb{K}^{n \times N}$ is row $s$-sparse with support $J_0 \subset [n]$.
Let $Y = A X_0 + W$.
Given $J \subsetneq J_0$, if $A$ satisfies the weak-1 RIP
\begin{equation}
\delta_{s+1}^\text{weak}(A;J_0) < \delta
\label{eq:prop:momp:cond1}
\end{equation}
for
\begin{align}
\delta {} & \leq \frac{\norm{X_0^{J_0 \setminus J}}_{2,\infty} - 2 \norm{A^* W}_{2,\infty}}{2 \norm{X_0^{J_0 \setminus J}}},
\label{eq:prop:momp:cond2}
\end{align}
then the next step of M-OMP will identify an element of $J_0$.
\label{prop:momp}
\end{proposition}
\begin{IEEEproof}
See Appendix~\ref{subsec:prop:momp}.
\end{IEEEproof}
\begin{remark}
Applying previous analysis \cite[Theorem~5]{GRSV08} of $p$-SOMP to the case $p = 2$ provides an alternative guarantee for M-OMP.
However, the conditions for the latter are stated in terms of the $p$-Babel function and its variations
rather than the weak-1 RIP as in our Proposition~\ref{prop:momp}.
The two guarantees are not directly comparable, but neither one is uniformly less or more demanding than the other in terms of the conditions required.
Because the form of the condition in Proposition~\ref{prop:momp} is amenable to our analysis, it is preferred in this paper.
\end{remark}
\begin{remark}
Unlike MUSIC, M-OMP is guaranteed only if the orthogonality between two vectors is nearly preserved through $A$.
Therefore, unlike MUSIC, M-OMP is not guaranteed by one direction of the weak-1 asymmetric RIP.
\end{remark}
\begin{remark}
For the noiseless case ($W = 0$), since
\begin{equation*}
\frac{\norm{X_0^{J_0 \setminus J}}_{2,\infty}}{\norm{X_0^{J_0 \setminus J}}} \geq \frac{1}{\sqrt{|J_0 \setminus J|}} \geq \frac{1}{\sqrt{s}},
\end{equation*}
it follows that Condition (\ref{eq:prop:momp:cond1})+(\ref{eq:prop:momp:cond2}) is implied by $\delta_{s+1}^\text{weak}(A;J_0) < \frac{1}{2\sqrt{s}}$,
which is a weaker requirement than
the previously known sufficient condition $\delta_{s+1}(A) < \frac{1}{3\sqrt{s}}$ \cite{DavWak10}.
\end{remark}

Next, the guarantee of SS-OMP is obtained as a corollary to Proposition~\ref{prop:momp}.
Its main utility is as a stepping stone to the guarantee for SA-MUSIC with partial support recovery by SS-OMP.
\begin{corollary}[SS-OMP]
Assume that $X_0 \in \mathbb{K}^{n \times N}$ is row $s$-sparse with support $J_0 \subset [n]$
and let $\widehat{S}$ be the estimated signal subspace of dimension $r \leq s$.
Suppose that there exists an $r$-dimensional subspace $\bar{S}$ of the signal subspace $S \triangleq \mathcal{R}(A_{J_0} X_0^{J_0})$
satisfying $\norm{P_{\widehat{S}} - P_{\bar{S}}} \leq \eta$.
Let $\Phi \in \mathbb{K}^{s \times r}$ satisfy $\Phi^*\Phi = I_r$, and $\bar{S} = \mathcal{R}(A_{J_0} \Phi)$.
Let $\rho_k(\Phi)$ denote the $k$-th largest $\ell_2$ norm of the rows of $\Phi$.
If $A$ satisfies the weak-1 RIP
\begin{equation}
\delta_{s+1}^\text{weak}(A;J_0) < \delta
\label{eq:prop:ssomp:cond1}
\end{equation}
for $\delta$ satisfying
\begin{equation}
\frac{\rho_k(\Phi)}{\sqrt{ 1+\delta}} - \frac{2\delta}{\sqrt{1-\delta}} \geq 2 \eta \norm{A^*}_{2,\infty},
\label{eq:prop:ssomp:cond2}
\end{equation}
then the first $k$ steps of SS-OMP will identify $k$ elements of $J_0$.
\label{cor:ssomp}
\end{corollary}
\begin{IEEEproof}
See Appendix~\ref{subsec:cor:ssomp}.
\end{IEEEproof}

The conditions for $\widehat{S}$ in terms of $\bar{S}$ and $\Phi$ appear rather technical,
but we will show that they are satisfied by our proposed subspace estimation scheme.
Furthermore, their implications are interpreted in the following remarks.

\begin{remark}
Since $\rho_k(\Phi)$ corresponds to the $\ell_2$ norm of a row of $\Phi$,
$\rho_k(\Phi)$ does not change by applying a rotation to the right of $\Phi$.
In other words, for fixed $\bar{S}$, $\rho_k(\Phi)$ is invariant to the choice of $\Phi$.
Now, since $\rho_k(\Phi)$ is monotonically decreasing in $k$,
the sufficient condition (\ref{eq:prop:ssomp:cond1})--(\ref{eq:prop:ssomp:cond2}) of Corollary~\ref{cor:ssomp}
gets more stringent (requires a smaller weak-1 RIC) as $k$ increases toward $s$.
This implies that the smaller the size of the partial support that is to be recovered by SS-OMP,
the less stringent the condition for the guarantee on the recovery.
Moreover, for $k > s-r$, the guarantee of Corollary~\ref{cor:ssomp} requires a more stringent condition
than the guarantee on the remaining step of SA-MUSIC in Theorem~\ref{thm:samusicoracle}, 
where a partial support of at least size $s-r$ is assumed to be known.
For these two reasons, switching to SA-MUSIC applied to the augmented subspace
after successful partial support recovery of size exactly $k = s-r$ by SS-OMP
is preferred rather than continuing any of the remaining steps of SS-OMP.
\end{remark}

\begin{figure}[htb]
\centerline{\includegraphics[height=55mm]{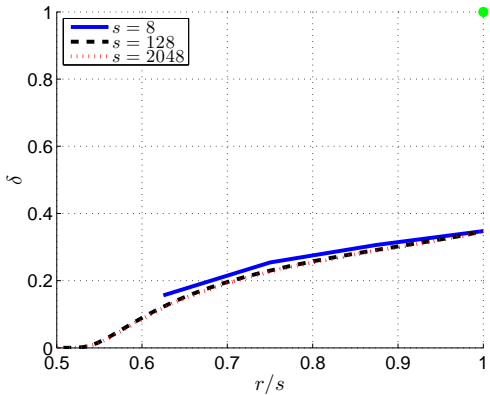}}
\caption{Required weak-1 RIC for the guarantee for SA-MUSIC+SS-OMP, rank-defective case in Theorem~\ref{thm:samusic} ($\eta = 0$).
The dot at the top right of the plot represents the weak-1 RIP $\alpha_{s+1}^\text{weak}(A, J_0) > 0$ required in the full rank case, $r=s$.}
\label{fig:gammadeltaMJd}
\end{figure}

Clearly, the smaller $\rho_{s-r}(\Phi)$, the more stringent the condition on $\delta_{s+1}^\text{weak}(A;J_0)$.
We therefore provide a deterministic lower bound on $\rho_{s-r}(\Phi)$.
The bound is based on the Cauchy-Binet formula, and holds for $r/s > 0.5$.
To facilitate the interpretation of the condition, the bound of Lemma~\ref{lemma:rhobndCB} is visualized in Fig.~\ref{fig:gammadeltaMJd}.

\begin{lemma}
For $r,s \in \mathbb{N}$ where $s/2 < r < s$,
let $\Phi \in \mathbb{K}^{s \times r}$ satisfy $\Phi^*\Phi = I_r$,
and let $\rho_k(\Phi)$ denote the $k$-th largest $\ell_2$ norm of the rows of $\Phi$.
Then,
\begin{equation}
\rho_{s-r}(\Phi) \geq \underline{\rho}(s,r) \triangleq \sup_{q > 0} \hat{\rho}(s,r,q)
\label{eq:lemma:rhobndCB:res}
\end{equation}
where $\hat{\rho}(s,r,q)$ is defined by
\begin{equation}
\begin{aligned}
\hat{\rho}(s,r,q)
\triangleq \left( \frac{\displaystyle \left(\frac{s!}{r!(s-r)!}\right)^{-q/2r} - \left( \frac{s}{r} - 1 \right)}{\displaystyle 2 - \frac{s}{r}} \right)^{1/q}.
\end{aligned}
\end{equation}
for $q > 0$.
\label{lemma:rhobndCB}
\end{lemma}
\begin{IEEEproof}
See Appendix~\ref{subsec:lemma:rhobndCB}.
\end{IEEEproof}

\begin{figure}[htb]
\centerline{\includegraphics[height=55mm]{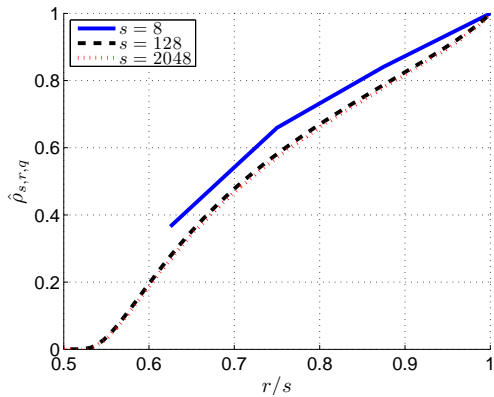}}
\caption{Lower bound $\hat{\rho}(s,r,q)$ on $\rho_{s-r}(U)$ by Lemma~\ref{lemma:rhobndCB} with $q = 10^{-3}$.}
\label{fig:rhobndsMJd}
\end{figure}

Combining Lemma~\ref{lemma:rhobndCB} and Corollary~\ref{cor:ssomp} (for the success of the first $s-r$ steps of SS-OMP) 
with Theorem~\ref{thm:samusicoracle}, we obtain another of our main results:
the following theorem provides a deterministic performance guarantee of SA-MUSIC combined with SS-OMP.
The proof is straightforward and therefore omitted.
\begin{theorem}[SA-MUSIC+SS-OMP, rank-defective case]
Assume that $X_0 \in \mathbb{K}^{n \times N}$ is row $s$-sparse with support $J_0 \subset [n]$ and $X_0^{J_0}$ is row-nondegenerate.
Let $\widehat{S}$ be the estimated signal subspace of dimension $r$, where $s/2 < r \leq s$.
Suppose that there exists an $r$-dimensional subspace $\bar{S}$ of the signal subspace $S \triangleq \mathcal{R}(A_{J_0} X_0^{J_0})$
satisfying $\norm{P_{\widehat{S}} - P_{\bar{S}}} \leq \eta$.
If $A$ satisfies the weak-1 RIP
\begin{equation}
\delta_{s+1}^\text{weak}(A;J_0) < \delta
\label{eq:thm:samusic:cond}
\end{equation}
for $\delta$ satisfying
\begin{equation}
1 - \sqrt{1 - \frac{1-\delta}{\norm{A^*}_{2,\infty}^2}}
\geq \frac{\eta\sqrt{1+\delta}}{\sqrt{1-\delta}-\eta\sqrt{1+\delta}},
\label{eq:thm:samusic:cond2}
\end{equation}
and
\begin{equation}
\frac{\underline{\rho}(s,r)}{\sqrt{1+\delta}} - \frac{2\delta}{\sqrt{1-\delta}} \geq 2 \eta \norm{A^*}_{2,\infty}
\label{eq:thm:samusic:cond3}
\end{equation}
where $\underline{\rho}(s,r)$ is defined in Lemma~\ref{lemma:rhobndCB},
then SA-MUSIC+SS-OMP applied to $\widehat{S}$ will identify $J_0$.
\label{thm:samusic}
\end{theorem}

As discussed in Remark~\ref{rem:u0spark}, the row-nondegeneracy condition on $X_0^{J_0}$ is a mild condition essential to SA-MUSIC
and likewise Condition (\ref{eq:thm:samusic:cond}) + (\ref{eq:thm:samusic:cond2}) is inherited from SA-MUSIC
(see Theorem~\ref{thm:samusicoracle} and Remark~\ref{rem:aw1rip2w1rip}).
Condition (\ref{eq:thm:samusic:cond}) + (\ref{eq:thm:samusic:cond3}), on the other hand,
is due to the use of SS-OMP, a suboptimal algorithm, for the partial support recovery (see Corollary~\ref{cor:ssomp}).
For the noiseless case ($\eta = 0$), with normalized $A$, (\ref{eq:thm:samusic:cond2}) reduces to $\delta \leq 1$,
which is a necessary condition for (\ref{eq:thm:samusic:cond3}).
Therefore, it suffices to satisfy (\ref{eq:thm:samusic:cond})+(\ref{eq:thm:samusic:cond3}),
where (\ref{eq:thm:samusic:cond3}) reduces to
\begin{equation}
\frac{\underline{\rho}(s,r)}{\sqrt{1+\delta}} - \frac{2\delta}{\sqrt{1-\delta}} \geq 0.
\label{eq:rhodeltanoiseless}
\end{equation}

As shown in Fig.~\ref{fig:gammadeltaMJd}, the weak-1 RIC $\delta$ required by (\ref{eq:rhodeltanoiseless}) increases with increasing $r/s$.
In other words, SA-MUSIC+SS-OMP benefits from higher dimension of the signal subspace $S$.
In particular, when $r = s$, SA-MUSIC reduces to MUSIC for the full rank case and hence it suffices to satisfy $\delta_{s+1}^\text{weak}(A;J_0) < \delta = 1$, which is plotted as the dot at the top right in Fig.~\ref{fig:gammadeltaMJd}.

\begin{figure}[htb]
\begin{center}
\begin{minipage}[htb]{0.45\linewidth}
\centerline{\includegraphics[height=55mm]{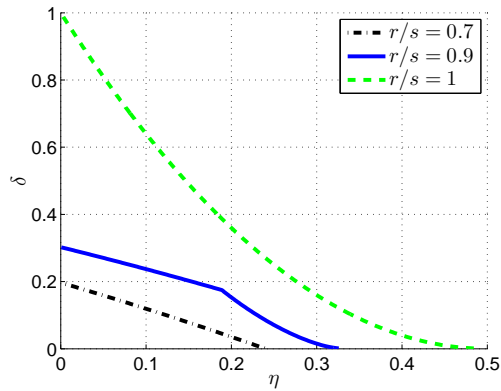}}
\end{minipage}
\end{center}
\caption{Trade-off between parameters $\delta$ (for the weak-1 RIP) and $\eta$ (for subspace estimate perturbation) for the guarantee of SA-MUSIC+SS-OMP in Theorem~\ref{thm:samusic}. Values $(\eta, \delta)$ in region below the curve provide a guarantee.}
\label{fig:tradeoff_samusic}
\end{figure}

Fig.~\ref{fig:tradeoff_samusic} visualizes the trade-off between $\delta$ and $\eta$
given by Conditions (\ref{eq:thm:samusic:cond2}) and (\ref{eq:thm:samusic:cond2}) for the noisy case.
Given $r/s$, any pair $(\eta,\delta)$ below the curve determined by $r/s$ guarantees SA-MUSIC+SS-OMP.
As $r/s$ increases, the curve shifts upward and the guarantee is given by larger $\delta$ and/or larger $\eta$.
In particular, when $r/s = 1$, SA-MUSIC reduces to MUSIC without the step of partial support recovery
and hence, by Theorem~\ref{thm:music}, the trade off is given by
\begin{equation*}
\eta \leq \frac{1 - \sqrt{\delta}}{2},
\end{equation*}
which is also plotted in Fig.~\ref{fig:tradeoff_samusic}.

%%%%%%
\subsection{SA-MUSIC with Partial Support Recovery by SS-OMSP}\label{subsec:ssomspw1a}
%%%%%%
\begin{proposition}[SS-OMSP, rank-defective case]
Assume that $X_0 \in \mathbb{K}^{n \times N}$ is row $s$-sparse with support $J_0 \subset [n]$ and $X_0^{J_0}$ is row-nondegenerate.
Let $\widehat{S}$ be the estimated signal subspace of dimension $r$, where $r \leq s$.
Suppose that there exists an $r$-dimensional subspace $\bar{S}$ of the signal subspace $S \triangleq \mathcal{R}(A_{J_0} X_0^{J_0})$
satisfying $\norm{P_{\widehat{S}} - P_{\bar{S}}} \leq \eta$.
Given $J \subsetneq J_0$, if $A$ satisfies the weak-1 asymmetric RIP
\begin{equation}
\alpha < \alpha_{s+1}^\text{weak}(A;J_0) \leq \beta_{s+1}^\text{weak}(A;J_0) < \beta
\label{eq:prop:ssomsp:cond2}
\end{equation}
for $\alpha$ and $\beta$ satisfying
\begin{equation}
\sqrt{\frac{\dim(P_{\mathcal{R}(A_J)}^\perp \bar{S})}{s-|J|}} \cdot \frac{\alpha}{\norm{A^*}_{2,\infty}}
- \sqrt{1 - \frac{\alpha^2}{\norm{A^*}_{2,\infty}^2}}
\geq \frac{2\eta\beta}{\alpha - \eta\beta},
\label{eq:prop:ssomsp:cond3}
\end{equation}
then the next step of SS-OMSP will identify an elements of $J_0 \setminus J$.
For the special case when $r = s$, Condition (\ref{eq:prop:ssomsp:cond3}) is replaced by the weaker condition
\begin{equation}
1 - \sqrt{1 - \frac{\alpha^2}{\norm{A^*}_{2,\infty}^2}}
\geq \frac{2\eta\beta}{\alpha - \eta\beta}.
\label{eq:prop:ssomsp:cond4}
\end{equation}
\label{prop:ssomsp}
\end{proposition}
\begin{IEEEproof}
See Appendix~\ref{subsec:prop:ssomsp}
\end{IEEEproof}

In the full row rank case, $X_0^{J_0}$ is trivially row-nondegenerate.
In the noiseless case, we have $\eta = 0$
and with normalized $A$, Condition (\ref{eq:prop:ssomsp:cond4}) reduces to $\alpha_{s+1}^\text{weak}(A;J_0) > 0$.
Therefore, for the full rank and noiseless case,
SS-OMSP is guaranteed by $\alpha_{s+1}^\text{weak}(A;J_0) > 0$,
which coincides with the condition for the guarantee of MUSIC in the same scenario (Proposition~\ref{prop:musicFenBre}).
In fact, in this noiseless and full row rank case, SS-OMSP is equivalent to the corresponding data domain algorithm, RA-ORMP.
The coincidence of the guarantees of MUSIC and of RA-ORMP in this special case has been shown before \cite{DavEld10}.
(Unlike the analysis of RA-ORMP \cite{DavEld10} though, Proposition~\ref{prop:ssomsp} also applies to the noisy and/or rank-defective cases.)

If $X_0^{J_0}$ is row-nondegenerate, then by Proposition~\ref{prop:augssp},
$\dim(\mathcal{R}(A_J) + \bar{S}) = s$ for any $J \subset J_0$ with $|J| = s-r$,
which implies $\dim(P_{\mathcal{R}(A_J)}^\perp \bar{S}) = r$.
Then, we also have $\dim(P_{\mathcal{R}(A_J)}^\perp \bar{S}) = r$ for any $|J| \leq s-r$.
Combining this result with Proposition~\ref{prop:augssp}, Proposition~\ref{prop:ssomsp}, and Theorem~\ref{thm:samusicoracle},
we obtain another main result of this paper: a guarantee for SA-MUSIC+SS-OMSP.

\begin{figure}[htb]
\begin{center}
\begin{minipage}[htb]{0.45\linewidth}
\centerline{\includegraphics[height=55mm]{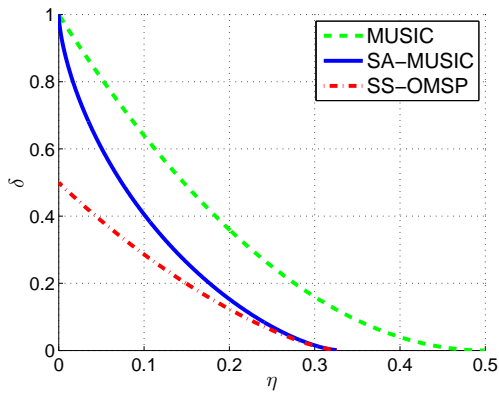}}
\end{minipage}
\end{center}
\caption{Comparison of SA-MUSIC and SS-OMSP when a partial support of size $s-r$ is given:
trade-off between parameter $\delta$ (for the weak-1 RIP) and $\eta$ (for subspace estimate perturbation).
The region below the curve provides a guarantee.}
\label{fig:tradeoff_withoracle}
\end{figure}

\begin{theorem}[SA-MUSIC+SS-OMSP, rank-defective case]
Assume that $X_0 \in \mathbb{K}^{n \times N}$ is row $s$-sparse with support $J_0 \subset [n]$ and $X_0^{J_0}$ is row-nondegenerate.
Let $\widehat{S}$ be the estimated signal subspace of dimension $r$, where $r \leq s$.
Suppose that there exists an $r$-dimensional subspace $\bar{S}$ of the signal subspace $S \triangleq \mathcal{R}(A_{J_0} X_0^{J_0})$
satisfying $\norm{P_{\widehat{S}} - P_{\bar{S}}} \leq \eta$.
If $A$ satisfies the weak-1 asymmetric RIP
\begin{equation}
\alpha < \alpha_{s+1}^\text{weak}(A;J_0) \leq \beta_{s+1}^\text{weak}(A;J_0) < \beta
\label{eq:thm:samusic2:cond2}
\end{equation}
for $\alpha$ and $\beta$ satisfying
\begin{equation}
1 - \sqrt{1 - \frac{\alpha^2}{\norm{A^*}_{2,\infty}^2}}
\geq \frac{2\eta\beta}{\alpha - \eta\beta},
\label{eq:thm:samusic2:cond3}
\end{equation}
and
\begin{equation}
\sqrt{\frac{r}{s}} \cdot \frac{\alpha}{\norm{A^*}_{2,\infty}}
- \sqrt{1 - \frac{\alpha^2}{\norm{A^*}_{2,\infty}^2}}
\geq \frac{2\eta\beta}{\alpha - \eta\beta},
\label{eq:thm:samusic2:cond4}
\end{equation}
then SA-MUSIC+SS-OMSP applied to $\widehat{S}$ will identify $J_0$.
\label{thm:samusic2}
\end{theorem}
\begin{remark}
With normalized $A$,
Condition (\ref{eq:thm:samusic2:cond3}) is implied by Condition (\ref{eq:thm:samusic2:cond4}),
which means that partial support recovery by SS-OMSP requires more stringent conditions than the subsequent MUSIC step in the guarantee of SA-MUSIC+SS-OMSP.
This results in the same guarantee for SA-MUSIC+SS-OMSP as for SS-OMSP alone.
However, in the numerical experiments in Section~\ref{sec:numerical_results}, the two algorithms exhibited substantially different performance.
To interpret this, we compare SA-MUSIC and SS-OMSP conditioned on the event that a correct partial support of size $s-r$ has been found.
By the row-nondegeneracy condition on $X_0^{J_0}$, it follows that
$\dim(P_{\mathcal{R}(A_J)}^\perp \bar{S}) = s-|J|$ for all $J \subset J_0$ with $|J| \geq s-r$.
Therefore, with known partial support, the remaining steps of SS-OMSP are guaranteed by
\begin{equation}
\frac{\alpha}{\norm{A^*}_{2,\infty}} - \sqrt{1 - \frac{\alpha^2}{\norm{A^*}_{2,\infty}^2}}
\geq \frac{2\eta\beta}{\alpha - \eta\beta},
\label{eq:ssomsporacle:eta}
\end{equation}
which is obtained from Condition (\ref{eq:prop:ssomsp:cond3}) with $\dim(P_{\mathcal{R}(A_J)}^\perp \bar{S}) = s-|J|$.
In contrast, as shown in Fig.~\ref{fig:tradeoff_withoracle}, when a partial support of size $s-r$ is given,
the condition in (\ref{eq:samusicoracle:eta}) for the guaranteed success of SA-MUSIC is substantially milder.
\end{remark}

\begin{figure}[htb]
\begin{center}
\begin{minipage}[htb]{0.45\linewidth}
\centerline{\includegraphics[height=55mm]{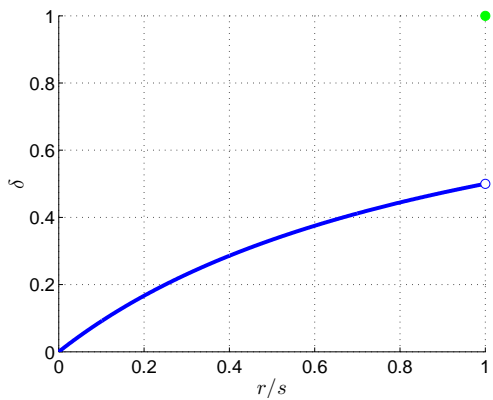}}
\end{minipage}
\end{center}
\caption{Required $\delta$ (for the weak-1 RIP) for the guarantee of SA-MUSIC+SS-OMSP in Theorem~\ref{thm:samusic2} for the noiseless case ($\eta = 0$). The dot at the top right of the plot represents the weak-1 RIP $\alpha_{s+1}^\text{weak}(A, J_0) > 0$ required in the full rank case, $r=s$.}
\label{fig:gammadeltasamusic2}
\end{figure}

\begin{remark}
With normalized $A$, Condition (\ref{eq:thm:samusic2:cond2})+(\ref{eq:thm:samusic2:cond3})+(\ref{eq:thm:samusic2:cond4})
is implied by the weak-1 RIP of $A$ given by $\delta_{s+1}^\text{weak}(A;J_0) < \delta$ for $\delta$ satisfying
\begin{equation*}
\eta \leq \sqrt{\frac{1-\delta}{1+\delta}} \cdot \frac{\sqrt{r/s} \sqrt{1-\delta} - \sqrt{\delta}}{2 + \sqrt{r/s}\sqrt{1-\delta} - \sqrt{\delta}}.
\end{equation*}
Furthermore, if we assume $\eta = 0$, then the guarantee for SA-MUSIC+SS-OMSP only requires
\begin{equation*}
\delta_{s+1}^\text{weak}(A;J_0) < \frac{r}{r+s},
\end{equation*}
which as shown in Fig.~\ref{fig:gammadeltasamusic2} becomes less demanding as $r/s$ increases.
\end{remark}

\begin{figure}[htb]
\begin{center}
\begin{minipage}[htb]{0.45\linewidth}
\centerline{\includegraphics[height=55mm]{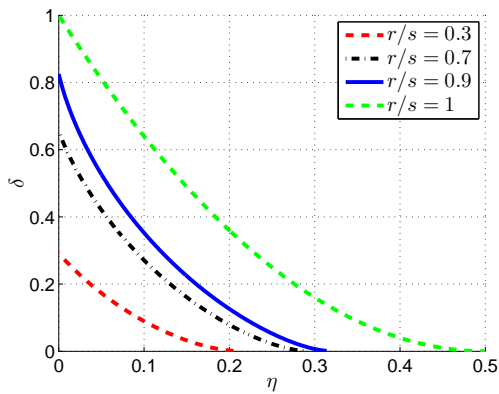}}
\end{minipage}
\end{center}
\caption{Trade-off between parameters $\delta$ (for the weak-1 RIP) and $\eta$ (for subspace estimate perturbation) for the guarantee of SA-MUSIC+SS-OMSP in Theorem~\ref{thm:samusic2}. Values $(\eta, \delta)$ in region below the curve provide a guarantee.}
\label{fig:tradeoff_samusic2}
\end{figure}

For the noisy case, Conditions (\ref{eq:thm:samusic2:cond3}) and (\ref{eq:thm:samusic2:cond4})
provide a trade-off between the parameters $\delta$ and $\eta$ for the guarantee of SA-MUSIC+SS-OMSP,
which is visualized in Fig.~\ref{fig:tradeoff_samusic2}.
As $r/s$ increases, SA-MUSIC+SS-OMSP benefits from higher dimension of the signal subspace $\widehat{S}$.
Compared to the guarantee of SA-MUSIC+SS-OMP in Fig.~\ref{fig:tradeoff_samusic},
SA-MUSIC+SS-OMSP is guaranteed by a weaker requirement on $A$ in terms of the weak-1 RIP.

%%%%%%
\subsection{Implication of the Guarantees in Weak-1 RIP to the Oversampling Factor}\label{subsec:translate}
%%%%%%
The results in the previous subsections were stated in terms of the weak-1 RIP.
Given an upper bound on $\delta_{s+1}^\text{weak}(A;J_0)$ (or bounds on $\alpha_{s+1}^\text{weak}(A;J_0)$ and $\beta_{s+1}^\text{weak}(A;J_0)$),
Section~\ref{sec:w1rip} then provides explicit conditions on the parameters $n,m,s$
that provide the weak-1 RIP for the matrices $A$ discussed there.

\noindent\textbf{Example~1 (i.i.d. Gaussian $A$, asymptotic case)}
In the first example, we consider asymptotic analysis with an i.i.d. Gaussian $A$.
By Proposition~\ref{prop:weak1aiidg2}, if $n$ and $s$ go to infinity while satisfying $n = o(e^s)$, \textit{i.e.},
$s$ grows faster than $\ln n$, and for $\gamma \in (0,1)$
\begin{equation*}
m > \left(\frac{1}{\gamma^2}\right) s,
\end{equation*}
then
\begin{equation*}
1-\gamma < \alpha_{s+1}^\text{weak}(A;J_0) \leq \beta_{s+1}^\text{weak}(A;J_0) < 1+\gamma
\end{equation*}
with probability 1.
Furthermore, in this asymptotic, $\norm{A^*}_{2,\infty} = 1$ with probability 1.
Assume that the estimated signal subspace $\widehat{S}$ is error-free ($\eta = 0$).
Then, for the full row rank case, \textit{i.e.}, $\dim(\widehat{S}) = s$,
all SA-MUSIC algorithms reduce to MUSIC without partial support recovery and are guaranteed by $m > s$,
which is also a necessary condition for the support recovery.
On the other hand, if $\dim(\widehat{S}) = r < s$, then SA-MUSIC+SS-OMSP is guaranteed by
\begin{equation*}
m > \left(1 + \frac{s}{r}\right)^2 s,
\end{equation*}
which unfortunately does not converges to $m > s$ as $r/s \rightarrow 1$.
The discontinuity at $r/s = 1$ is the limitation of the current analysis in this paper.
However, it provides a valid upper bound on the oversampling factor $m/s$ in the given asymptotic.

\noindent\textbf{Example~2 (structured $A$, non-asymptotic case)}
In the second example, we perform the analysis with matrices that arise in practical applications
(\textit{e.g.}, spectrum blind sampling or DOA estimation).
We use the results in Section~\ref{sec:w1rip} for incoherent unit-norm tight frames and partial Fourier matrices.
Fortunately, these matrices have normalized columns and we do not need to worry about $\norm{A^*}_{2,\infty}$ any longer.
Given $\delta$, the weak-1 RIP, $\delta_{s+1}^\text{weak}(A;J_0) < \delta$
(or the weak-1 asymmetric RIP, $\alpha_{s+1}^\text{weak}(A;J_0) < \alpha$) holds with probability $1 - \epsilon$ if
\begin{equation}
m \geq C_1 (s + C_2).
\label{eq:mns}
\end{equation}
For an incoherent unit-norm tight frame, $C_1$ is a constant that depends only on $\delta$.
However, for the random partial Fourier case, $C_1$ also depends on $\ln(s+1)$, $\ln(n-s)$, and $\epsilon$.
We summarize the explicit formulae for $C_1$ and $C_2$:
{\allowdisplaybreaks
\begin{align*}
\intertext{$\bullet$ Random partial Fourier (Proposition~\ref{prop:weak1randfourier})}
C_1 {} & = \frac{2(3+\delta)}{3\delta^2} \left[ \ln\left(\frac{2(n-s)}{\epsilon}\right) + \ln(s+1) \right], \\
C_2 {} & = 1.
\intertext{$\bullet$ Incoherent unit-norm tight frame with random support (Proposition~\ref{prop:weak1untf})}
C_1 {} & = \frac{4\sqrt{e}}{\delta^2}, \\
C_2 {} & = 288 K^2 \ln\left(\frac{2(n-s)}{\epsilon}\right) + 1
\end{align*}} %
where $K$ is determined by the coherence of $A$ (\textit{e.g.}, $K = 1$ if $A$ achieves the Welch bound).

Substituting $\delta$ (for the weak-1 RIP) into these expressions
determines the explicit scaling of $m$ versus $s$ and the other problem parameters that will provide guaranteed recovery.
As $r/s$ increases, $\delta$ increases and hence the oversampling factor $C_1$ in (\ref{eq:mns}) decreases.
In particular, when $r = s$, SA-MUSIC reduces to MUSIC without need of any partial support recovery
and hence is guaranteed by the weak-1 asymmetric RIP $\alpha_{s+1}^\text{weak}(A;J_0) > 0$,
which corresponds to $C_1 = 1$ and $C_2 = 1$ in (\ref{eq:mns}).

In the presence of perturbation ($\eta > 0$), Figs.~\ref{fig:tradeoff_samusic} and \ref{fig:tradeoff_samusic2} provide the required $\delta$ for the guarantees and the oversampling factor is computed from $\delta$ similarly.
The relation between $\eta$ and the number of snapshots $N$ will be investigated in the next section.

%%%%%%%%%%%%
\section{Analysis of Signal Subspace Estimation}\label{sec:estss}
%%%%%%%%%%%%
Unlike the previous works in sensor array processing \cite{StoNeh89} which relies on asymptotics,
we analyze the perturbation in the estimate of the signal subspace with finitely many observations.
Combined with the results in Section~\ref{sec:analysisww1rip},
this analysis provides non-asymptotic guarantees in the noisy case for the new proposed algorithms directly in terms of the measurement noise.
The results also enable us to extend the previous performance guarantees
\cite{FenBre96}, \cite{Feng97}, \cite{BreFen96} of MUSIC to the noisy and finite snapshot case, which was missing before.

\noindent\textbf{Assumption~1 (Noise)}
Given row $s$-sparse $X_0 \in \mathbb{K}^{n \times N}$ with support $J_0$,
the snapshot matrix $Y \in \mathbb{K}^{m \times N}$ is obtained with sensing matrix $A \in \mathbb{K}^{m \times n}$ as
\begin{equation*}
Y = A_{J_0} X_0^{J_0} + W
\end{equation*}
where the columns of $W$ are independent realizations of Gaussian vector $w \in \mathbb{K}^m$ with $\mathbb{E} w = 0$
and $\mathbb{E} ww^* = \sigma_w^2 I_m$.
For the complex field case, we assume circular Gaussian distribution.
We assume that $A_{J_0}$ has full column rank.

\noindent\textbf{Assumption~2 (Number of snapshots)}
We assume that the number of snapshots $N$ is large but finite, more specifically, $N$ satisfies $N \geq m$.
It is also assumed that $m > s$ which is required for support recovery by any method.

Our assumption on the number of snapshots is motivated by the following considerations.
In compressed sensing, the goal is usually to minimize the number of expensive measurements.
Now, in certain applications of joint sparse recovery, taking many snapshots is a rather trivial task
compared to acquiring many measurements in a single snapshot.
For example, in spectrum-blind sampling \cite{FenBre96}, the number $m$ of measurements per snapshot $m$ determines the sampling rate,
the increase of which is usually expensive and limited by hardware.
In contrast, taking many snapshots only results in delay in the support recovery and is usually less expensive than raising the sampling rate.
Similarly, in DOA estimation \cite{MalCW05} and in distributed sensor networks \cite{BDWSB09},
increasing $m$ requires more sensors, which is expensive,
whereas increasing the number of snapshots $N$ corresponds to delay in estimation, which is relatively less expensive.
This motivates the setting $N \geq m$ in the analysis of this subsection.

\noindent\textbf{Assumption~3 (Signal)}
We assume that the nonzero rows of $X_0$ follow the \textit{mixed multichannel model} given by
\begin{equation*}
X_0^{J_0} = \Psi \Lambda \Phi
\end{equation*}
where $\Psi \in \mathbb{K}^{s \times M}$ with $M \leq s$ is a mixing matrix that has full column rank,
$\Lambda \in \mathbb{R}^{M \times M}$ is a deterministic, positive, and diagonal matrix,
and the elements of $\Phi \in \mathbb{K}^{M \times N}$ are independent zero mean and unit variance Gaussian random variables.
Note that $\rank(X_0^{J_0}) = \rank(\Phi) = M$ with probability 1.
We assume that $\Phi$ is independent of $W$.
The rows of $\Lambda \Phi$ correspond to realizations of $M$ statistically independent sources,
where the diagonal entries of $\Lambda$ represent the magnitudes of the sources.
The columns of $X_0^{J_0}$ in this model are independent realizations of Gaussian vector $x \in \mathbb{K}^s$ with $\mathbb{E} x = 0$
and $\mathbb{E} xx^* = \Psi \Lambda^2 \Psi^*$.

The mixed multichannel model generalizes the \textit{multichannel model} \cite{GRSV08}
proposed for the average case analysis of various methods for joint sparse recovery.
With $\Psi = I_s$, the mixed multichannel model reduces to the multichannel model.
However, with a rectangular mixing matrix $\Psi \in \mathbb{K}^{s \times M}$ for $M < s$,
the mixed multichannel model can describe the ``rank defect'',
which is due to the correlation between the mixed sources, \textit{i.e.}, between the rows of $X_0^{J_0}$.
Such correlation, which often arises in the above mentioned applications,
cannot be represented by the multichannel model, in which $X_0^{J_0}$ has full row rank with probability one for $N \geq s$.

With the mixed multichannel model, the $N$ columns of $A_{J_0} X_0^{J_0} \in \mathbb{K}^{m \times N}$
are independently distributed Gaussian vectors with zero mean and covariance matrix
\begin{equation*}
\Gamma \triangleq A_{J_0} \Psi \Lambda^2 \Psi^* A_{J_0}^*.
\end{equation*}

\noindent\textbf{Assumption~4 (Covariance matrix)}
We assume that there exists a significant gap between at least one pair of consecutive eigenvalues of $\Gamma$, more specifically,
the covariance matrix $\Gamma$ satisfies the following conditions given by the parameters $r \in \mathbb{N}$ and $\tau, \nu, \theta \in (0,1)$:
\begin{align}
{} & (1-\theta)\lambda_r(\Gamma) - (1+\theta)\lambda_{r+1}(\Gamma) \nonumber \\
{} & \geq (1+\theta)(1+\nu)\tau\lambda_1(\Gamma) \label{eq:gapGamma1} \\
{} & (1+\theta)\lambda_k(\Gamma) - (1-\theta)\lambda_{k+1}(\Gamma) \nonumber \\
{} & < (1-\theta)(1-\nu)\tau\lambda_1(\Gamma), \quad \forall k > r. \label{eq:gapGamma2}
\end{align}
Condition (\ref{eq:gapGamma1}) asserts that there exists a significant gap between two consecutive eigenvalues $\lambda_r(\Gamma)$ and $\lambda_{r+1}(\Gamma)$.
Condition (\ref{eq:gapGamma2}) asserts that there does not exist a significant gap between any two consecutive eigenvalues smaller than $\lambda_r(\Gamma)$.
Together, the two conditions imply that $r$ is the maximal value that satisfies (\ref{eq:gapGamma1})
(a gap can not be both big enough and small enough at the same time).

When $\Gamma$ is well conditioned, such that its condition number $\kappa(\Gamma)$ satisfies
\begin{equation*}
\kappa^2(\Gamma) \triangleq \frac{\lambda_1(\Gamma)}{\lambda_M(\Gamma)} \leq \frac{1-\theta}{(1+\theta)(1+\nu)\tau}.
\end{equation*}
then $r$ that satisfies (\ref{eq:gapGamma1})--(\ref{eq:gapGamma2})) will assume its maximal value of $\rank(\Gamma)=M$.
In this case $\lambda_{r+1}(\Gamma) = 0$, and (\ref{eq:gapGamma2}) is trivially satisfied.
Otherwise, we consider that $\Gamma$ is ill-conditioned with one or more insignificant eigenvalues
and set $r$ to the index of the smallest eigenvalue larger than those considered insignificant.
In this case, (\ref{eq:gapGamma2}) implies that $\lambda_{r+1}(\Gamma)$ is bounded from above by
\begin{equation*}
\lambda_{r+1}(\Gamma) < (1-\nu)\tau\lambda_1(\Gamma)\sum_{\ell=1}^{M-r}\left(\frac{1-\theta}{1+\theta}\right)^\ell.
\end{equation*}

\begin{proposition}
Suppose that Assumptions A1--A4 hold and define
\begin{equation}
C_{\eta,\nu,\theta,\tau} \triangleq (1+\theta)\tau \min\left\{ \frac{(1+\nu)\eta}{3} ,~ \frac{\nu}{2+\tau} \right\}.
\label{eq:prop:estss_aca:const}
\end{equation}
Let $\bar{S}$ be the subspace spanned by the $r$ dominant eigenvectors of $\Gamma_S$ defined by
\begin{equation*}
\Gamma_S \triangleq \frac{A_{J_0} X_0^{J_0} (X_0^{J_0})^* A_{J_0}^*}{N}.
\end{equation*}
Let $\epsilon,\eta \in (0,1)$.
If the number of snapshots $N$ satisfies
\begin{align}
N {} & > 2(m+s), \label{eq:prop:estss_aca:cond1} \\
N {} & \geq \left(\frac{36}{\theta^2}\right) \left[s + \ln\left(\frac{8}{\epsilon}\right)\right], \label{eq:prop:estss_aca:cond2} \\
N {} & \geq \left(\frac{144}{C_{\eta,\nu,\theta,\tau}^2}\right) \left(\frac{\sigma_w^2}{\lambda_1(\Gamma)} + 2\sqrt{\frac{\sigma_w^2}{\lambda_1(\Gamma)}}\right) \nonumber \\
{} & \quad \cdot \left[m+s+\ln\left(\frac{8}{\epsilon}\right)\right], \label{eq:prop:estss_aca:cond3}
\end{align}
then with probability $1 - \epsilon$, Algorithm~\ref{alg:estss} with parameter $\tau$ computes an $r$-dimensional subspace $\widehat{S}$
such that
\begin{equation}
\norm{P_{\widehat{S}} - P_{\bar{S}}} \leq \eta.
\label{prop:estss_aca:res}
\end{equation}
\label{prop:estss_aca}
\end{proposition}
\begin{IEEEproof}
See Appendix~\ref{subsec:prop:estss_aca}.
\end{IEEEproof}

When the noise variance $\sigma_w^2$ is small compared to $\lambda_1(\Gamma)$,
Condition (\ref{eq:prop:estss_aca:cond3}) dominates the first two and is simplified as
\begin{equation*}
N = O\left( \frac{(\sigma_w^2/\lambda_1(\Gamma))^{1/2}[m+s+\ln(8/\epsilon)]}{\eta^2 }\right)
\end{equation*}
so that the number of snapshots required for the guarantee scales linearly in $m$.
Alternatively, in the same scenario, Proposition~\ref{prop:estss_aca} implies that (\ref{prop:estss_aca:res})
holds for
\begin{equation*}
\eta = O\left( \left(\frac{\sigma_w^2}{\lambda_1(\Gamma)}\right)^{1/4} \sqrt{\frac{m+s+\ln(8/\epsilon)}{N}}\right).
\end{equation*}

Define the average per-sample SNR as the ratio of the powers of the measured signal and noise,
\begin{equation*}
\text{SNR} \triangleq \frac{\mathbb{E} \norm{A_{J_0} X_0^{J_0}}_F^2}{\mathbb{E} \norm{W}_F^2}
= \frac{\sum_{k=1}^M\lambda_k(\Gamma)}{m \sigma_w^2}.
\end{equation*}
Then, $\lambda_1(\Gamma)/\sigma_w^2$ is related to the SNR by
\begin{equation*}
\frac{\lambda_1(\Gamma)}{\sigma_w^2} = \left(\frac{\lambda_1(\Gamma)}{\frac{1}{m} \mathrm{tr}(\Gamma)}\right) \cdot \text{SNR}.
\end{equation*}
For fixed $\Gamma$ and SNR, $\eta$ scales proportionally to $N^{-1/2}$.
With more snapshots, SA-MUSIC algorithms access an estimate of signal subspace with higher accuracy (smaller $\eta$)
and hence, as shown in Figs.~\ref{fig:tradeoff_samusic} and \ref{fig:tradeoff_samusic2},
the admissible $\delta$ increases, which results in decrease of the required oversampling factor $m/s$.
Eventually, as $N$ goes to infinity, the performance converges to that in the noiseless case.

%%%%%%%%%%%%
\section{Numerical Experiments}\label{sec:numerical_results}
%%%%%%%%%%%%

We compared the performance of the two proposed SA-MUSIC algorithms:
SA-MUSIC + SS-OMP and SS-MUSIC + SS-OMSP
versus MUSIC \cite{FenBre96}, M-BP\footnote{The noise variance is given to the M-BP algorithm in the experiment.},
and the two subspace greedy algorithms proposed in this paper for partial support recovery: SS-OMP and SS-OMSP.
As an upper bound on the performance of SA-MUSIC, we included in the comparison SA-MUSIC with known (``oracle'') partial support.
The sensing matrix $A$ was generated as randomly selected $m$ rows of the $n \times n$ DFT matrix.
The snapshot matrix $Y = A X_0 + Z$ was corrupted by additive i.i.d. circular complex Gaussian noise $Z$.

The algorithms were tested on random $X_0^{J_0}$ of rank less than $s$.
Singular vectors $U_0$ and $V_0$ of $X_0^{J_0} = U_0 \Sigma_0 V_0^*$ were generated as random orthonormal columns.
In order to observe the effect of the rank-defect rather than ill-conditioning,
the singular values of $X_0^{J_0}$ are set to a common value in the first experiment.
The performance is assessed by the rate of successful support recovery
\footnote{
M-BP does not produce an $s$-sparse solution in the presence of noise.
In this case, the solution by M-BP has been approximated to the nearest $s$-sparse vector
and the support is computed as that of the $s$-sparse approximation.
}.
\begin{figure*}[htb]
\begin{center}
\begin{minipage}[htb]{0.45\linewidth}
\centerline{\includegraphics[height=55mm]{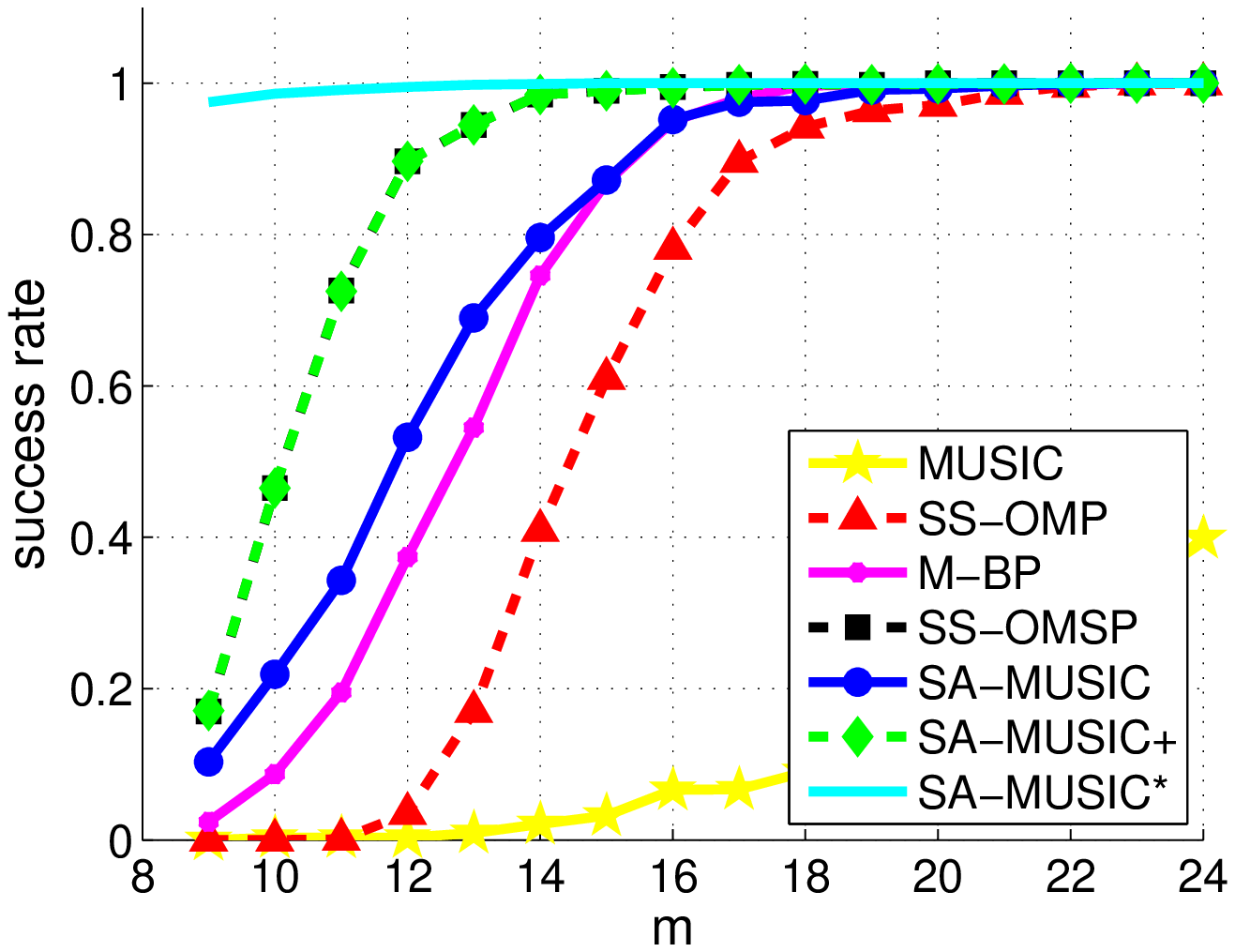}}
\centering{\footnotesize(a)}
\end{minipage}
\begin{minipage}[htb]{0.45\linewidth}
\centerline{\includegraphics[height=55mm]{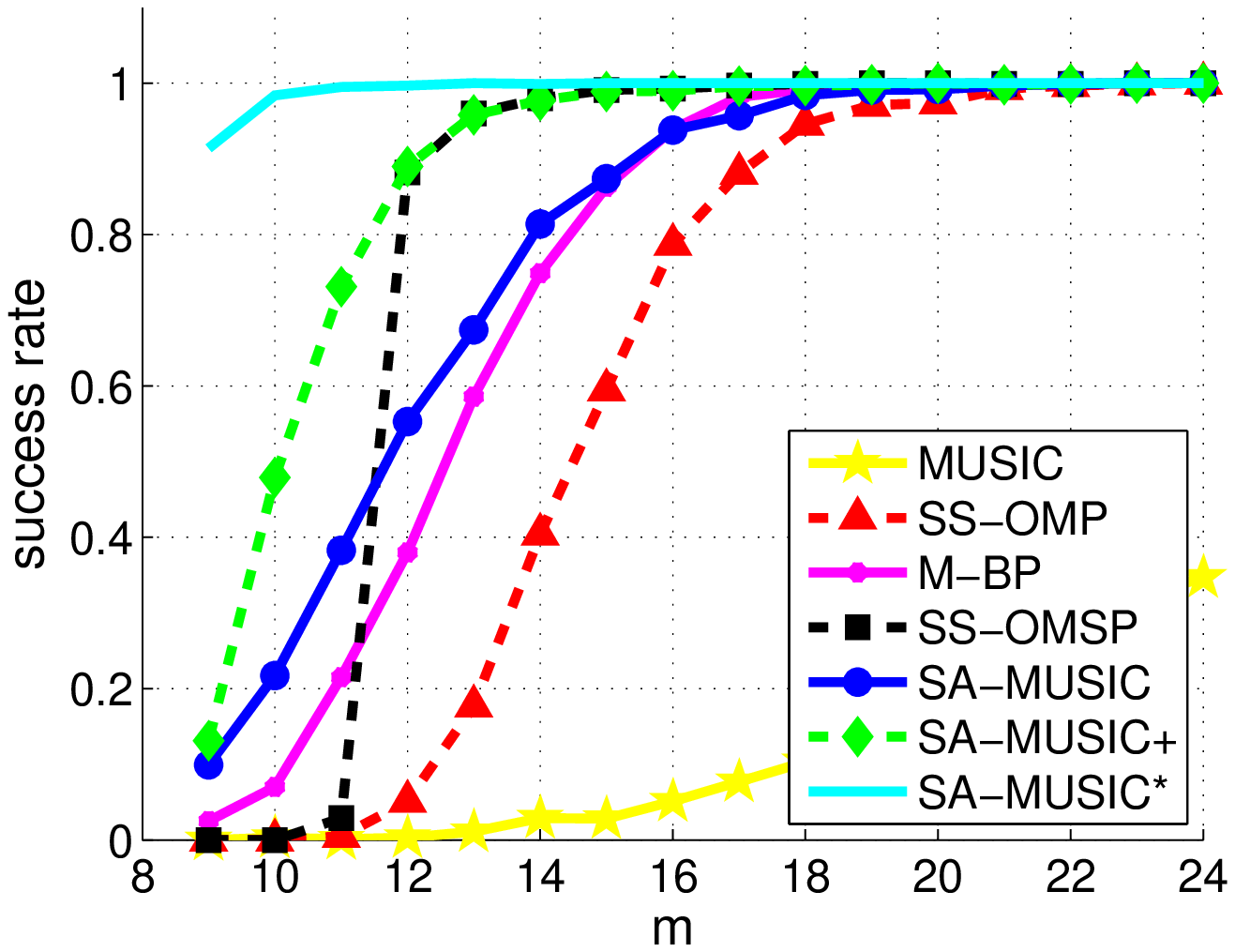}}
\centering{\footnotesize(b)}
\end{minipage}
\begin{minipage}[htb]{0.45\linewidth}
\centerline{\includegraphics[height=55mm]{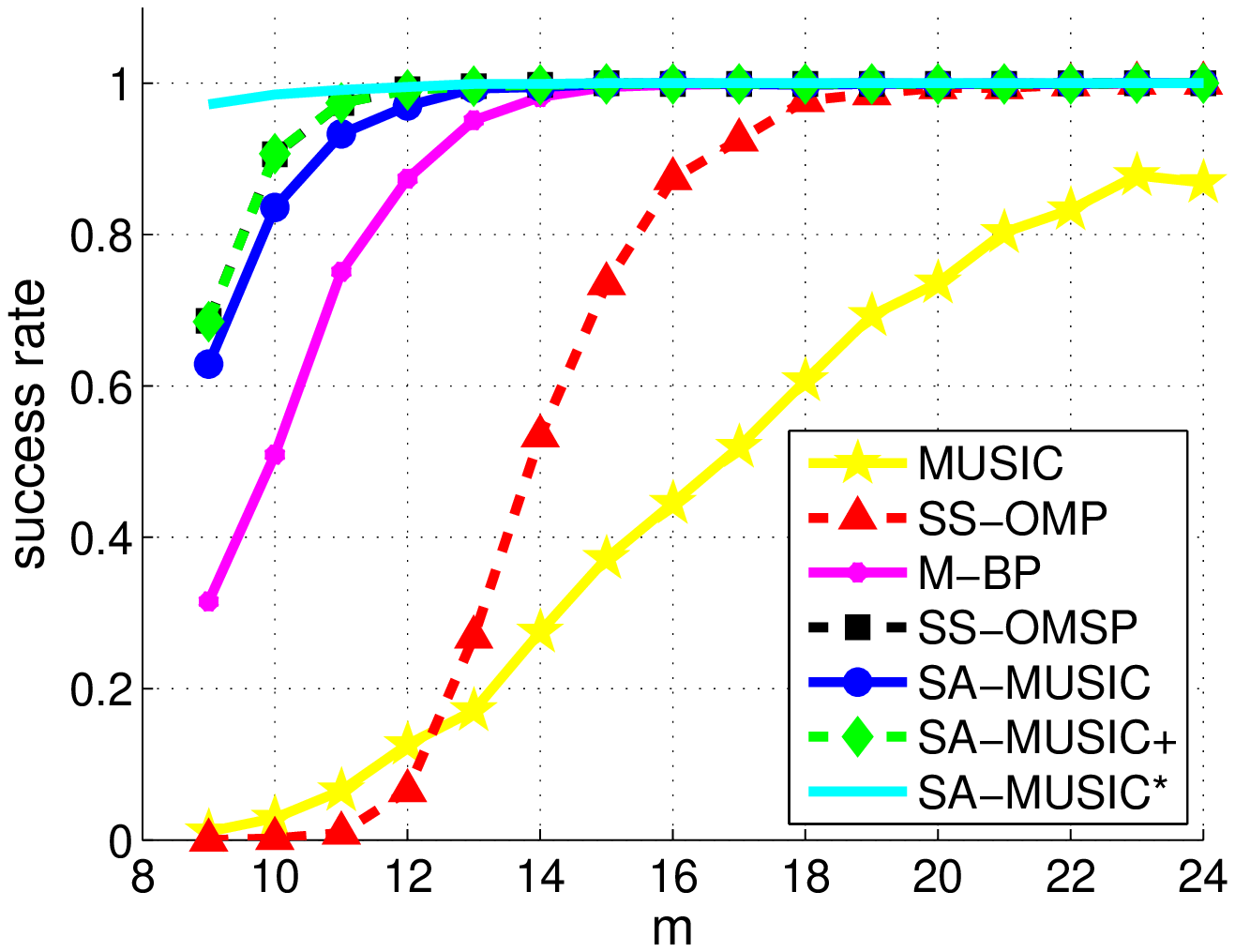}}
\centering{\footnotesize(c)}
\end{minipage}
\begin{minipage}[htb]{0.45\linewidth}
\centerline{\includegraphics[height=55mm]{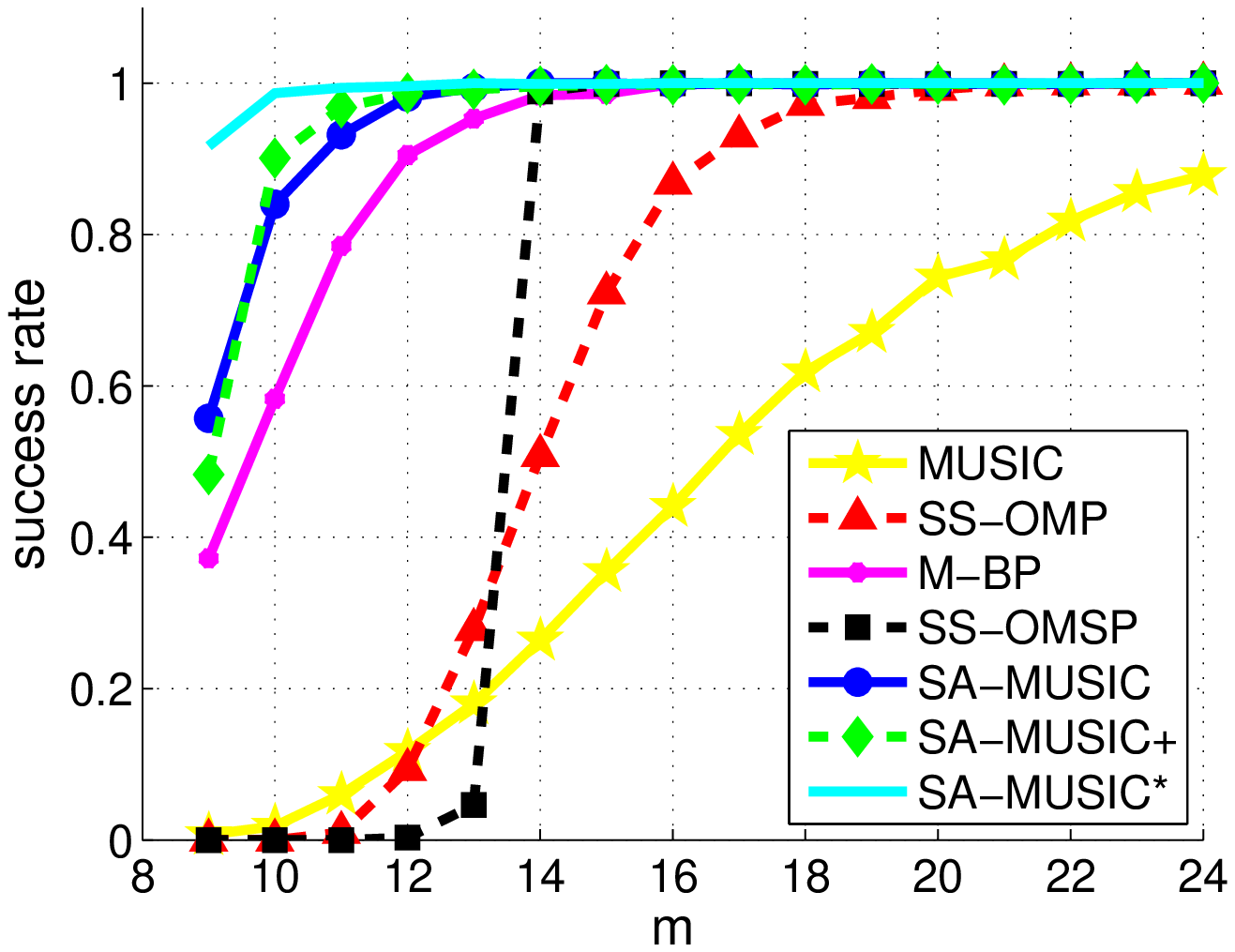}}
\centering{\footnotesize(d)}
\end{minipage}
\begin{minipage}[htb]{0.45\linewidth}
\centerline{\includegraphics[height=55mm]{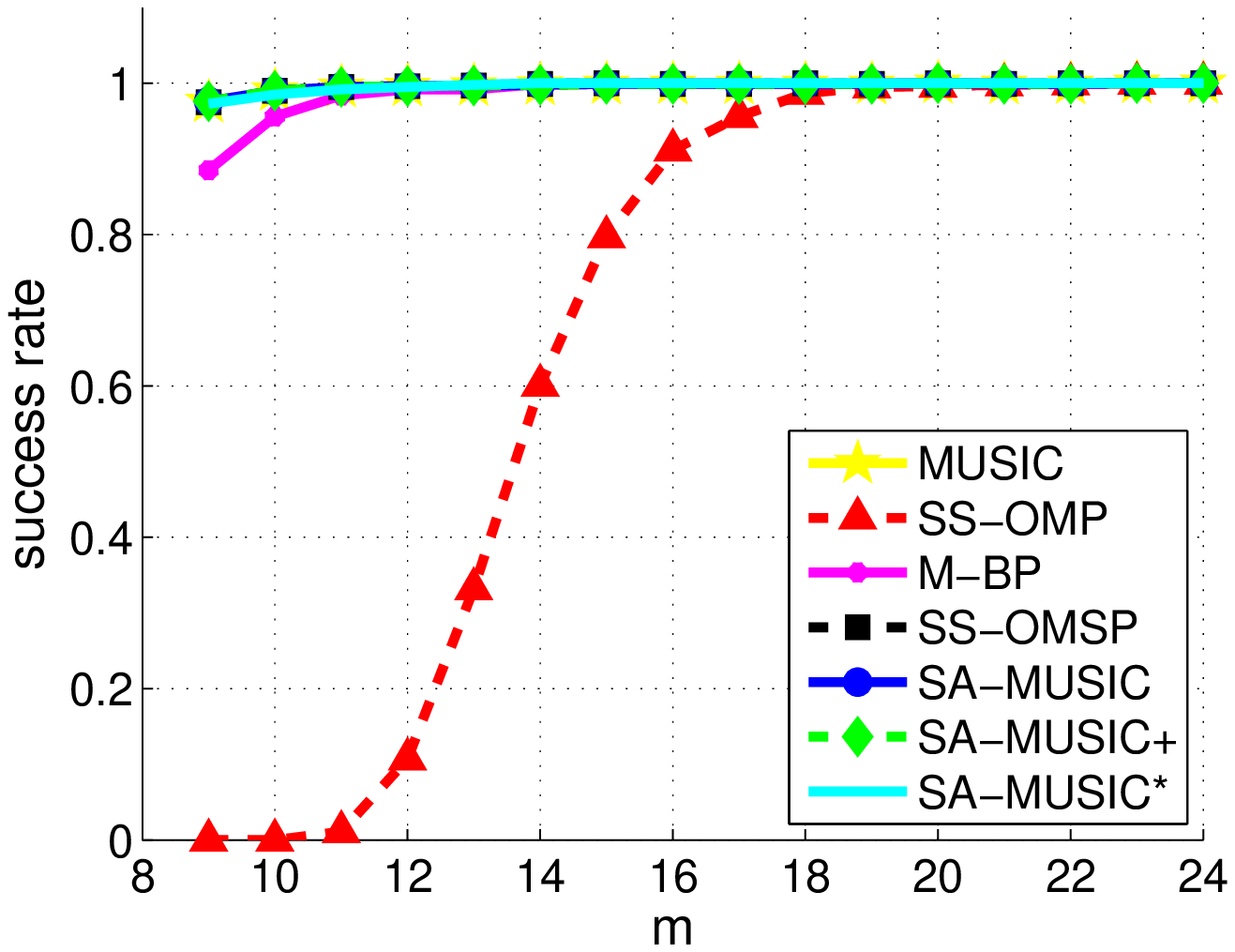}}
\centering{\footnotesize(e)}
\end{minipage}
\begin{minipage}[htb]{0.45\linewidth}
\centerline{\includegraphics[height=55mm]{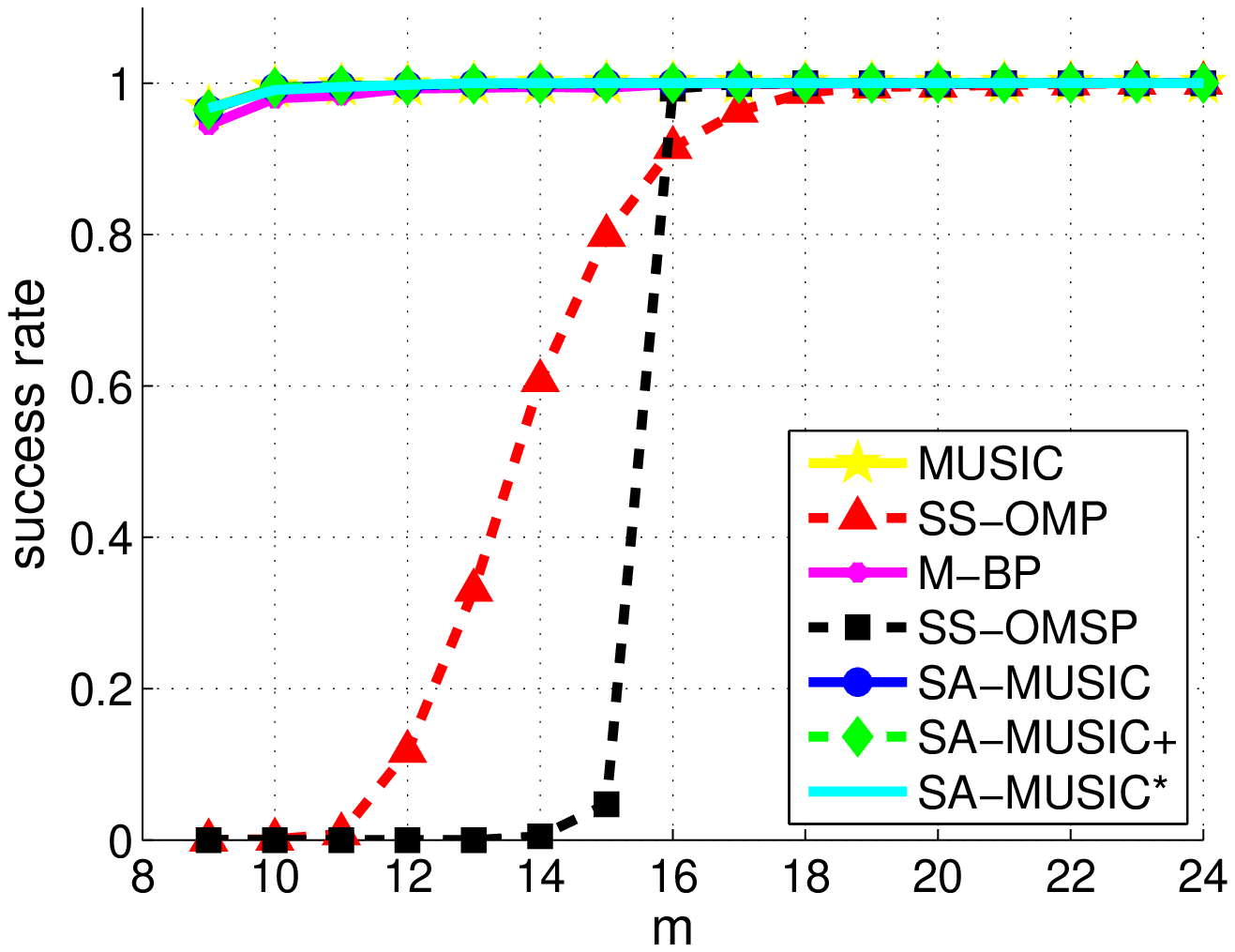}}
\centering{\footnotesize(f)}
\end{minipage}
\end{center}
\caption{Test on rank-defect, $n = 128$, $s = 8$, $N = 256$,
Left columns (noiseless): (a) $\rank(X_0^{J_0}) = 4$. (c) $\rank(X_0^{J_0}) = 6$.
(e) $\rank(X_0^{J_0}) = 8$ (full row rank).
Right columns (SNR = 30 dB): (b) $\rank(X_0^{J_0}) = 4$. (d) $\rank(X_0^{J_0}) = 6$.
(f) $\rank(X_0^{J_0}) = 8$ (full row rank). }
\label{fig:res1R}
\end{figure*}
As shown in Fig.~\ref{fig:res1R}, MUSIC fails when $\rank(X_0^{J_0}) < s$.
SS-OMP is little affected by the rank-defect,
but its performance does not improve much with increasing $r$.

The performance of SA-MUSIC algorithms with greedy partial support recovery improves with increasing $r$.
SA-MUSIC + SS-OMP and SA-MUSIC + SS-OMSP, labeled as ``SA-MUSIC'' and ``SA-MUSIC+'', respectively, in Fig.~\ref{fig:res1R}, 
performed better than M-BP in this experiment at much lower computational cost.
SA-MUSIC with known partial support of size $s-r$ is labeled as ``SA-MUSIC with oracle'' in Fig.~\ref{fig:res1R}
and shows perfect recovery when $m > s+1$, which is nearly the necessary condition $m > s$.
This suggests that the success of partial support recovery is more critical than the subsequent steps
and leaves room for improving SA-MUSIC by combining with a better algorithm for partial support recovery than SS-OMP or SS-OMSP.

For the noiseless case, the performance of SA-MUSIC+SS-OMSP and SS-OMSP coincides.
However, with noise, the performance SS-OMSP severely degrades even for the full row rank case.
For the full row rank case, all algorithms except SS-OMP and SS-OMSP (noisy case) were successful in terms of nearly achieving the necessary condition $m > s$.
Again, SS-OMSP is sensitive to the perturbation in the estimate of signal subspace in this case.

\begin{figure*}[htb]
\begin{center}
\begin{minipage}[htb]{0.49\linewidth}
\centerline{\includegraphics[height=55mm]{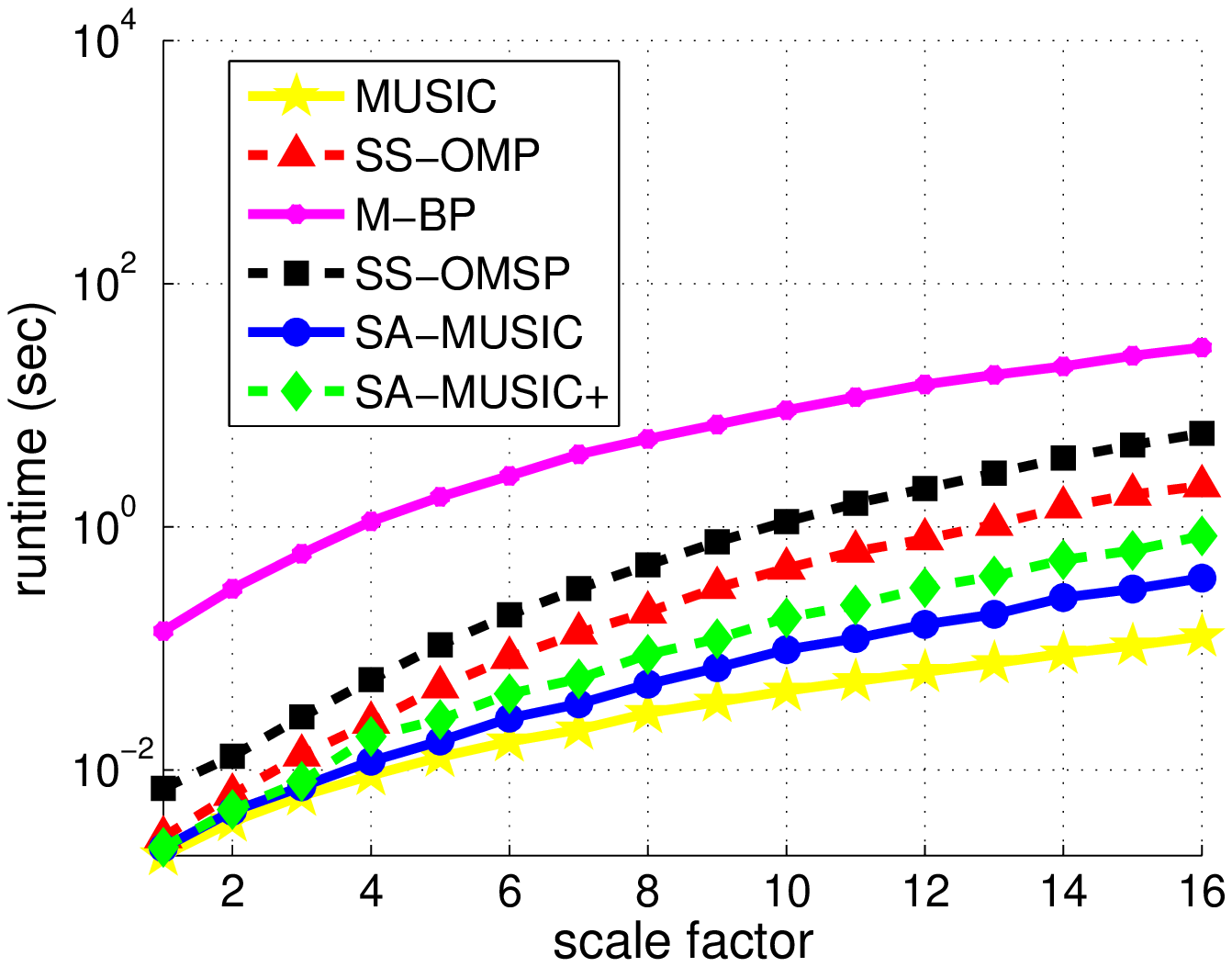}}
\end{minipage}
\end{center}
\caption{Comparison of runtime $N = 256$, SNR = 30 dB.}
\label{fig:runtime}
\end{figure*}

Regarding the computation, we compared the runtime of each algorithm by increasing the size of the problem.
In this experiment, fixing $n = (\text{scale factor}) \times 64$, we set the other parameters to $s = n/16$, $r = \lceil 7s/8 \rceil$, and $m = 2s$.
As shown in Fig.~\ref{fig:runtime}, SA-MUSIC is about a 100 times faster than M-BP
\footnote{
For M-BP, we used an efficient implementation \texttt{SPGL1} \cite{BerFei08}, \cite{spgl1}.
On the other hand, the other methods were implemented as plain Matlab script.
Therefore, the speed comparison does not unfairly favor SA-MUSIC.}.

In order to see the effect of ill-conditioning,
in the second experiment, we tested the algorithms on
a random matrix $X_0^{J_0} \in \mathbb{C}^{s \times N}$ with full row rank that has geometrically decaying singular values.
The $k$-th largest singular value of $X_0^{J_0}$ is set as $\sigma_k(X_0^{J_0}) = \kappa^{-(k-1)/(s-1)}$
for $k = 1,\ldots,s$ so that the condition number of $X_0^{J_0}$ becomes $\kappa$.
The singular vectors were generated randomly as in the first experiment.
\begin{figure*}[htb]
\begin{center}
\begin{minipage}[htb]{0.45\linewidth}
\centerline{\includegraphics[height=55mm]{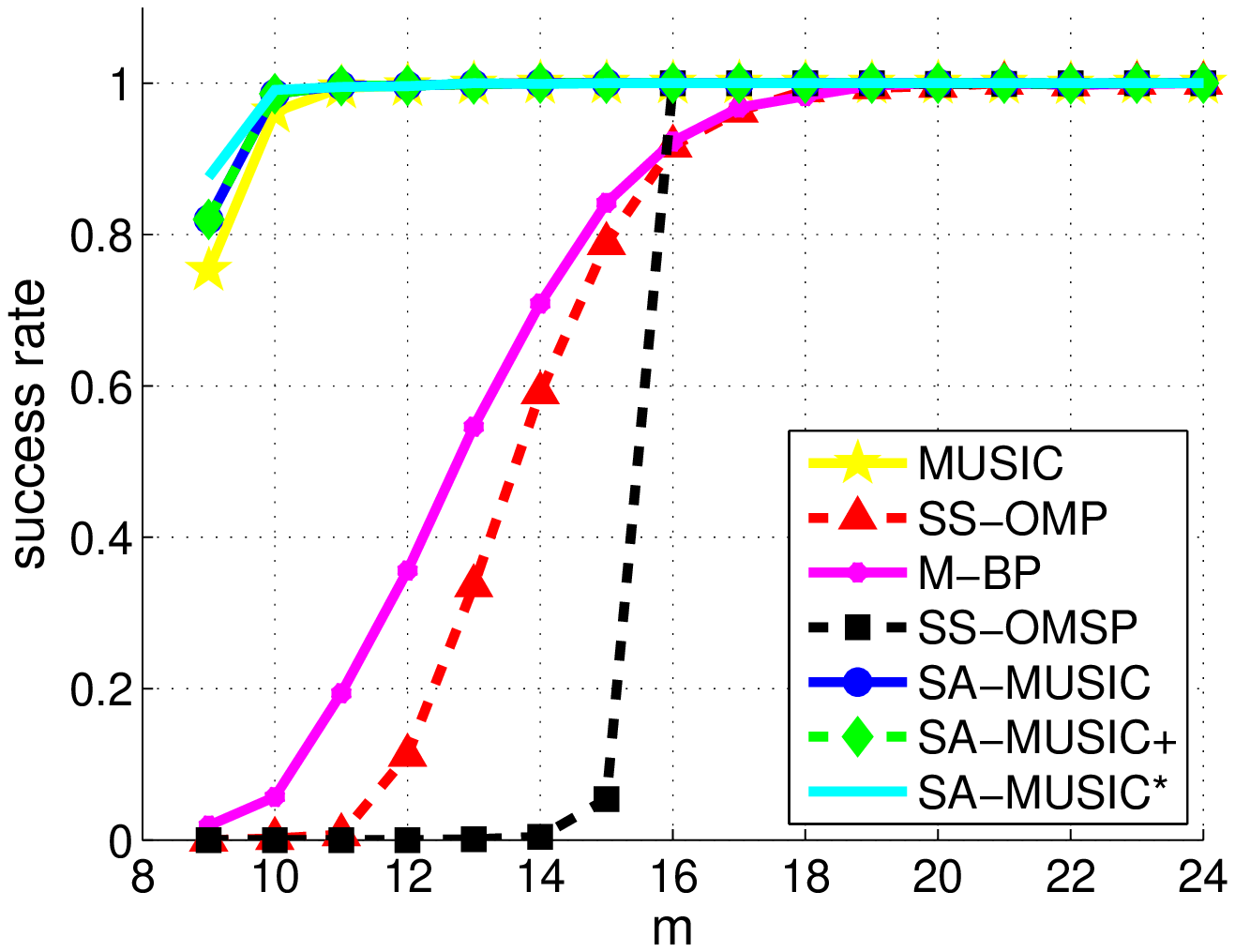}}
\centering{\footnotesize(a)}
\end{minipage}
\begin{minipage}[htb]{0.45\linewidth}
\centerline{\includegraphics[height=55mm]{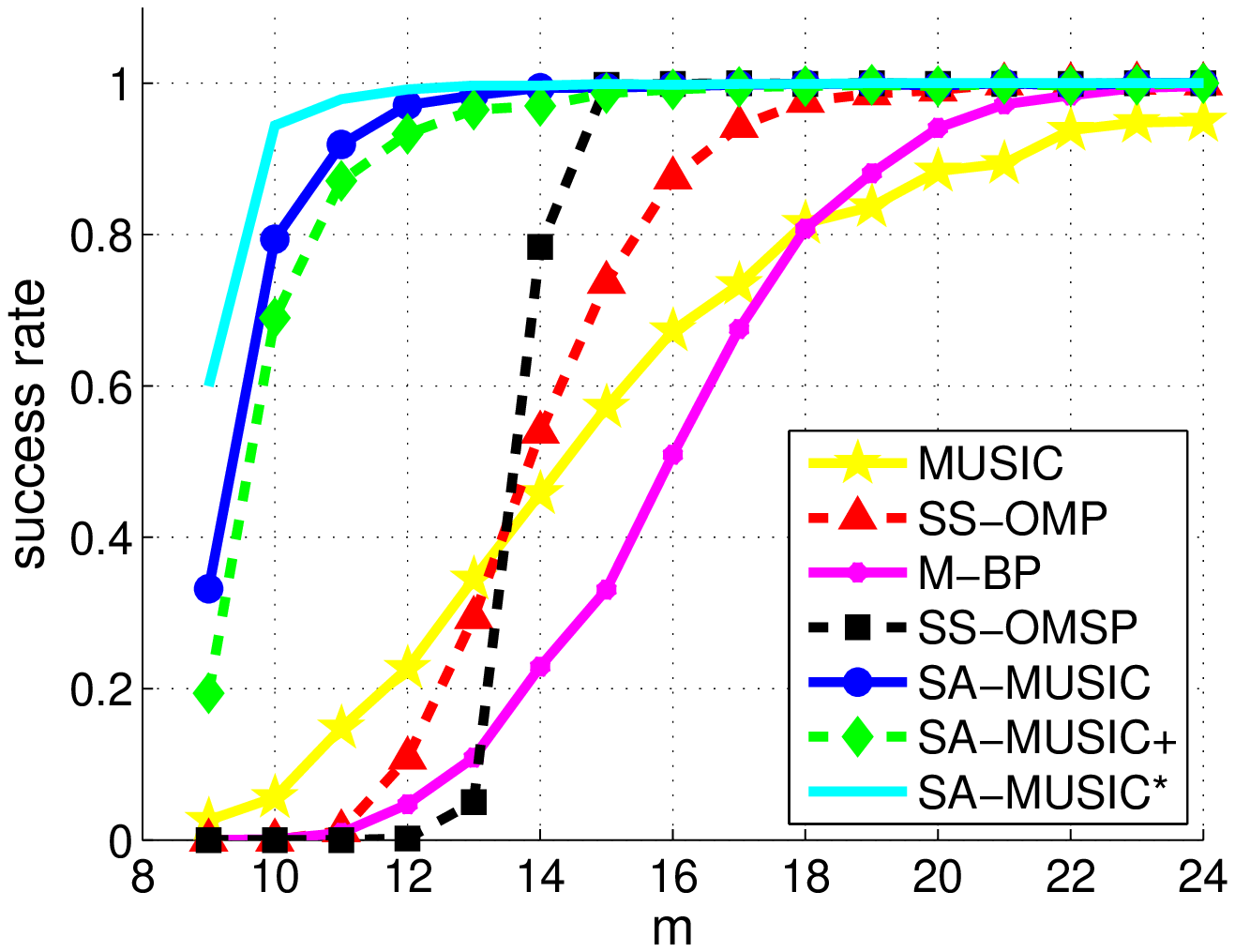}}
\centering{\footnotesize(b)}
\end{minipage}
\end{center}
\caption{Test on large condition number, $n = 128$, $s = 8$, $\rank(X_0^{J_0}) = s$ (full row rank), $N = 256$.
(a) $\kappa = 10$, SNR = 30 dB. (b) $\kappa = 50$, SNR = 30 dB.}
\label{fig:res2R}
\end{figure*}
Fig.~\ref{fig:res2R} compares the performance of the algorithms for the weak noise case.
We note that M-BP is sensitive to the ill-conditioning of $X_0^{J_0}$.
When $X_0^{J_0}$ is well conditioned with $\kappa = 10$,
the dimension of the estimated signal subspace is equal to the row rank of $X_0^{J_0}$ ($=s$)
and hence SA-MUSIC coincides with MUSIC without any SS-OMP step.
However, when $\kappa = 50$, the estimated rank $r$ by using (\ref{eq:estss:thresrank}) in Algorithm~\ref{alg:estss} with $\tau = 10^3$ is smaller than $\rank(X_0^{J_0})$ and hence MUSIC suffers from the rank-defect while SA-MUSIC provides consistent performance.
The performance of M-OMP is invariant to the rank-defect but is poor compared to that of SA-MUSIC.

%%%%%%%%%%%%
\section{Discussion}\label{sec:discussion}
%%%%%%%%%%%%

\subsection{Comparison to Compressive MUSIC}
An algorithm similar to SA-MUSIC named ``Compressive MUSIC'' (CS-MUSIC) has been independently proposed by Kim \textit{et al.} \cite{KimLY10}.
Although the main ideas in SA-MUSIC and compressive MUSIC are similar, in fact,
the two papers differ in the following ways.
First, the algorithms considered are different, in particular in the step of partial support recovery.
Second, the analyses in the two papers are fundamentally different.
The analysis of Kim \textit{et al.} \cite{KimLY10} depends heavily on
the assumption that $A$ is an i.i.d. Gaussian matrix and the size of the problem goes to infinity satisfying certain scaling laws.
The authors showed that, under certain conditions,
the probability of failure in the support recovery converges to 0 in their ``large system model''.
However, since no convergence rate is shown, the analysis provides no guarantee on any finite dimensional problem.
In contrast, the guarantees in this paper are non-asymptotic and based on the weak-1 RIP.
Our guarantees provide explicit formulae for the required $m$ as functions of $s$ and $n$,
for various sensing matrices $A$ including i.i.d. Gaussian, random partial Fourier, and incoherent unit-norm tight frame,
whereas the analysis in \cite{KimLY10} only applies to an i.i.d. Gaussian $A$.

%%%%%%
\subsection{Comparison to the Guarantee of M-BP with the Multichannel Model}
%%%%%%
Various practical algorithms including $p$-SOMP, $p$-thresholding, and M-BP,
have been analyzed under the multichannel model \cite{GRSV08}, \cite{EldRau09}.
Although it is restricted to the noiseless case,
the average case guarantees of M-BP with the multichannel model \cite{EldRau09}
has been shown to be better than the other guarantees of the same kind for other algorithms.
Therefore, we compare the guarantees of SA-MUSIC algorithms to that of M-BP.

For this comparison, we too assume that the snapshots are noise-free, \textit{i.e.}, $Y = A_{J_0} X_0^{J_0}$.
Nevertheless, the guarantee of SA-MUSIC algorithms in this paper is restricted neither to the noiseless case nor to the multichannel model.

In the noiseless case, the signal subspace estimation is perfect, $\widehat{S} = S \triangleq \mathcal{R}(A_{J_0} X_0^{J_0})$,
with $r \triangleq \dim(\widehat{S}) = \rank(X_0^{J_0})$.
If $N \geq s$ where $s$ is the sparsity level, then $X_0^{J_0}$ following the multichannel model has full row rank with probability 1.
In the full row rank case, any SA-MUSIC algorithm reduces to MUSIC and provides the best possible guarantee with the minimal requirement $\alpha_{s+1}^\text{weak}(A;J_0) > 0$, which reduces to $m > s$ for certain matrices such as i.i.d. Gaussian $A$.
This completes the comparison in the case $N \geq s$.
Therefore, to compare the performance of SA-MUSIC and M-BP in the rank-defective case, we assume that $N < s$.
The rank of $X_0^{J_0}$ is then determined by the number of snapshots, \textit{i.e.}, $\rank(X_0^{J_0}) = N$ and hence $r = N$.

Previous work \cite[Theorem~4.4]{EldRau09} showed that M-BP is guaranteed with probability $1-\epsilon$
if $A$ satisfies the weak-1 RIP
\begin{equation}
\delta_{s+1}^\text{weak}(A;J_0) < \delta
\label{eq:mbp:cond1}
\end{equation}
for $\delta$ satisfying
\begin{equation}
\left(\frac{\delta}{1-\delta}\right)^{-2} + 2 \ln \left(\frac{\delta}{1-\delta}\right) \geq \frac{2\ln(n/\epsilon)}{N} + 1.
\label{eq:mbp:cond2}
\end{equation}

\begin{figure*}[htb]
\begin{center}
\begin{minipage}[htb]{0.45\linewidth}
\centerline{\includegraphics[height=55mm]{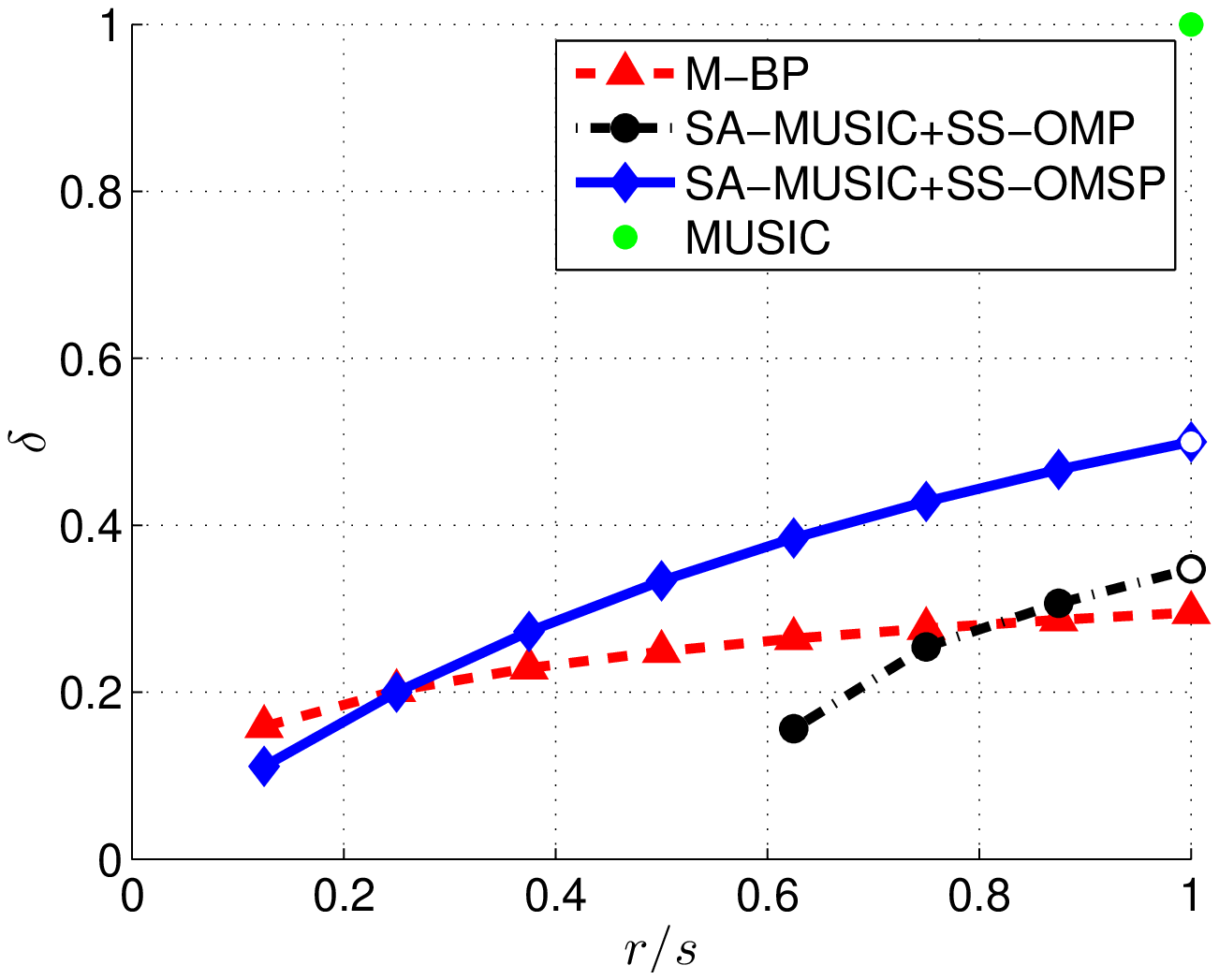}}
\centering{\footnotesize(a)}
\end{minipage}
\begin{minipage}[htb]{0.45\linewidth}
\centerline{\includegraphics[height=55mm]{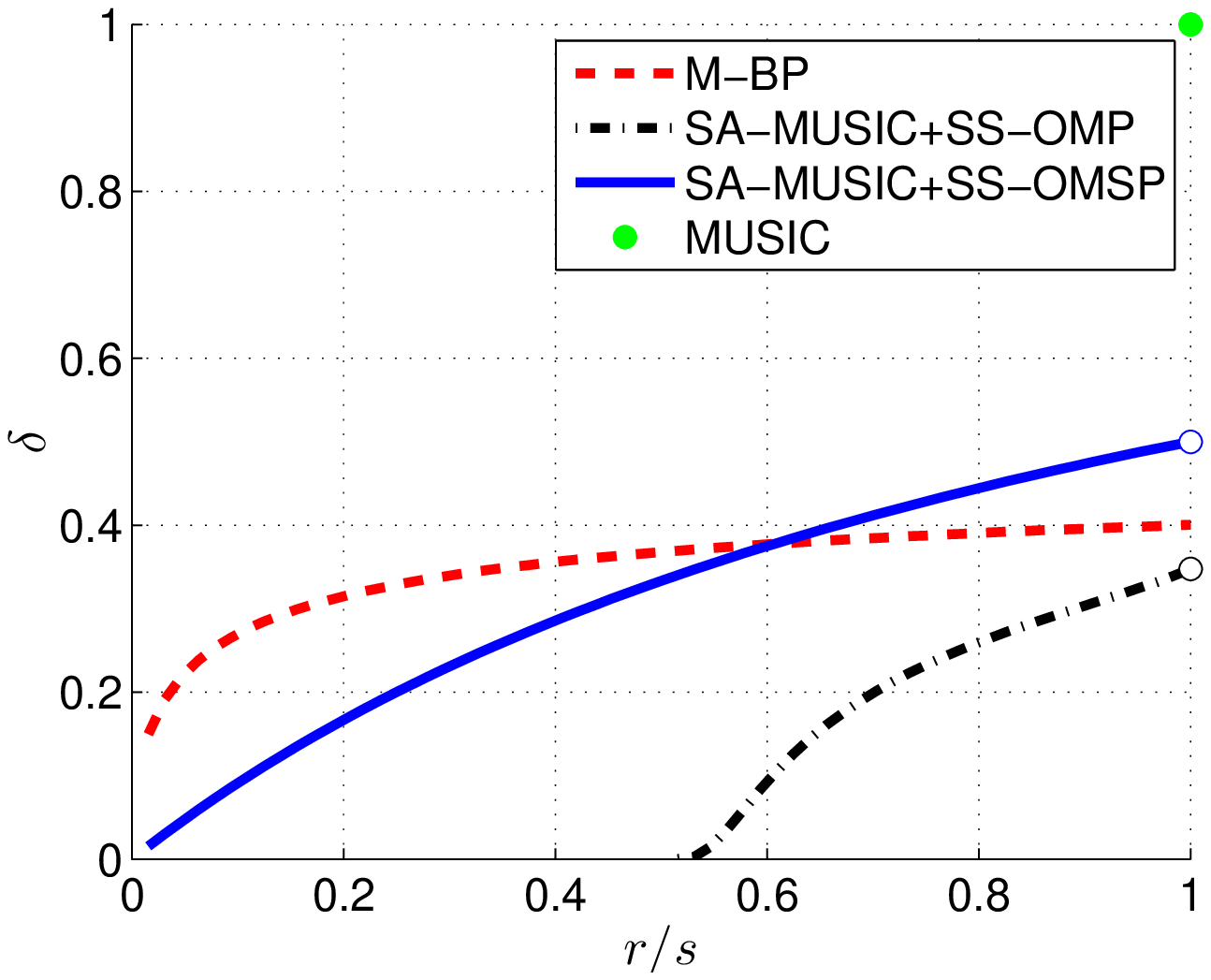}}
\centering{\footnotesize(b)}
\end{minipage}
\end{center}
\caption{Required weak-1 RIC for the guarantees of M-BP (the average case analysis with the multichannel model with error probability $\epsilon = 10^{-3}$), and SA-MUSIC (worst case analysis) for the noiseless case (a) $n = 128, s = 8$. (b) $n = 1024, s = 64$.}
\label{fig:gammadelta}
\end{figure*}

SA-MUSIC+SS-OMP and SA-MUSIC+SS-OMSP are guaranteed by Theorems~\ref{thm:samusic} and \ref{thm:samusic2}, respectively.
In particular, when $X_0^{J_0}$ follows the multichannel model, it is row-nondegenerate with probability 1.
Therefore, we need only compare the weak-1 RIP conditions in Theorems~\ref{thm:samusic} and \ref{thm:samusic2} to the weak-1 RIP given by (\ref{eq:mbp:cond1})+(\ref{eq:mbp:cond2}).
Fig.~\ref{fig:gammadelta} displays this comparison.

For all three algorithms, as $r$ increases, $\delta$ required for the guarantee increases
and hence the guarantee is obtained subject to a milder condition.
Fig~\ref{fig:gammadelta} (a) shows that SA-MUSIC+SS-OMSP requires larger RIC and hence requires reduced oversampling factor $m/s$
compared to M-BP when the size of the problem is small ($n = 128$).
Fig~\ref{fig:gammadelta} (b) shows that SA-MUSIC+SS-OMSP provides a better guarantee (larger RIC) than M-BP in the regime $r/s \geq 0.6$ when $n = 1024$.

The theoretical guarantee not withstanding, in our simulations,
the recovery rate of the SA-MUSIC algorithms was always higher than that of M-BP and often substantially so.

%%%%%%
\subsection{Comparison to the Analysis of Group LASSO in High Dimension}
%%%%%%
The guarantee of Group LASSO by Obozinski \textit{et al.} \cite{OboWJ11}
is quite tight and, in particular, achieves the optimal guarantee by the minimal requirement ($m > s$) for certain scenarios.
However, their guarantee is asymptotic and only applies to Gaussian $A$.
In contrast, although our guarantee of SA-MUSIC+SS-OMP is not as tight as that of Group LASSO \cite{OboWJ11},
the guarantee is non-asymptotic, \textit{i.e.}, valid for any finite problems,
and applies to wider class of matrices that arise in practical applications, including the partial Fourier case.

%%%%%%
\subsection{Comparison to Compressed Sensing with Block Sparsity}
%%%%%%
The joint sparse recovery problem can be cast as a special case of compressed sensing with block sparsity \cite{MisEld09}.
The block structure in the sparsity pattern in the latter problem has been exploited to improve the performance of the sparse recovery
(\textit{cf.} \cite{BDWSB09}, \cite{MisEld09}, \cite{BCDH10}, \cite{LPTG10}).
For example, the analysis \cite{LPTG10} showed improvement of the structured sparse recovery over the unstructured original problem.
However, the special case where the sensing matrix is a block diagonal with repeated blocks,
which corresponds to the joint sparse recovery problem, is not covered by the proposed theory \cite{LPTG10}.

\section*{Acknowledgements}
We thank Wei Dai and Olgica Milenkovic for the discussion that helped improve an earlier version of the draft,
and Yonina Eldar for pointing out her paper \cite{EldRau09} that we find very useful in improving the analysis.
We also appreciate the effort of the anonymous reviewers and their useful comments and suggestions.

\appendix

\numberwithin{equation}{section}
\renewcommand{\theequation}{A.\arabic{equation}}

\subsection{Lemmata}
We list a few lemmata that will be used for the proofs in the appendix.
\begin{lemma}[{\cite[Corollary~III.1.5]{Bha97}}]
Let $A_1 \in \mathbb{K}^{m \times n_1}$ and $A_2 \in \mathbb{K}^{m \times n_2}$.
Let $A \in \mathbb{K}^{m \times n}$ be the concatenation of $A_1$ and $A_2$, \textit{i.e.}, $A = [A_1,~ A_2]$ where $n = n_1 + n_2$.
Then, there is an interlacing relation between the singular values of $A$ and $A_1$ given by
\begin{equation}
\sigma_k(A) \geq \sigma_k(A_1) \geq \sigma_{k+n_2}(A)
\label{eq:interlacing_concat}
\end{equation}
for $k = 1,\ldots,n_1$.
\label{lemma:interlacing_submatrix}
\end{lemma}

\begin{lemma}
Let $A \in \mathbb{K}^{m \times n}$ and let $J, J_0 \subset [n]$.
Then, it follows that, for $k = 1,\ldots,|J_0 \setminus J|$,
\begin{equation}
\begin{aligned}
{} & \lambda_k(A_{J_0 \cup J}^* A_{J_0 \cup J}) \\
{} & \geq \lambda_k(A_{J_0 \setminus J}^* P_{\mathcal{R}(A_J)}^\perp A_{J_0 \setminus J}) \\
{} & \geq \lambda_{k+|J|}(A_{J_0 \cup J}^* A_{J_0 \cup J}),
\label{eq:interlacing_schurcomp}
\end{aligned}
\end{equation}
which is equivalent to
{\allowdisplaybreaks
\begin{align}
\sigma_k(A_{J_0 \cup J})
\geq \sigma_k(P_{\mathcal{R}(A_J)}^\perp A_{J_0 \setminus J})
\geq \sigma_{k+|J|}(A_{J_0 \cup J}^* A_{J_0 \cup J}).
\label{eq:interlacing_schurcomp_s}
\end{align}} %
\label{lemma:interlacing_schur}
\end{lemma}

\begin{IEEEproof}
Note that $A_{J_0 \setminus J}^* P_{\mathcal{R}(A_J)}^\perp A_{J_0 \setminus J}$ is
the Schur complement of the block $A_J^* A_J$ of the matrix
\begin{equation*}
\left[
\begin{array}{cc}
A_{J_0 \setminus J}^* A_{J_0 \setminus J} & A_{J_0 \setminus J}^* A_J \\
A_J^* A_{J_0 \setminus J} & A_J^* A_J
\end{array}
\right]
= \Pi^* A_{J_0 \cup J}^* A_{J_0 \cup J} \Pi
\end{equation*}
where $\Pi$ is a permutation matrix that satisfies
\begin{equation*}
A_{J_0 \cup J} \Pi = [A_{J_0 \setminus J},~ A_J].
\end{equation*}
Application of the interlacing relation of the eigenvalues of the Schur complement \cite{Smi92} completes the proof.
\end{IEEEproof}

\begin{lemma}
Let $A \in \mathbb{K}^{m \times n}$ and $B \in \mathbb{K}^{n \times p}$ where $m \geq n$ and $p \geq n$.
Then,
\begin{equation*}
\norm{AB} \geq \sigma_{n-k+1}(A) \cdot \sigma_k(B)
\end{equation*}
for $k = 1,\ldots,n$.
\label{lemma:lidskii}
\end{lemma}

\begin{IEEEproof}
Let $A = U_1 \Sigma_1 V_1^*$ and $A = U_2 \Sigma_2 V_2^*$ denote the extended SVD of $A$ and $B$, respectively,
where $\Sigma_1,V_1,U_2,\Sigma_2 \in \mathbb{K}^{n \times n}$.
Let $k \in [n]$. Then,
{\allowdisplaybreaks
\begin{align*}
\norm{AB}
{} & = \norm{U_1 \Sigma_1 V_1^* U_2 \Sigma_2 V_2^*} \\
{} & = \norm{\Sigma_1 V_1^* U_2 \Sigma_2} \\
{} & = \norm{V_1 \Sigma_1 V_1^* U_2 \Sigma_2 U_2^*} \\
{} & \geq \lambda_{n-k+1}(V_1 \Sigma_1 V_1^*) \cdot \lambda_k(U_2 \Sigma_2 U_2^*) \\
{} & = \sigma_{n-k+1}(A) \cdot \sigma_k(B)
\end{align*}} %
where the inequality follows from the Gel'fand-Naimark theorem \cite[Theorem~III.4.5]{Bha97}.
\end{IEEEproof}

\subsection{Proof of Proposition~\ref{prop:augssp}}
\label{subsec:prop:augssp}
We use the following lemma to prove Proposition~\ref{prop:augssp}.
\begin{lemma}
Suppose that $\Phi \in \mathbb{K}^{s \times r}$ where $r \leq s$ satisfies
\begin{equation*}
\krank(\Phi^*) = r
\end{equation*}
and that $\Psi \in \mathbb{K}^{s \times k}$ for $k < r$ spans a $k$-dimensional subspace of $\mathcal{R}(\Phi)$.
Then,
\begin{equation*}
\krank(\Psi^*) = k.
\end{equation*}
\label{lemma:fullkrank}
\end{lemma}
\begin{IEEEproof}
There exists $R \in \mathbb{K}^{r \times k}$ such that $R$ has full rank and $\Psi = \Phi R$.
Let $K$ be a set of $k$ indices from $[s]$.
Since $\krank(\Phi^*) = r$ implies $\rank((\Phi^*)_K) = k$,
$\rank((\Psi^*)_K) = \rank(R^* (\Phi^*)_K) = \rank((\Phi^*)_K) = k$.
Since $K$ was arbitrary, we have $\krank(\Psi^*) = k$.
\end{IEEEproof}

\begin{IEEEproof}[Proof of Proposition~\ref{prop:augssp}]
By the projection update formula, we have
\begin{equation*}
P_{\mathcal{R}(A_{J_1}) + \bar{S}} = P_{\mathcal{R}(A_{J_1})} + P_{P_{\mathcal{R}(A_{J_1})}^\perp \bar{S}}.
\end{equation*}
Note that $\mathcal{R}(A_{J_1})$ and $P_{\mathcal{R}(A_{J_1})}^\perp \bar{S}$ are orthogonal to each other
and both are subspaces of $\mathcal{R}(A_{J_0})$.
Furthermore, by assumption, $\rank(A_{J_0}) = s$ and hence $\rank(A_{J_1}) = |J_1| = s-r$.
Therefore, it suffices to show that
\begin{equation}
\dim(P_{\mathcal{R}(A_{J_1})}^\perp \bar{S}) = r.
\label{eq:proof:prop:augssp:eq1}
\end{equation}
Let $U \in \mathbb{K}^{n \times r}$ satisfy $S = \mathcal{R}(A_{J_0} U^{J_0})$ and $U^{[n] \setminus J_0} = 0$.
Then, (\ref{eq:proof:prop:augssp:eq1}) is equivalent to
\begin{equation*}
\rank(P_{\mathcal{R}(A_{J_1})}^\perp A_{J_0} X_0^{J_0}) = r,
\end{equation*}
which holds since
{\allowdisplaybreaks
\begin{align*}
{} & \sigma_r(P_{\mathcal{R}(A_{J_1})}^\perp A_{J_0} U^{J_0}) \\
{} & = \sigma_r(P_{\mathcal{R}(A_{J_1})}^\perp A_{J_0 \setminus J_1} U^{J_0 \setminus J_1}) \\
{} & \overset{\text{(a)}}{\geq} \sigma_r(P_{\mathcal{R}(A_{J_1})}^\perp A_{J_0 \setminus J_1}) \sigma_r(U^{J_0 \setminus J_1}) \\
{} & \overset{\text{(b)}}{\geq} \sigma_s(A_{J_0}) \sigma_r(U^{J_0 \setminus J_1}) > 0
\end{align*}} %
where (a) and (b) follow from Lemma~\ref{lemma:lidskii}, which provides a lower bound on the minimum singular value of the product, and Lemma~\ref{lemma:interlacing_schur}, respectively
and the last step follows from the assumption that $A_{J_0}$ has full column rank and $\krank((U^{J_0})^*) = r$,
which holds by Lemma~\ref{lemma:fullkrank} because $X_0^{J_0}$ is row-nondegenerate.
\end{IEEEproof}%[Proof of Proposition~\ref{prop:augssp}]

\subsection{Proof of Proposition~\ref{prop:weak1iidg2}}
\label{subsec:prop:weak1iidg2}
The proof of Proposition~\ref{prop:weak1iidg2} is based on the following theorem by Davidson and Szarek \cite{DavSza01}.
\begin{theorem}[{\cite[Theorem~II.13]{DavSza01}}]
Given $m,n \in \mathbb{N}$ with $m \leq n$, consider the random matrix $G \in \mathbb{R}^{n \times m}$
whose entries are i.i.d. Gaussian following $\mathcal{N}(0,\frac{1}{n})$.
Then, for any $t > 0$,
\begin{align*}
\mathbb{P}\left(\sigma_1(G) \geq 1 + \sqrt{\frac{m}{n}} + t\right) {} & \leq \exp\left(-\frac{nt^2}{2}\right), \\
\mathbb{P}\left(\sigma_m(G) \leq 1 - \sqrt{\frac{m}{n}} - t\right) {} & \leq \exp\left(-\frac{nt^2}{2}\right).
\end{align*}
\label{thm:davsza}
\end{theorem}

\begin{IEEEproof}[Proof of Proposition~\ref{prop:weak1iidg2}]
Note that (\ref{eq:prop:weak1iidg2:cond1}) implies
\begin{equation*}
1 - \sqrt{1-\delta} \geq \sqrt{1+\delta} - 1 \geq 2 \sqrt{\frac{s+1}{m}}.
\end{equation*}

Let $j \in [n] \setminus J$.
Theorem~\ref{thm:davsza} implies
{\allowdisplaybreaks
\begin{align*}
{} & \mathbb{P}\left(\sigma_1(A_{J \cup \{j\}}) \geq \sqrt{1+\delta} \right) \\
{} & \leq \exp\left(-\frac{m}{2} \left( \sqrt{1+\delta} - 1 - \sqrt{\frac{s+1}{m}} \right)^2 \right)
\intertext{and}
{} & \mathbb{P}\left(\sigma_{s+1}(A_{J \cup \{j\}}) \leq \sqrt{1-\delta} \right) \\
{} & \leq \exp\left(-\frac{m}{2} \left( 1 - \sqrt{1-\delta} - \sqrt{\frac{s+1}{m}} \right)^2 \right).
\end{align*}}
Since
\begin{equation*}
1 - \sqrt{1-\delta} - \sqrt{\frac{s+1}{m}} \geq \sqrt{1+\delta} - 1 - \sqrt{\frac{s+1}{m}} > 0,
\end{equation*}
it follows that
{\allowdisplaybreaks
\begin{align*}
{} & \mathbb{P}\left(\norm{A_{J \cup \{j\}}^* A_{J \cup \{j\}} - I_{s+1}} \geq \delta \right) \\
{} & \leq 2 \exp\left(-\frac{m}{2} \left( \sqrt{1+\delta} - 1 - \sqrt{\frac{s+1}{m}} \right)^2 \right).
\end{align*}} %
By considering the union of the events $(\norm{A_{J \cup \{j\}}^* A_{J \cup \{j\}} - I_{s+1}} \geq \delta)$ for all $j \in [n] \setminus J$,
we obtain
\begin{equation}
\begin{aligned}
{} & \mathbb{P}\left(\delta_{s+1}^\text{weak}(A;J) \geq \delta \right) \\
{} & \leq 2 (n-s) \exp\left(-\frac{m}{2} \left( \sqrt{1+\delta} - 1 - \sqrt{\frac{s+1}{m}} \right)^2 \right).
\end{aligned}
\label{eq:proof:prop:weak1iidg2:eq1}
\end{equation}
The RHS of (\ref{eq:proof:prop:weak1iidg2:eq1}) is bounded from above by $\epsilon$
if
\begin{equation*}
(\sqrt{1+\delta} - 1) \sqrt{m} \geq \sqrt{s+1} + \sqrt{2 \ln\left(\frac{2(n-s)}{\epsilon}\right)},
\end{equation*}
which is implied by
\begin{equation}
(\sqrt{1+\delta} - 1) \sqrt{m} \geq \sqrt{2} \sqrt{(s+1) + 2 \ln\left(\frac{2(n-s)}{\epsilon}\right)}
\label{eq:proof:prop:weak1iidg2:eq2}
\end{equation}
where we used the concavity of the square root function.
Noting that (\ref{eq:proof:prop:weak1iidg2:eq2}) coincides with (\ref{eq:prop:weak1iidg2:cond1}) completes the proof.
\end{IEEEproof} %{Proof of Proposition~\ref{prop:weak1iidg2}}

\subsection{Proof of Proposition~\ref{prop:weak1aiidg2}}
\label{subsec:prop:weak1aiidg2}
%\begin{IEEEproof}[Proof of Proposition~\ref{prop:weak1aiidg2}]
Let $j \in [n] \setminus J$.
Theorem~\ref{thm:davsza} implies
\begin{align*}
{} & \mathbb{P}\left(\sigma_{s+1}(A_{J \cup \{j\}}) \leq 1-\gamma \right) \\
{} & \leq \exp\left(-\frac{m}{2} \left( \gamma - \sqrt{\frac{s+1}{m}} \right)^2 \right).
\end{align*}
By considering the union of the events $(\sigma_{s+1}(A_{J \cup \{j\}}) \leq 1-\gamma)$ for all $j \in [n] \setminus J$,
we obtain
\begin{equation}
\begin{aligned}
{} & \mathbb{P}\left(\alpha_{s+1}^\text{weak}(A;J) \leq 1-\gamma \right) \\
{} & \leq (n-s) \exp\left(-\frac{m}{2} \left( \gamma - \sqrt{\frac{s+1}{m}} \right)^2 \right).
\end{aligned}
\label{eq:proof:prop:weak1aiidg2:eq1}
\end{equation}
Condition (\ref{eq:prop:weak1aiidg2:cond1}) implies that
the RHS of (\ref{eq:proof:prop:weak1aiidg2:eq1}) is bounded from above by $\epsilon$.
%\end{IEEEproof} %{Proof of Proposition~\ref{prop:weak1aiidg2}}

\subsection{Proof of Proposition~\ref{prop:weak1randfourier}}
\label{subsec:prop:weak1randfourier}
%\begin{IEEEproof} %{Proof of Proposition~\ref{prop:weak1randfourier}}
Let $j \in [n] \setminus J$.
Then, by \cite[Lemma~2.1]{CanPla10ripless}, $A_{J \cup \{j\}}$ satisfies
\begin{equation*}
\begin{aligned}
{} & \mathbb{P}(\norm{A_{J \cup \{j\}}^* A_{J \cup \{j\}} - I_{s+1}} \geq \delta) \\
{} & \quad \leq 2 (s+1) \exp \left( - \frac{m}{s} \cdot \frac{\delta^2}{2(1 + \delta/3)} \right).
\end{aligned}
\end{equation*}
Therefore, by considering the union of the events $(\norm{A_{J \cup \{j\}}^* A_{J \cup \{j\}} - I_{s+1}} \geq \delta)$ for all $j \in [n] \setminus J$,
we obtain
\begin{equation}
\begin{aligned}
{} & \mathbb{P}(\delta_{s+1}^\text{weak}(A;J) \geq \delta) \\
{} & \quad \leq 2 (n-s) (s+1) \exp \left( - \frac{m}{s} \cdot \frac{\delta^2}{2(1 + \delta/3)} \right).
\label{eq:proof:prop:weak1randfourier:eq1}
\end{aligned}
\end{equation}
Since
(\ref{eq:prop:weak1randfourier:cond}) implies that
the RHS of (\ref{eq:proof:prop:weak1randfourier:eq1}) is less than $\epsilon$,
the proof is complete.
%\end{IEEEproof} %{Proof of Proposition~\ref{prop:weak1randfourier}}

\subsection{Proof of Proposition~\ref{prop:weak1untf}}
\label{subsec:prop:weak1untf}
%\begin{IEEEproof}[Proof of Proposition~\ref{prop:weak1untf}]
Let $j \in [n] \setminus J$.
First, we derive a probabilistic upper bound on $\norm{A_{J \cup \{j\}}^* A_{J \cup \{j\}} - I_{s+1}}$
where the probability is with respect to the choice of $J$.
To this end, we use the relevant result in \cite{Tro08}.
\cite[Theorem~12]{Tro08} claims that, for $s \geq 3$ and $\alpha \geq 1$,
\begin{equation}
\mathbb{P}(\norm{A_{J \cup \{j\}}^* A_{J \cup \{j\}} - I_{s+1}} > \delta) \leq \left(\frac{s+1}{2}\right)^{-\alpha}
\label{eq:strip_tro08}
\end{equation}
if
\begin{equation}
\sqrt{144 \mu^2 (s+1) \ln\left(\frac{s+1}{2}+1\right) \alpha} + \frac{2(s+1)}{n} \norm{A}^2 \leq e^{-1/4} \delta.
\label{eq:condstrip_tro08}
\end{equation}
We assumed that $\mu \leq \frac{K}{\sqrt{m}}$ and $\norm{A} = \sqrt{\frac{n}{m}}$.
To bound the RHS of (\ref{eq:strip_tro08}) from above by $\frac{\epsilon}{n-s}$, let
\begin{equation*}
\alpha = \frac{\ln\left(\frac{n-s}{\epsilon}\right)}{\ln\left(\frac{s}{2}\right)}.
\end{equation*}
By the concavity of the square root function, Condition (\ref{eq:condstrip_tro08}) is implied by
{\allowdisplaybreaks
\begin{align}
{} & \frac{144 K^2 (s+1)}{m} \frac{\ln\left(\frac{s+1}{2}+1\right)}{\ln\left(\frac{s+1}{2}\right)} \ln\left(\frac{n-s}{\epsilon}\right) + \frac{2(s+1)}{m} \nonumber \\
{} & \quad \leq \frac{\delta^2}{2\sqrt{e}}.
\label{eq:proof:prop:weak1untf:eq1}
\end{align}} %
Since $\frac{\ln(s+1)}{\ln s} \leq 2$ for all $s \geq 3$, (\ref{eq:proof:prop:weak1untf:eq1}) is implied by
\begin{equation*}
m \geq \frac{4\sqrt{e}}{\delta^2} \left[ (s+1) + 288K^2 \ln\left( \frac{n-s}{\epsilon} \right) \right]
\end{equation*}

By considering the union of the events $(\norm{A_{J \cup \{j\}}^* A_{J \cup \{j\}} - I_{s+1}} \geq \delta)$ for all $j \in [n] \setminus J$,
we obtain
\begin{equation*}
\mathbb{P}(\delta_{s+1}^\text{weak}(A;J) \geq \delta) \leq \sum_{j \in [n] \setminus J} \mathbb{P}(\norm{A_{J \cup \{j\}}^* A_{J \cup \{j\}} - I_{s+1}}
\leq \epsilon.
\end{equation*}
%\end{IEEEproof} %{Proof of Proposition~\ref{prop:weak1untf}}

\subsection{Proof of Theorem~\ref{thm:music}}
\label{subsec:thm:music}
%\begin{IEEEproof} %{Proof of Theorem~\ref{thm:music}}
MUSIC finds $J_0$ if
\begin{equation}
\min_{k \in J_0} \frac{\norm{P_{\widehat{S}} a_k}_2}{\norm{a_k}_2} > \max_{k \in [n] \setminus J_0} \frac{\norm{P_{\widehat{S}} a_k}_2}{\norm{a_k}_2}.
\label{eq:proof:thm:music:eq1}
\end{equation}

Since $\alpha_{s+1}^\text{weak}(A;J_0) > 0$, it follows that all columns of $A_{J_0}$ are linearly independent.
Furthermore, since $\rank(X_0^{J_0}) = s$, we have
\begin{equation}
S = \mathcal{R}(A_{J_0}).
\label{eq:proof:thm:music:eq2}
\end{equation}
By the triangle inequality,
{\allowdisplaybreaks
\begin{align}
{} & \left| \frac{\norm{P_{\widehat{S}} a_k}_2}{\norm{a_k}_2} - \frac{\norm{P_S a_k}_2}{\norm{a_k}_2} \right| \nonumber \\
{} & \leq \frac{\norm{(P_{\widehat{S}} - P_S) a_k}_2}{\norm{a_k}_2} \nonumber \\
{} & = \norm{P_{\widehat{S}} - P_S} \leq \eta. \label{eq:proof:thm:music:eq3}
\end{align}} %
Then, for all $k \in J_0$,
{\allowdisplaybreaks
\begin{align*}
\frac{\norm{P_{\widehat{S}} a_k}_2}{\norm{a_k}_2}
{} & \overset{\text{(a)}}{\geq} \frac{\norm{P_S a_k}_2}{\norm{a_k}_2} - \eta \\
{} & \overset{\text{(b)}}{=} \frac{\norm{P_{\mathcal{R}(A_{J_0})} a_k}_2}{\norm{a_k}_2} - \eta \\
{} & = 1 - \eta,
\end{align*}} %
where (a) and (b) follow from (\ref{eq:proof:thm:music:eq3}) and (\ref{eq:proof:thm:music:eq2}), respectively.
Then, we obtain a lower bound on the LHS of (\ref{eq:proof:thm:music:eq1}) given by
{\allowdisplaybreaks
\begin{align}
\min_{k \in J_0} \frac{\norm{P_{\widehat{S}} a_k}_2}{\norm{a_k}_2} \geq 1 - \eta. \label{eq:proof:thm:music:eq4}
\end{align}} %
Similarly, by (\ref{eq:proof:thm:music:eq2}) and (\ref{eq:proof:thm:music:eq3}), we have
{\allowdisplaybreaks
\begin{equation*}
\frac{\norm{P_{\widehat{S}} a_k}_2}{\norm{a_k}_2}
\leq \frac{\norm{P_{\mathcal{R}(A_{J_0})} a_k}_2}{\norm{a_k}_2} + \eta
\end{equation*}
for all $k \in [n] \setminus J_0$ where $\norm{P_{\mathcal{R}(A_{J_0})} a_k}_2$ is further bounded from above by
{\allowdisplaybreaks
\begin{align*}
\frac{\norm{P_{\mathcal{R}(A_{J_0})} a_k}_2^2}{\norm{a_k}_2^2}
{} & = 1 - \frac{\norm{P_{\mathcal{R}(A_{J_0})}^\perp a_k}_2^2}{\norm{a_k}_2^2} \\
{} & = 1 - \frac{\sigma_1^2(P_{\mathcal{R}(A_{J_0})}^\perp a_k)}{\norm{a_k}_2^2} \\
{} & \overset{\text{(c)}}{\leq} 1 - \frac{\sigma_{s+1}^2(A_{J_0 \cup \{k\}})}{\norm{a_k}_2^2} \\
{} & \overset{\text{(d)}}{\leq} 1 - \frac{\left\{\alpha_{s+1}^\text{weak}(A;J_0)\right\}^2}{\norm{a_k}_2^2} \\
{} & \leq 1 - \frac{\left\{\alpha_{s+1}^\text{weak}(A;J_0)\right\}^2}{\norm{A^*}_{2,\infty}^2}
\end{align*}} %
where (c) holds by Lemma~\ref{lemma:interlacing_schur} and (d) follows by the definition of the weak-1 asymmetric RIP.
Then, we obtain an upper bound on the RHS of (\ref{eq:proof:thm:music:eq1}) given by
{\allowdisplaybreaks
\begin{align}
\max_{k \in [n] \setminus J_0} \norm{P_{\widehat{S}} a_k}_2
\leq \left[ 1 - \frac{\left\{\alpha_{s+1}^\text{weak}(A;J_0)\right\}^2}{\norm{A^*}_{2,\infty}^2} \right]^{1/2} + \eta.
\label{eq:proof:thm:music:eq5}
\end{align}} %

Combining (\ref{eq:proof:thm:music:eq4}) and (\ref{eq:proof:thm:music:eq5}),
we note that MUSIC is guaranteed if $A$ satisfies the weak-1 asymmetric RIP given by
\begin{equation*}
\alpha_{s+1}^\text{weak}(A;J_0) > \alpha
\end{equation*}
for $\alpha > 0$ satisfying
\begin{equation*}
\alpha \geq \norm{A^*}_{2,\infty} \left\{1 - \left(1 - 2 \eta\right)^2\right\}^{1/2}.
\end{equation*}
%\end{IEEEproof} %{Proof of Theorem~\ref{thm:music}}

\subsection{Proof of Proposition~\ref{prop:errinaugssp}}
\label{subsec:prop:errinaugssp}
%\begin{IEEEproof}[Proof of Proposition~\ref{prop:errinaugssp}]
Let $d \triangleq \dim(P_{\mathcal{R}(A_J)}^\perp \bar{S})$.
By the row-nondegeneracy condition on $X_0^{J_0}$ and Lemma~\ref{lemma:fullkrank}, it follows that $d \geq 1$.
There exists $Q \in \mathbb{K}^{(s-k) \times d}$ where $k = |J|$ such that $Q^* Q = I_d$
and $\mathcal{R}(A_{J_0 \setminus J} Q)$ is a subspace of $\bar{S}$.
Then, it follows that
\begin{align}
\underbrace{\sigma_d(P_{\mathcal{R}(A_J)}^\perp A_{J_0 \setminus J} Q)}_{(\star)}
{} & = \sigma_d(P_{\mathcal{R}(A_J)}^\perp P_{\bar{S}} A_{J_0 \setminus J} Q) \nonumber \\
{} & \leq \sigma_d(P_{\mathcal{R}(A_J)}^\perp P_{\bar{S}}) \cdot \norm{A_{J_0 \setminus J} Q} \nonumber \\
{} & \leq \sigma_d(P_{\mathcal{R}(A_J)}^\perp P_{\bar{S}}) \cdot \norm{A_{J_0}}. \label{eq:proof:prop:errinaugssp:eq1}
\end{align}
By the variational characterization of the singular values, $(\star)$ is bounded from below by
\begin{equation}
(\star) \geq \sigma_{s-k}(P_{\mathcal{R}(A_J)}^\perp A_{J_0 \setminus J}) \geq \sigma_s(A_{J_0})
\label{eq:proof:prop:errinaugssp:eq2}
\end{equation}
where the last step follows by Lemma~\ref{lemma:interlacing_schur}.

Let $\kappa \triangleq \sigma_1(A_{J_0})/\sigma_s(A_{J_0})$.
Combining (\ref{eq:proof:prop:errinaugssp:eq1}) and (\ref{eq:proof:prop:errinaugssp:eq2}),
we obtain
\begin{equation}
\sigma_d(P_{\mathcal{R}(A_J)}^\perp P_{\bar{S}}) \geq \kappa^{-1}.
\label{eq:proof:prop:errinaugssp:eq3}
\end{equation}

Since $\dim(\bar{S}) = \dim(\widehat{S}) = r$, it follows that
\begin{align}
\norm{P_{\widehat{S}} - P_{\bar{S}}}
{} & = \sup_{\begin{subarray}{c} \bar{x} \in \bar{S} \\ \norm{\bar{x}}_2 = 1 \end{subarray}}
\inf_{\begin{subarray}{c} \hat{x} \in \widehat{S} \\ \norm{\hat{x}}_2 = 1 \end{subarray}} \norm{\bar{x} - \hat{x}}_2 \nonumber \\
{} & = \sup_{\begin{subarray}{c} \hat{x} \in \widehat{S} \\ \norm{\hat{x}}_2 = 1 \end{subarray}}
\inf_{\begin{subarray}{c} \bar{x} \in \bar{S} \\ \norm{\bar{x}}_2 = 1 \end{subarray}} \norm{\hat{x} - \bar{x}}_2
\label{eq:proof:prop:errinaugssp:eq4}
\end{align}
and hence $\norm{P_{\widehat{S}} - P_{\bar{S}}} \leq \eta$ implies the followings:
for all $\bar{x} \in \bar{S}$, there exists $\hat{x} \in \widehat{S}$ such that $\norm{\bar{x} - \hat{x}}_2 \leq \eta \norm{\bar{x}}_2$.
Similarly, for all $\hat{x} \in \widehat{S}$, there exists $\bar{x} \in \bar{S}$ such that $\norm{\hat{x} - \bar{x}}_2 \leq \eta \norm{\hat{x}}_2$.

The following identity is well known (see \textit{e.g.}, \cite{BotSpi10}):
given two subspace $S_1$ and $S_2$,
\begin{equation}
\norm{P_{S_1} - P_{S_2}} = \max\{ \norm{P_{S_1}^\perp P_{S_2}} , \norm{P_{S_2}^\perp P_{S_1}} \}.
\label{eq:proof:prop:errinaugssp:id}
\end{equation}

Note, by (\ref{eq:proof:prop:errinaugssp:id}), that
\begin{align}
{} & \norm{P_{P_{\mathcal{R}(A_J)}^\perp \widehat{S}} - P_{P_{\mathcal{R}(A_J)}^\perp \bar{S}}} \nonumber \\
{} & = \max\Big\{ \underbrace{\norm{P_{P_{\mathcal{R}(A_J)}^\perp \widehat{S}}^\perp P_{P_{\mathcal{R}(A_J)}^\perp \bar{S}}}}_{(\ast)}
,~ \\
{} & \qquad \underbrace{\norm{P_{P_{\mathcal{R}(A_J)}^\perp \bar{S}}^\perp P_{P_{\mathcal{R}(A_J)}^\perp \widehat{S}}}}_{(\ast\ast)} \Big\}.
\label{eq:proof:prop:errinaugssp:eq5}
\end{align}

First, we derive an upper bound on $(\ast)$, which is equivalently rewritten as
\begin{equation}
(\ast) = \sup_{\begin{subarray}{c} z \in P_{\mathcal{R}(A_J)}^\perp \bar{S} \\ \norm{z}_2 = 1 \end{subarray}} \inf_{y \in P_{\mathcal{R}(A_J)}^\perp \widehat{S}} \norm{z-y}_2.
\label{eq:proof:prop:errinaugssp:eq6}
\end{equation}
Let $z \in P_{\mathcal{R}(A_J)}^\perp \bar{S}$ satisfy $\norm{z}_2 = 1$.
Then, by (\ref{eq:proof:prop:errinaugssp:eq3}), there exists $\bar{x} \in \bar{S}$ such that $P_{\mathcal{R}(A_J)}^\perp \bar{x} = z$ and $\norm{\bar{x}}_2 \leq \kappa$.
By the argument following (\ref{eq:proof:prop:errinaugssp:eq4}),
there exists $\hat{x} \in \widehat{S}$ such that $\norm{\bar{x} - \hat{x}}_2 \leq \eta \norm{\bar{x}}_2 \leq \eta \kappa$.
Then, it follows that
\begin{equation*}
\norm{\underbrace{P_{\mathcal{R}(A_J)}^\perp \bar{x}}_{=z} - \underbrace{P_{\mathcal{R}(A_J)}^\perp \hat{x}}_{\in P_{\mathcal{R}(A_J)}^\perp \widehat{S}}}_2 \leq \norm{\bar{x} - \hat{x}}_2 \leq \eta \kappa
\end{equation*}
and hence
\begin{equation*}
\inf_{y \in P_{\mathcal{R}(A_J)}^\perp \widehat{S}} \norm{z-y}_2 \leq \eta \kappa.
\end{equation*}
Since $z$ was an arbitrary unit-norm element in $P_{\mathcal{R}(A_J)}^\perp \bar{S}$, we obtain
\begin{equation}
(\ast) \leq \eta \kappa.
\label{eq:proof:prop:errinaugssp:eq7}
\end{equation}

Next, we derive an upper bound on $(\ast\ast)$, which is equivalently rewritten as
\begin{equation}
(\ast\ast) = \sup_{\begin{subarray}{c} z \in P_{\mathcal{R}(A_J)}^\perp \widehat{S} \\ \norm{z}_2 = 1 \end{subarray}} \inf_{y \in P_{\mathcal{R}(A_J)}^\perp \bar{S}} \norm{z-y}_2.
\label{eq:proof:prop:errinaugssp:eq8}
\end{equation}
Since
\begin{align*}
\sigma_d(P_{\mathcal{R}(A_J)}^\perp P_{\bar{S}})
{} & \leq \sigma_d(P_{\mathcal{R}(A_J)}^\perp P_{\widehat{S}}) + \norm{P_{\mathcal{R}(A_J)}^\perp(P_{\widehat{S}} - P_{\bar{S}})} \\
{} & \leq \sigma_d(P_{\mathcal{R}(A_J)}^\perp P_{\widehat{S}}) + \norm{P_{\widehat{S}} - P_{\bar{S}}} \\
{} & \leq \sigma_d(P_{\mathcal{R}(A_J)}^\perp P_{\widehat{S}}) + \eta
\end{align*}
by (\ref{eq:proof:prop:errinaugssp:eq3}), we obtain a lower bound on $\sigma_d(P_{\mathcal{R}(A_J)}^\perp P_{\widehat{S}})$ given by
\begin{equation}
\sigma_d(P_{\mathcal{R}(A_J)}^\perp P_{\widehat{S}}) \geq \kappa^{-1}-\eta = \frac{1-\eta\kappa}{\kappa}.
\label{eq:proof:prop:errinaugssp:eq9}
\end{equation}

Let $z \in P_{\mathcal{R}(A_J)}^\perp \widehat{S}$ satisfy $\norm{z}_2 = 1$.
Then, by (\ref{eq:proof:prop:errinaugssp:eq9}), there exists $\hat{x} \in \widehat{S}$ such that $P_{\mathcal{R}(A_J)}^\perp \hat{x} = z$ and $\norm{\hat{x}}_2 \leq \frac{\kappa}{1-\eta\kappa}$.
By the argument following (\ref{eq:proof:prop:errinaugssp:eq4}),
there exists $\bar{x} \in \bar{S}$ such that $\norm{\hat{x} - \bar{x}}_2 \leq \eta \norm{\hat{x}}_2 \leq \frac{\eta\kappa}{1-\eta\kappa}$.
Then, it follows that
\begin{equation*}
\norm{\underbrace{P_{\mathcal{R}(A_J)}^\perp \hat{x}}_{=z} - \underbrace{P_{\mathcal{R}(A_J)}^\perp \bar{x}}_{\in P_{\mathcal{R}(A_J)}^\perp \bar{S}}}_2 \leq \norm{\hat{x} - \bar{x}}_2 \leq \frac{\eta\kappa}{1-\eta\kappa}
\end{equation*}
and hence
\begin{equation*}
\inf_{y \in P_{\mathcal{R}(A_J)}^\perp \bar{S}} \norm{z-y}_2 \leq \frac{\eta\kappa}{1-\eta\kappa}.
\end{equation*}
Since $z$ was an arbitrary unit-norm element in $P_{\mathcal{R}(A_J)}^\perp \widehat{S}$, we obtain
\begin{equation}
(\ast\ast) \leq \frac{\eta\kappa}{1-\eta\kappa}.
\label{eq:proof:prop:errinaugssp:eq10}
\end{equation}

Applying (\ref{eq:proof:prop:errinaugssp:eq7}) and (\ref{eq:proof:prop:errinaugssp:eq10}) to (\ref{eq:proof:prop:errinaugssp:eq5})
completes the proof.
%\end{IEEEproof} %[Proof of Proposition~\ref{prop:errinaugssp}]

\subsection{Proof of Theorem~\ref{thm:samusicoracle}}
\label{subsec:thm:samusicoracle}
%\begin{IEEEproof}[Proof of Theorem~\ref{thm:samusicoracle}]
MUSIC applied to $\widetilde{S}$ is successful if
\begin{equation}
\min_{k \in J_0 \setminus J_1} \frac{\norm{P_{\widetilde{S}} a_k}_2}{\norm{a_k}_2}
> \max_{k \in [n] \setminus J_0} \frac{\norm{P_{\widetilde{S}} a_k}_2}{\norm{a_k}_2}.
\label{eq:proof:thm:samusicoracle:eq1}
\end{equation}

Since (\ref{eq:thm:samusicoracle:cond2}) implies $\rank(A_{J_0}) = s$ and $X_0^{J_0}$ is row-nondegenerate,
by Proposition~\ref{prop:augssp}, we obtain
\begin{equation}
\bar{S} + \mathcal{R}(A_{J_1}) = \mathcal{R}(A_{J_0})
\label{eq:proof:thm:samusicoracle:eq2}
\end{equation}
and hence, by the projection update formula,
\begin{equation}
P_{\mathcal{R}(A_{J_0})} = P_{\mathcal{R}(A_{J_1})} + P_{P_{\mathcal{R}(A_{J_1})}^\perp \bar{S}}.
\label{eq:proof:thm:samusicoracle:eq3}
\end{equation}

Since the augmented subspace is given by $\widetilde{S} = \widehat{S} + \mathcal{R}(A_{J_1})$,
by the projection update formula, we have
\begin{equation}
P_{\widetilde{S}} = P_{\mathcal{R}(A_{J_1})} + P_{P_{\mathcal{R}(A_{J_1})}^\perp \widehat{S}}.
\label{eq:proof:thm:samusicoracle:eq4}
\end{equation}

By the triangle inequality, it follows that
{\allowdisplaybreaks
\begin{align}
{} & \left| \frac{\big\|P_{\widetilde{S}} a_k\big\|_2}{\norm{a_k}_2}
- \frac{\big\|P_{\mathcal{R}(A_{J_0})} a_k\big\|_2}{\norm{a_k}_2} \right| \nonumber \\
{} & \leq \frac{\big\|(P_{\widetilde{S}} - P_{\mathcal{R}(A_{J_0})}) a_k\big\|_2}{\norm{a_k}_2} \nonumber
\intertext{By (\ref{eq:proof:thm:samusicoracle:eq2}) we continue by}
{} & = \frac{\big\|(P_{\mathcal{R}(A_{J_1})+\widehat{S}} - P_{\mathcal{R}(A_{J_1})+\bar{S}}) a_k\big\|_2}{\norm{a_k}_2} \nonumber \\
{} & \overset{\text(a)}{=} \frac{\big\|(P_{P_{\mathcal{R}(A_{J_1})}^\perp \widehat{S}} - P_{P_{\mathcal{R}(A_{J_1})}^\perp \bar{S}}) a_k\big\|_2}{\norm{a_k}_2} \nonumber \\
{} & \leq \big\|P_{P_{\mathcal{R}(A_{J_1})}^\perp \widehat{S}} - P_{P_{\mathcal{R}(A_{J_1})}^\perp \bar{S}}\big\| \nonumber \\
{} & \overset{\text(b)}{\leq} \frac{\eta\sigma_1(A_{J_0})}{\sigma_s(A_{J_0}) - \eta\sigma_1(A_{J_0})} \nonumber \\
{} & < \frac{\eta\beta}{\alpha - \eta\beta} \label{eq:proof:thm:samusicoracle:eq5}
\end{align}} %
where (a) follows from (\ref{eq:proof:thm:samusicoracle:eq3}) and (\ref{eq:proof:thm:samusicoracle:eq4}),
and (b) follows from Proposition~\ref{prop:errinaugssp}
because $\sigma_s(A_{J_0}) > \eta \sigma_1(A_{J_0})$ is implied by (\ref{eq:thm:samusicoracle:cond2}) and (\ref{eq:thm:samusicoracle:cond3}).

By (\ref{eq:proof:thm:samusicoracle:eq5}),
it holds for all $k \in J_0 \setminus J_1$ that
{\allowdisplaybreaks
\begin{align*}
\frac{\norm{P_{\widetilde{S}} a_k}_2}{\norm{a_k}_2}
{} & > \frac{\norm{P_{\mathcal{R}(A_{J_0})} a_k}_2}{\norm{a_k}_2} - \frac{\eta\beta}{\alpha - \eta\beta} \\
{} & = 1 - \frac{\eta\beta}{\alpha - \eta\beta}.
\end{align*}} %
This yields a lower bound on the LHS of (\ref{eq:proof:thm:samusicoracle:eq1}) given by
{\allowdisplaybreaks
\begin{align}
\min_{k \in J_0 \setminus J_1} \frac{\norm{P_{\widetilde{S}} a_k}_2^2}{\norm{a_k}_2^2}
> 1 - \frac{\eta\beta}{\alpha - \eta\beta}. \label{eq:proof:thm:samusicoracle:eq6}
\end{align}} %

Similarly, by (\ref{eq:proof:thm:samusicoracle:eq5}),
it holds for all $k \in [n] \setminus J_0$ that
{\allowdisplaybreaks
\begin{align*}
\frac{\norm{P_{\widetilde{S}} a_k}_2}{\norm{a_k}_2}
{} & < \underbrace{\frac{\norm{P_{\mathcal{R}(A_{J_0})} a_k}_2}{\norm{a_k}_2}}_{(\ast)}
+ \frac{\eta\beta}{\alpha - \eta\beta}
\end{align*}} %
where $(\ast)$ is further bounded from above by
{\allowdisplaybreaks
\begin{align*}
(\ast)^2 {} & = \frac{\norm{a_k}_2^2 - \norm{P_{\mathcal{R}(A_{J_0})}^\perp a_k}_2^2}{\norm{a_k}_2^2} \\
{} & = 1 - \frac{\sigma_1^2(P_{\mathcal{R}(A_{J_0})}^\perp a_k)}{\norm{a_k}_2^2} \\
{} & \overset{(c)}{\leq} 1 - \frac{\sigma_{s+1}^2(A_{J_0 \cup \{k\}})}{\norm{a_k}_2^2} \\
{} & \leq 1 - \frac{[\alpha_{s+1}^\text{weak}(A;J_0)]^2}{\norm{a_k}_2^2} \\
{} & < 1 - \frac{\alpha^2}{\norm{a_k}_2^2}
\end{align*}} %
where (c) follows by Lemma~\ref{lemma:interlacing_schur}.
This yields an upper bound on the RHS of (\ref{eq:proof:thm:samusicoracle:eq1}) given by
{\allowdisplaybreaks
\begin{align}
\max_{k \in [n] \setminus J_0} \norm{P_{\widetilde{S}} a_k}_2^2
< \sqrt{1 - \frac{\alpha^2}{\norm{A^*}_{2,\infty}^2}}
+ \frac{\eta\sigma_1^2(A_{J_0})}{\sigma_s^2(A_{J_0}) - \eta\sigma_1^2(A_{J_0})}. \label{eq:proof:thm:samusicoracle:eq7}
\end{align}} %

Finally, by applying the bounds in (\ref{eq:proof:thm:samusicoracle:eq5}) and (\ref{eq:proof:thm:samusicoracle:eq6}) to (\ref{eq:proof:thm:samusicoracle:eq1}),
we note that (\ref{eq:proof:thm:samusicoracle:eq1}) is implied by the weak-1 asymmetric RIP given by
\begin{equation*}
1 - \frac{\eta\beta}{\alpha - \eta\beta}
\geq \sqrt{1 - \frac{\alpha^2}{\norm{A^*}_{2,\infty}^2}}
+ \frac{\eta\beta}{\alpha - \eta\beta},
\end{equation*}
which is equivalent to (\ref{eq:thm:samusicoracle:cond3}).
This completes the proof.
%\end{IEEEproof}%[Proof of Theorem~\ref{thm:samusicoracle}]

\subsection{Proof of Proposition~\ref{prop:momp}}
\label{subsec:prop:momp}
The following lemma is used in the proof of Proposition~\ref{prop:momp}.
\begin{lemma}
Suppose that $A \in \mathbb{K}^{m \times n}$ and $P \in \mathbb{K}^{n \times n}$ is an orthogonal projector in $\mathbb{K}^n$.
Then, for all $x,y \in \mathbb{K}^n$,
\begin{equation}
\Big| \left|\langle A P x, A P y \rangle\right| - \left|\langle P x, P y \rangle\right| \Big|
\leq \norm{P A^* A P - P} \cdot \norm{x}_2 \cdot \norm{y}_2.
\end{equation}
\label{lemma:IPdev}
\end{lemma}

\begin{IEEEproof}[Proof of Proposition~\ref{prop:momp}]
Given $J \subsetneq J_0$, the next step of M-OMP will be also successful if
\begin{equation}
\max_{j \in J_0 \setminus J} \norm{a_j^* P_{\mathcal{R}(A_J)}^\perp Y}_2
> \max_{j \in [n] \setminus J_0} \norm{a_j^* P_{\mathcal{R}(A_J)}^\perp Y}_2.
\label{eq:comparenextMOMP}
\end{equation}
We derive a sufficient condition for (\ref{eq:comparenextMOMP}).
First, we derive a lower bound of the LHS of (\ref{eq:comparenextMOMP}).
{\allowdisplaybreaks
\begin{align}
{} & \max_{j \in J_0 \setminus J} \norm{a_j^* P_{\mathcal{R}(A_J)}^\perp Y}_2 \nonumber \\
{} & \overset{\text{$\Delta$-ineq}}{\geq} \max_{j \in J_0 \setminus J} \norm{a_j^* P_{\mathcal{R}(A_J)}^\perp A_{J_0} X_0^{J_0}}_2 - \norm{a_j^* P_{\mathcal{R}(A_J)}^\perp W}_2 \nonumber \\
{} & \overset{~~~~~~}{\geq}
\max_{j \in J_0 \setminus J} \norm{a_j^* P_{\mathcal{R}(A_J)}^\perp A_{J_0} X_0^{J_0}}_2 - \norm{A_{J_0 \setminus J}^* P_{\mathcal{R}(A_J)}^\perp}_{2,\infty} \norm{W} \nonumber \\
{} & \overset{~~~~~~}{\geq}
\max_{j \in J_0 \setminus J} \underbrace{ \norm{a_j^* P_{\mathcal{R}(A_J)}^\perp A_{J_0} X_0^{J_0}}_2 }_{(\ast)} - \norm{A^*}_{2,\infty} \norm{W}.
\label{eq:proof:prop:momp:eq1}
\end{align}} %
Let $\Pi_J \in \mathbb{K}^{n \times n}$ denote the coordinate projection that satisfies
$\Pi_J e_j = e_j$ for $j \in J$ and $\Pi_J e_j = 0$ for $j \in [n] \setminus J$.
Then, for all $j \in J_0 \setminus J$, we bound the term $(\ast)$ in (\ref{eq:proof:prop:momp:eq1}) from below by
{\allowdisplaybreaks
\begin{align}
(\ast) {} & = \norm{a_j^* P_{\mathcal{R}(A_J)}^\perp A_{J_0 \setminus J} X_0^{J_0 \setminus J}}_2 \nonumber \\
{} & = \sup_{\norm{z}_2 = 1} \left| \langle a_j, ~  P_{\mathcal{R}(A_J)}^\perp A_{J_0 \setminus J} X_0^{J_0 \setminus J} z \rangle \right| \nonumber \\
{} & = \sup_{\norm{z}_2 = 1} \left| \langle P_{\mathcal{R}(A_J)}^\perp a_j, ~ P_{\mathcal{R}(A_J)}^\perp A_{J_0 \setminus J} X_0^{J_0 \setminus J} z \rangle \right| \nonumber \\
{} & = \sup_{\norm{z}_2 = 1} \underbrace{\left| \langle P_{\mathcal{R}(A_J)}^\perp A \Pi_{J_0 \setminus J} e_j, ~ P_{\mathcal{R}(A_J)}^\perp A \Pi_{J_0 \setminus J} X_0 z \rangle \right|}_{(\ast\ast)}. \label{eq:proof:prop:momp:eq2}
\end{align}} %
Note that
{\allowdisplaybreaks
\begin{align}
(\ast\ast) {} & \overset{\text(a)}{\geq} \left|\langle \Pi_{J_0 \setminus J} e_j, ~ \Pi_{J_0 \setminus J} X_0 z \rangle\right| \nonumber \\
{} & \quad - \norm{\Pi_{J_0 \setminus J} A^* P_{\mathcal{R}(A_J)}^\perp A \Pi_{J_0 \setminus J} - \Pi_{J_0 \setminus J}} \nonumber \\
{} & \qquad \cdot \norm{\Pi_{J_0 \setminus J} e_j}_2 \norm{\Pi_{J_0 \setminus J} X_0 z}_2 \nonumber \\
{} & = \left|\langle \Pi_{J_0 \setminus J} e_j, ~ \Pi_{J_0 \setminus J} X_0 z \rangle\right| \nonumber \\
{} & \quad - \norm{A_{J_0 \setminus J}^* P_{\mathcal{R}(A_J)}^\perp A_{J_0 \setminus J} - I_{|J_0 \setminus J|}} ~ \norm{\Pi_{J_0 \setminus J} X_0 z}_2 \nonumber \\
{} & \overset{\text(b)}{>} \left|\langle \Pi_{J_0 \setminus J} e_j, ~ \Pi_{J_0 \setminus J} X_0 z \rangle\right| - \delta \norm{\Pi_{J_0 \setminus J} X_0 z}_2 \label{eq:proof:prop:momp:eq2a}
\end{align}} %
where (a) follows from Lemma~\ref{lemma:IPdev} and (b) holds since
\begin{align*}
{} & \norm{A_{J_0 \setminus J}^* P_{\mathcal{R}(A_J)}^\perp A_{J_0 \setminus J} - I_{|J_0 \setminus J|}} \\
{} & = \max\Big\{ 1 - \lambda_{|J_0 \setminus J|}(A_{J_0 \setminus J}^* P_{\mathcal{R}(A_J)}^\perp A_{J_0 \setminus J}) ,~ \\
{} & \qquad \lambda_1(A_{J_0 \setminus J}^* P_{\mathcal{R}(A_J)}^\perp A_{J_0 \setminus J}) - 1 \Big\} \\
{} & \overset{\text(c)}{\leq} \max\Big\{ 1 - \lambda_s(A_{J_0}^* A_{J_0}) ,~ \lambda_1(A_{J_0}^* A_{J_0}) - 1 \Big\} \\
{} & = \norm{A_{J_0}^* A_{J_0} - I_s} \overset{\text(d)}{<} \delta
\end{align*}
where (c) follows from Lemma~\ref{lemma:interlacing_schur} and (d) is implied by (\ref{eq:prop:momp:cond1}).

We then continue (\ref{eq:proof:prop:momp:eq2}) by using (\ref{eq:proof:prop:momp:eq2a})
{\allowdisplaybreaks
\begin{align}
(\ast) {} & > \sup_{\norm{z}_2 = 1} \left|\langle \Pi_{J_0 \setminus J} e_j, ~ \Pi_{J_0 \setminus J} X_0 z \rangle\right| \nonumber \\
{} & \quad - \sup_{\norm{z}_2 = 1} \delta \norm{\Pi_{J_0 \setminus J} X_0 z}_2 \nonumber \\
{} & = \norm{e_j^* \Pi_{J_0 \setminus J} X_0}_2 - \delta \norm{\Pi_{J_0 \setminus J} X_0} \nonumber \\
{} & = \norm{e_j^* \Pi_{J_0 \setminus J} X_0}_2 - \delta \norm{X_0^{J_0 \setminus J}} \label{eq:proof:prop:momp:eq3}.
\end{align}} %
Maximizing the lower bound on $(\ast)$ in (\ref{eq:proof:prop:momp:eq3}) over $j \in J_0 \setminus J$ yields
{\allowdisplaybreaks
\begin{align*}
{} & \max_{j \in J_0 \setminus J} \norm{a_j^* P_{\mathcal{R}(A_J)}^\perp A_{J_0} X_0^{J_0}}_2 \\
{} & > \max_{j \in J_0 \setminus J} \norm{e_j^* \Pi_{J_0 \setminus J} X_0}_2 - \delta \norm{X_0^{J_0 \setminus J}} \\
{} & = \norm{X_0^{J_0 \setminus J}}_{2,\infty} - \delta \norm{X_0^{J_0 \setminus J}}
\end{align*}} %
and hence the LHS of (\ref{eq:comparenextMOMP}) is bounded from below by
\begin{equation}
\begin{aligned}
{} & \max_{j \in J_0 \setminus J} \norm{a_j^* P_{\mathcal{R}(A_J)}^\perp Y}_2 \\
{} & > \norm{X_0^{J_0 \setminus J}}_{2,\infty} - \delta \norm{X_0^{J_0 \setminus J}} - \norm{A^*}_{2,\infty} \norm{W}.
\end{aligned}
\label{eq:proof:prop:momp:bndlhs}
\end{equation}
Next, we derive an upper bound on the RHS of (\ref{eq:comparenextMOMP}).
{\allowdisplaybreaks
\begin{align}
{} & \max_{j \in [n] \setminus J_0} \norm{a_j^* P_{\mathcal{R}(A_J)}^\perp Y}_2 \nonumber \\
{} & \overset{\text{$\Delta$-ineq}}{\leq} \max_{j \in [n] \setminus J_0} \norm{a_j^* P_{\mathcal{R}(A_J)}^\perp A_{J_0} X_0^{J_0}}_2
+ \norm{a_j^* P_{\mathcal{R}(A_J)}^\perp W}_2 \nonumber \\
{} & \overset{~~~~~~}{\leq}
\max_{j \in [n] \setminus J_0} \norm{a_j^* P_{\mathcal{R}(A_J)}^\perp A_{J_0} X_0^{J_0}}_2 + \norm{P_{\mathcal{R}(A_J)}^\perp a_j}_2 \norm{W} \nonumber \\
{} & \overset{~~~~~~}{\leq}
\max_{j \in [n] \setminus J_0} \underbrace{ \norm{a_j^* P_{\mathcal{R}(A_J)}^\perp A_{J_0} X_0^{J_0}}_2 }_{(\star)} + \norm{A^*}_{2,\infty} \norm{W}. \label{eq:proof:prop:momp:eq4}
\end{align}} %
Similarly to the previous case, we derive an upper bound on $(\star)$ for all $j \in [n] \setminus J_0$.
{\allowdisplaybreaks
\begin{align}
(\star) {} & = \norm{e_j^* A^* P_{\mathcal{R}(A_J)}^\perp A_{J_0} X_0^{J_0}}_2 \nonumber \\
{} & = \norm{e_j^* A^* P_{\mathcal{R}(A_J)}^\perp A X_0}_2 \nonumber \\
{} & = \norm{e_j^* \Pi_{(J_0 \setminus J) \cup \{j\}} A^* P_{\mathcal{R}(A_J)}^\perp A \Pi_{(J_0 \setminus J) \cup \{j\}} X_0}_2 \nonumber \\
{} & = \sup_{\norm{z}_2 = 1} \Big|\langle P_{\mathcal{R}(A_J)}^\perp A \Pi_{(J_0 \setminus J) \cup \{j\}} e_j,~ \nonumber \\
{} & \qquad P_{\mathcal{R}(A_J)}^\perp A \Pi_{(J_0 \setminus J) \cup \{j\}} X_0 z \rangle\Big| \nonumber \\
\intertext{By Lemma~\ref{lemma:IPdev}, the last term is at most}
{} & \leq \sup_{\norm{z}_2 = 1} \left|\langle \Pi_{(J_0 \setminus J) \cup \{j\}} e_j, ~ \Pi_{(J_0 \setminus J) \cup \{j\}} X_0 z \rangle\right| \nonumber \\
{} & \quad + \sup_{\norm{z}_2 = 1} \norm{A_{(J_0 \setminus J) \cup \{j\}}^* A_{(J_0 \setminus J) \cup \{j\}} - I_{s-|J|+1}} \nonumber \\
{} & \qquad \cdot \norm{\Pi_{(J_0 \setminus J) \cup \{j\}} X_0 z}_2 \nonumber \\
{} & = \sup_{\norm{z}_2 = 1} \underbrace{|\langle e_j, ~ X_0 z \rangle|}_{=0} \nonumber \\
{} & \quad + \sup_{\norm{z}_2 = 1} \norm{A_{(J_0 \setminus J) \cup \{j\}}^* A_{(J_0 \setminus J) \cup \{j\}} - I_{s-|J|+1}} \nonumber \\
{} & \qquad \cdot \norm{\Pi_{(J_0 \setminus J) \cup \{j\}} X_0 z}_2 \nonumber \\
{} & = \norm{A_{(J_0 \setminus J) \cup \{j\}}^* A_{(J_0 \setminus J) \cup \{j\}} - I_{s-|J|+1}} \nonumber \\
{} & \qquad \cdot \norm{\Pi_{(J_0 \setminus J) \cup \{j\}} X_0} \nonumber \\
{} & = \norm{A_{(J_0 \setminus J) \cup \{j\}}^* A_{(J_0 \setminus J) \cup \{j\}} - I_{s-|J|+1}} \cdot \norm{X_0^{J_0 \setminus J}} \nonumber \\
{} & \leq \norm{A_{J_0 \cup \{j\}}^* A_{J_0 \cup \{j\}} - I_{s+1}} \cdot \norm{X_0^{J_0 \setminus J}} \nonumber \\
{} & < \delta \norm{X_0^{J_0 \setminus J}}. \label{eq:proof:prop:momp:eq5}
\end{align}} %
Maximizing the upper bound on $(\star)$ in (\ref{eq:proof:prop:momp:eq5}) yields
an upper bound on the RHS of (\ref{eq:comparenextMOMP}) given by
\begin{equation}
\begin{aligned}
{} & \max_{j \in [n] \setminus J_0} \norm{a_j^* P_{\mathcal{R}(A_J)}^\perp Y}_2 \\
{} & < \delta \norm{X_0^{J_0 \setminus J}} + \norm{A^*}_{2,\infty} \norm{W}.
\end{aligned}
\label{eq:proof:prop:momp:bndrhs}
\end{equation}
Applying the bounds in (\ref{eq:proof:prop:momp:bndlhs}) and (\ref{eq:proof:prop:momp:bndrhs}) to (\ref{eq:comparenextMOMP}),
we conclude that, for the success of the next step, it suffices to satisfy
{\allowdisplaybreaks
\begin{align*}
{} & \norm{X_0^{J_0 \setminus J}}_{2,\infty} - 2 \delta \norm{X_0^{J_0 \setminus J}} \geq 2 \norm{A^*}_{2,\infty} \norm{W},
\end{align*}} %
which is Condition (\ref{eq:prop:momp:cond2}).
This completes the proof.
\end{IEEEproof} %[Proof of Proposition~\ref{prop:momp}]

\begin{IEEEproof}[Proof of Lemma~\ref{lemma:IPdev}]
The proof follows from the properties of an inner product:
{\allowdisplaybreaks
\begin{align*}
{} & \Big| \left|\langle A P x, A P y \rangle\right| - \left|\langle P x, P y \rangle\right| \Big| \\
{} & \overset{\text{(a)}}{\leq} \left| \langle A P x, A P y \rangle - \langle P x, P y \rangle \right| \\
{} & \overset{\text{(b)}}{=} \left| \langle x, P A^* A P y \rangle - \langle x, P y \rangle \right| \\
{} & = \left| \langle x, (P A^* A P - P) y \rangle \right| \\
{} & \leq \norm{P A^* A P - P} ~ \norm{x}_2 \norm{y}_2
\end{align*}} %
where (a) follows from the triangle inequality, (b) follows since from the idempotence of $P$.
\end{IEEEproof} %[Proof of Lemma~\ref{lemma:IPdev}]

\subsection{Proof of Corollary~\ref{cor:ssomp}}
\label{subsec:cor:ssomp}
%\begin{IEEEproof}[Proof of Corollary~\ref{cor:ssomp}]
Recall that SS-OMP is M-OMP applied to $P_{\widehat{S}}$ instead of $Y$.
Since we consider the first $k$ steps of SS-OMP, we assume that $J \subsetneq J_0$ satisfies $|J| \leq k-1$.
Since (\ref{eq:prop:ssomp:cond1}) implies $\norm{A_{J_0}^* A_{J_0} - I_s} < 1$, all columns of $A_{J_0}$ are linearly independent.
Therefore, the projection $P_{\bar{S}}$ is written as
\begin{equation*}
P_{\bar{S}} = A_{J_0} \Phi \left(\Phi^* A_{J_0}^* A_{J_0} \Phi \right)^{-1} \Phi^* A_{J_0}^*.
\end{equation*}

Since $\Phi^* \Phi = I_r$, it follows that
\begin{align}
{} & \norm{\Phi^* A_{J_0}^* A_{J_0} \Phi} \nonumber \\
{} & = \norm{\Phi^*} \cdot \norm{A_{J_0}^* A_{J_0}} \cdot \norm{\Phi} \nonumber \\
{} & \leq \norm{A_{J_0}^* A_{J_0}} \nonumber \\
{} & = \norm{A_{J_0}}^2
\label{eq:proof:cor:ssomp:eq1}
\end{align}
and
{\allowdisplaybreaks
\begin{align}
{} & \sigma_r(\Phi^* A_{J_0}^* A_{J_0} \Phi) \nonumber \\
{} & = \sigma_r(\Phi \Phi^* A_{J_0}^* A_{J_0} \Phi \Phi^*) \nonumber \\
{} & \geq \sigma_s(A_{J_0}^* A_{J_0}) \nonumber \\
{} & = \sigma_s^2(A_{J_0}) \label{eq:proof:cor:ssomp:eq2}
\end{align}} %
where the inequality follows from the variational characterization of the eigenvalue
since $\Phi \Phi^*$ is a projection onto an $r$-dimensional subspace.

Note that
{\allowdisplaybreaks
\begin{align}
{} & \max\{ 1 - \sigma_s^2(A_{J_0}), \norm{A_{J_0}}^2 - 1 \} \nonumber \\
{} & = \norm{A_{J_0}^* A_{J_0} - I_s} \nonumber \\
{} & \leq \delta_{s+1}^\text{weak}(A;J_0) \overset{\text{(a)}}{<} \delta
\label{eq:proof:cor:ssomp:eq3}
\end{align}} %
where (a) follows from (\ref{eq:prop:ssomp:cond1}).

Let $\Upsilon = \left(\Phi^* A_{J_0}^* A_{J_0} \Phi \right)^{-1/2}$.
Then, by (\ref{eq:proof:cor:ssomp:eq2}) and (\ref{eq:proof:cor:ssomp:eq3}),
\begin{equation*}
\lambda_1(\Upsilon) \leq \frac{1}{\sigma_s(A_{J_0})} < \frac{1}{\sqrt{1-\delta}}
\end{equation*}
and similarly, by (\ref{eq:proof:cor:ssomp:eq1}) and (\ref{eq:proof:cor:ssomp:eq3}),
\begin{equation*}
\lambda_r(\Upsilon) \geq \frac{1}{\sigma_1(A_{J_0})} > \frac{1}{\sqrt{1+\delta}}.
\end{equation*}

Let $U \in \mathbb{K}^{n \times r}$ satisfy $U^{[n] \setminus J_0} = 0$ and $U^{J_0} = \Phi$.
Then, $P_{\bar{S}}$ is also written as
\begin{equation*}
P_{\bar{S}} = A (U \Upsilon V^*)
\end{equation*}
where $V$ is given by $V = A_{J_0} \Phi \Upsilon$ and satisfies $V^* V = I_r$.

Since $P_{\widehat{S}}$ satisfies
\begin{equation*}
\norm{P_{\widehat{S}} - P_{\bar{S}}} \leq \eta,
\end{equation*}
we can write $P_{\widehat{S}}$ as
\begin{equation*}
P_{\widehat{S}} = A (U \Upsilon V^*) + Z
\end{equation*}
for some $Z$ satisfying $\norm{Z} \leq \eta$.

The following two inequalities applied to Proposition~\ref{prop:momp} complete the proof.
{\allowdisplaybreaks
\begin{align*}
{} & \norm{U^{J_0 \setminus J} \Upsilon V^*} = \norm{U^{J_0 \setminus J} \Upsilon} \\
{} & \leq \norm{U \Upsilon} \leq \norm{U} ~ \norm{\Upsilon} \\
{} & = \norm{\Phi} ~ \norm{\Upsilon} < \frac{1}{\sqrt{1-\delta}}
\end{align*}} %
and
{\allowdisplaybreaks
\begin{align*}
{} & \norm{U^{J_0 \setminus J} \Upsilon V^*}_{2,\infty} = \norm{U^{J_0 \setminus J} \Upsilon}_{2,\infty} \\
{} & > \frac{\norm{U^{J_0 \setminus J}}_{2,\infty}}{\sqrt{1+\delta}}
\geq \frac{\rho_{|J|+1}\Phi}{\sqrt{1+\delta}} \\
{} & \geq \frac{\rho_k\Phi}{\sqrt{1+\delta}}
\end{align*}} %
where the last step follows since $|J| \leq k-1$.
%\end{IEEEproof} %[Proof of Corollary~\ref{cor:ssomp}]

\subsection{Proof of Lemma~\ref{lemma:rhobndCB}}
\label{subsec:lemma:rhobndCB}
%\begin{IEEEproof}[Proof of Lemma~\ref{lemma:rhobndCB}]
By the Cauchy-Binet formula, it follows that
\begin{align*}
1 {} & = \det(\Phi^* \Phi) \\
{} & = \sum_{J \in [s], |J| = r} \det\left((\Phi^J)^* \Phi^J\right) \\
{} & = \sum_{J \in [s], |J| = r} \left|\det(\Phi^J)\right|^2.
\end{align*}
The determinant of $\Phi^J$ is bounded from above by
\begin{equation*}
\left|\det(\Phi^J)\right| \leq \prod_{k=1}^r \norm{\Phi^J e_k}_2 \leq \left( \frac{1}{r} \sum_{k=1}^r \norm{\Phi^J e_k}_2^q \right)^{r/q}
\end{equation*}
where the first inequality follows from Hadamard's inequality and the second inequality follows from
the inequality of arithmetic and geometric means.
Therefore,
\begin{align*}
1 {} & \leq \sum_{J \in [s], |J| = r} \left( \frac{1}{r} \sum_{k=1}^r \norm{\Phi^J e_k}_2^q \right)^{2r/q} \\
{} & \leq \frac{s!}{r!(s-r)!} \max_{J \in [s], |J| = r} \left( \frac{1}{r} \sum_{k=1}^r \norm{\Phi^J e_k}_2^q \right)^{2r/q} \\
{} & \leq \frac{s!}{r!(s-r)!} \left( \frac{1}{r} \sum_{k=1}^r (\rho_k(\Phi))^q \right)^{2r/q},
\end{align*}
which implies
\begin{equation}
\left(\frac{s!}{r!(s-r)!}\right)^{-q/2r} \leq \frac{1}{r} \sum_{k=1}^r (\rho_k(\Phi))^q.
\label{eq:proof:lemma:rhobndCB:eq1}
\end{equation}
Note that
\begin{align}
\sum_{j=1}^r (\rho_k(\Phi))^q
{} & \leq \sum_{k=1}^{s-r} (\rho_k(\Phi))^q + \sum_{k=s-r+1}^r (\rho_k(\Phi))^q \nonumber \\
{} & \leq (s-r) + (2r-s) (\rho_{s-r}(\Phi))^q.
\label{eq:proof:lemma:rhobndCB:eq2}
\end{align}
Combining (\ref{eq:proof:lemma:rhobndCB:eq1}) and (\ref{eq:proof:lemma:rhobndCB:eq2}), we obtain
\begin{equation}
\rho_{s-r}(\Phi) \geq \left( \frac{\displaystyle \left(\frac{s!}{r!(s-r)!}\right)^{-q/2r} - \frac{s}{r} + 1}{\displaystyle 2 - \frac{s}{r}} \right)^{1/q}.
\label{eq:proof:lemma:rhobndCB:eq3}
\end{equation}
%\end{IEEEproof} %Proof of Lemma~\ref{lemma:rhobndCB}

\subsection{Proof of Proposition~\ref{prop:ssomsp}}
\label{subsec:prop:ssomsp}
% \begin{IEEEproof}[Proof of Proposition~\ref{prop:ssomsp}]
Assume that $J \subsetneq J_0$ is given from the previous steps where $k = |J| < s$.

Define
\begin{equation}
q_j \triangleq \frac{P_{\mathcal{R}(A_J)}^\perp a_j}{\norm{P_{\mathcal{R}(A_J)}^\perp a_j}_2}
\label{eq:proof:prop:ssomsp:defqj}
\end{equation}
for $j \in [n]$. Then, $\norm{q_j}_2 = 1$ for all $j \in [n]$.

For the guarantee of the next step of SS-OMSP, we need to show that
\begin{equation}
\max_{j \in J_0 \setminus J} \norm{(P_{P_{\mathcal{R}(A_J)}^\perp \widehat{S}}) q_j}
> \max_{j \in [n] \setminus J_0} \norm{(P_{P_{\mathcal{R}(A_J)}^\perp \widehat{S}}) q_j}.
\label{eq:proof:prop:ssomsp:eq0}
\end{equation}

By the triangle inequality, it follows that
{\allowdisplaybreaks
\begin{align*}
{} & \Big|
\norm{(P_{P_{\mathcal{R}(A_J)}^\perp \widehat{S}}) q_j}
- \norm{(P_{P_{\mathcal{R}(A_J)}^\perp \bar{S}}) q_j}
\Big| \\
{} & \leq \norm{(P_{P_{\mathcal{R}(A_J)}^\perp \widehat{S}} - P_{P_{\mathcal{R}(A_J)}^\perp \bar{S}}) q_j} \\
{} & \leq \norm{P_{P_{\mathcal{R}(A_J)}^\perp \widehat{S}} - P_{P_{\mathcal{R}(A_J)}^\perp \bar{S}}} \\
{} & \leq \frac{\eta\beta}{\alpha - \eta\beta}
\end{align*}} %
where the last step follows from Proposition~\ref{prop:errinaugssp}
since $X_0^{J_0}$ is row-nondegenerate and (\ref{eq:prop:ssomsp:cond3}) implies $\alpha > \sqrt{\eta}\beta$.

Then, (\ref{eq:proof:prop:ssomsp:eq0}) is implied by
\begin{equation}
\max_{j \in J_0 \setminus J} \norm{(P_{P_{\mathcal{R}(A_J)}^\perp \bar{S}}) q_j}
> \max_{j \in [n] \setminus J_0} \norm{(P_{P_{\mathcal{R}(A_J)}^\perp \bar{S}}) q_j} + \frac{2\eta\beta}{\alpha - \eta\beta}.
\label{eq:proof:prop:ssomsp:eq1}
\end{equation}

Let $Q = [q_1,\ldots,q_n] \in \mathbb{K}^{m \times n}$ where $q_j$ for each $j \in [n]$ is defined in (\ref{eq:proof:prop:ssomsp:defqj}).
Then, $Q_{J_0 \setminus J}$ satisfies
{\allowdisplaybreaks
\begin{align}
\sigma_{s-k}(Q_{J_0 \setminus J})
{} & \geq \frac{\sigma_{s-k}(P_{\mathcal{R}(A_J)}^\perp A_{J_0 \setminus J})}{\max_{j \in J_0 \setminus J} \norm{P_{\mathcal{R}(A_J)}^\perp a_j}_2} \nonumber \\
{} & \overset{\text{(a)}}{\geq} \frac{\sigma_s(A_{J_0})}{\max_{j \in J_0 \setminus J} \norm{a_j}} \nonumber \\
{} & \overset{\text{(b)}}{\geq} \frac{\sigma_s(A_{J_0})}{\norm{A^*}_{2,\infty}} \nonumber \\
{} & \overset{\text{(c)}}{>} \frac{\alpha}{\norm{A^*}_{2,\infty}}
\label{eq:proof:prop:ssomsp:eq2}
\end{align}} %
where (a) and (b) follow by Lemma~\ref{lemma:interlacing_schur} and Lemma~\ref{lemma:interlacing_submatrix}, respectively,
and (c) is implied by (\ref{eq:prop:ssomsp:cond2}).

First, we bound the LHS of (\ref{eq:proof:prop:ssomsp:eq1}) from below by
{\allowdisplaybreaks
\begin{align}
{} & \max_{j \in J_0 \setminus J} \norm{(P_{P_{\mathcal{R}(A_J)}^\perp \bar{S}}) q_j} \nonumber \\
{} & = \max_{j \in J_0 \setminus J} \norm{q_j^* P_{P_{\mathcal{R}(A_J)}^\perp \bar{S}}} \nonumber \\
{} & = \norm{Q_{J_0 \setminus J}^* P_{P_{\mathcal{R}(A_J)}^\perp \bar{S}}}_{2,\infty} \nonumber \\
{} & \geq \frac{\norm{Q_{J_0 \setminus J}^* P_{P_{\mathcal{R}(A_J)}^\perp \bar{S}}}_F}{\sqrt{s-k}} \nonumber \\
{} & \overset{\text{(d)}}{\geq} \underbrace{\left(\frac{\sum_{\ell=1}^{\dim(P_{\mathcal{R}(A_J)}^\perp \bar{S})} \sigma_{s-k+1-\ell}^2(Q_{J_0 \setminus J})}{s-k}\right)^{1/2}}_{(\ast)}
\label{eq:proof:prop:ssomsp:eq3}
\end{align}
where (d) follows from the variational characterization of the Ky-Fan norm \cite{Bha97}.
For the special case when $r = s$, $\bar{S}$ is given by $\bar{S} = S = \mathcal{R}(A_{J_0})$ and hence
\begin{equation}
\dim(P_{\mathcal{R}(A_J)}^\perp \bar{S}) = \rank(P_{\mathcal{R}(A_J)}^\perp A_{J_0 \setminus J}) = s - k,
\label{eq:proof:prop:ssomsp:eq3z}
\end{equation}
which follows since $\sigma_{s-k}(P_{\mathcal{R}(A_J)}^\perp A_{J_0 \setminus J}) \geq \sigma_s(A_{J_0}) > 0$ by Lemma~\ref{lemma:interlacing_schur}.
In this case, the RHS of (\ref{eq:proof:prop:ssomsp:eq3}) reduces to
\begin{align}
(\ast) {} & = \left(\frac{\sum_{\ell=1}^{s-k} \sigma_{s-k+1-\ell}^2(Q_{J_0 \setminus J})}{s-k}\right)^{1/2} \nonumber \\
{} & = \left(\frac{\norm{Q_{J_0 \setminus J}}_F^2}{s-k}\right)^{1/2}
= \left(\frac{\sum_{\ell=1}^{s-k} \norm{q_j}_2^2}{s-k}\right)^{1/2} = 1. \label{eq:proof:prop:ssomsp:eq3a}
\end{align}

Otherwise, if $r < s$, we derive a lower bound on the RHS of (\ref{eq:proof:prop:ssomsp:eq3}) by
\begin{align}
(\ast) {} & \geq \sqrt{\frac{\dim(P_{\mathcal{R}(A_J)}^\perp \bar{S})}{s-k}} \cdot \sigma_{s-k}(Q_{J_0 \setminus J}) \nonumber \\
{} & \overset{\text{(e)}}{>} \frac{\alpha}{\norm{A^*}_{2,\infty}} \cdot \sqrt{\frac{\dim(P_{\mathcal{R}(A_J)}^\perp \bar{S})}{s-|J|}}
\label{eq:proof:prop:ssomsp:eq3b}
\end{align}} %
where (e) follows from (\ref{eq:proof:prop:ssomsp:eq2}).

Next, we derive an upper bound on the RHS of (\ref{eq:proof:prop:ssomsp:eq1}).
Assume that $j \in [n] \setminus J_0$.
Since
{\allowdisplaybreaks
\begin{align*}
P_{\mathcal{R}(A_J)}^\perp \bar{S}
{} & \subset P_{\mathcal{R}(A_J)}^\perp \mathcal{R}(A_{J_0}) \\
{} & = \mathcal{R}(P_{\mathcal{R}(A_J)}^\perp A_{J_0})
= \mathcal{R}(P_{\mathcal{R}(A_J)}^\perp A_{J_0 \setminus J}),
\end{align*}} %
it follows that
\begin{equation}
\norm{(P_{P_{\mathcal{R}(A_J)}^\perp \bar{S}})q_j}_2
\leq \norm{(P_{\mathcal{R}(P_{\mathcal{R}(A_J)}^\perp A_{J_0 \setminus J})})q_j}_2.
\label{eq:proof:prop:ssomsp:eq4}
\end{equation}

Since $P_{\mathcal{R}(A_J)}^\perp A_{J_0 \setminus J}$ has full column rank by (\ref{eq:proof:prop:ssomsp:eq3z}), we can define
\begin{equation*}
R \triangleq \left[A_{J_0 \setminus J}^* P_{\mathcal{R}(A_J)}^\perp A_{J_0 \setminus J}\right]^{-1/2}
\end{equation*}
and
\begin{equation*}
\Phi \triangleq P_{\mathcal{R}(A_J)}^\perp A_{J_0 \setminus J} R.
\end{equation*}
Then, $\Phi$ is an orthonormal basis for $\mathcal{R}(P_{\mathcal{R}(A_J)}^\perp A_{J_0 \setminus J})$.

Since $A_J$ has full column rank, we can also define
\begin{equation*}
\Psi \triangleq A_J (A_J^* A_J)^{-1/2}.
\end{equation*}
Then, $\Psi$ is an orthonormal basis for $\mathcal{R}(A_J)$.

We bound the RHS of (\ref{eq:proof:prop:ssomsp:eq4}) from above by
{\allowdisplaybreaks
\begin{align}
{} & \norm{(P_{\mathcal{R}(P_{\mathcal{R}(A_J)}^\perp A_{J_0 \setminus J})})q_j}_2 \nonumber \\
{} & = \norm{\Phi^* q_j}_2 \nonumber \\
{} & \overset{\text{(f)}}{\leq} \Big\| [q_j, \Phi]^* [q_j, \Phi] - I_{s-k+1} \Big\| \nonumber \\
{} & = \Big\| [q_j, \Phi]^* P_{\mathcal{R}(A_J)}^\perp [q_j, \Phi] - I_{s-k+1} \Big\| \nonumber \\
{} & = \max\Big\{1 - \lambda_{s-k+1}\left([q_j, \Phi]^* P_{\mathcal{R}(A_J)}^\perp [q_j, \Phi]\right) ,~ \nonumber \\
{} & \qquad \Big\|[q_j, \Phi]^* P_{\mathcal{R}(A_J)}^\perp [q_j, \Phi]\Big\| - 1 \Big\} \nonumber \\
{} & \overset{\text{(g)}}{\leq} \max\Big\{1 - \lambda_s\Big([q_j, \Phi, \Psi]^* [q_j, \Phi, \Psi]\Big) ,~ \nonumber \\
{} & \qquad \Big\|[q_j, \Phi, \Psi]^* [q_j, \Phi, \Psi]\Big\| - 1 \Big\} \nonumber \\
{} & = \Big\| [q_j, \Phi, \Psi]^* [q_j, \Phi, \Psi] - I_{s+1} \Big\|
\label{eq:proof:prop:ssomsp:eq7}
\end{align}} %
where (f) and (g) follow from Lemma~\ref{lemma:interlacing_submatrix} and Lemma~\ref{lemma:interlacing_schur}, respectively.

Note that $[\Phi,\Psi]$ is an orthonormal basis of $\mathcal{R}(A_{J_0})$.
Therefore, the RHS of (\ref{eq:proof:prop:ssomsp:eq7}) is bounded from above by
{\allowdisplaybreaks
\begin{align*}
{} & \Big\| [q_j, \Phi, \Psi]^* [q_j, \Phi, \Psi] - I_{s+1} \Big\|^2 \\
{} & = \Bigg\|
\left[
\begin{array}{cc}
0 & q_j^* [\Phi, \Psi] \\
{[\Phi, \Psi]^* q_j} & 0
\end{array}
\right]
\Bigg\|^2 \\
{} & = \Bigg\|
\left[
\begin{array}{cc}
0 & \norm{[\Phi, \Psi]^* q_j}_2^2 \\
{\norm{[\Phi, \Psi]^* q_j}_2^2} & 0
\end{array}
\right]
\Bigg\| \\
{} & = \norm{[\Phi, \Psi]^* q_j}_2^2 \\
{} & = \norm{P_{\mathcal{R}(A_{J_0})} q_j}_2^2 \\
{} & = 1 - \norm{P_{\mathcal{R}(A_{J_0})}^\perp q_j}_2^2 \\
{} & = 1 - \frac{\norm{P_{\mathcal{R}(A_{J_0})}^\perp a_j}_2^2}{\norm{P_{\mathcal{R}(A_{J})}^\perp a_j}_2^2} \\
{} & \overset{\text{(h)}}{\leq} 1 - \frac{\sigma_{s+1}^2(A_{J_0 \cup \{j\}})}{\norm{a_j}_2^2} \\
{} & \overset{\text{(i)}}{<} 1 - \frac{\alpha^2}{\norm{A^*}_{2,\infty}^2}
\end{align*}} %
where (h) follows from Lemma~\ref{lemma:interlacing_schur} and (i) is implied by (\ref{eq:prop:ssomsp:cond2}).

Since $j$ was arbitrary in $[n] \setminus J_0$, (\ref{eq:proof:prop:ssomsp:eq7}) implies that
\begin{equation}
\max_{j \in [n] \setminus J} \norm{(P_{\mathcal{R}(P_{\mathcal{R}(A_J)}^\perp A_{J_0 \setminus J})})q_j}_2
< \sqrt{1 - \frac{\alpha^2}{\norm{A^*}_{2,\infty}^2}}.
\label{eq:proof:prop:ssomsp:eq8}
\end{equation}

Applying (\ref{eq:proof:prop:ssomsp:eq3b}), and (\ref{eq:proof:prop:ssomsp:eq8}) to (\ref{eq:proof:prop:ssomsp:eq1}),
we obtain the following sufficient condition for (\ref{eq:proof:prop:ssomsp:eq0})
\begin{equation*}
\sqrt{\frac{\dim(P_{\mathcal{R}(A_J)}^\perp \bar{S})}{s-|J|}} \cdot \frac{\alpha}{\norm{A^*}_{2,\infty}}
- \sqrt{1 - \frac{\alpha^2}{\norm{A^*}_{2,\infty}^2}}
\geq \frac{2\eta\beta}{\alpha - \eta\beta},
\end{equation*}
which is equivalent to (\ref{eq:prop:ssomsp:cond3}).
For the full row rank case,
applying (\ref{eq:proof:prop:ssomsp:eq3a}), and (\ref{eq:proof:prop:ssomsp:eq8}) to (\ref{eq:proof:prop:ssomsp:eq1}),
we obtain Condition (\ref{eq:prop:ssomsp:cond4}), which implies (\ref{eq:proof:prop:ssomsp:eq0}).
%\end{IEEEproof}

\subsection{Proof of Proposition~\ref{prop:estss_aca}}
\label{subsec:prop:estss_aca}
To prove Proposition~\ref{prop:estss_aca}, we use the following lemmata.
The proof of Lemma~\ref{lemma:davsza:complex} is deferred after the proof of Proposition~\ref{prop:estss_aca}.

\begin{lemma}[{A case of \cite[Theorem~VII.3.3]{Bha97}}]
For positive semi-definite matrices $\Gamma_1, \Gamma_2 \in \mathbb{K}^{m \times m}$ and $r \in \{1,\ldots,m-1\}$,
let $S_1$ and $S_2$ be the subspaces spanned by the $r$-dominant eigenvectors of $\Gamma_1$ and $\Gamma_2$, respectively.
Then,
\begin{align*}
3 \norm{\Gamma_1 - \Gamma_2}
{} & \geq \norm{P_{S_1}^\perp P_{S_2}} \cdot \left[\lambda_r(\Gamma_2) - \lambda_{r+1}(\Gamma_1)\right].
\end{align*}
\label{lemma:iss2}
\end{lemma}

\begin{lemma}
Given $m,n \in \mathbb{N}$ with $n > 2m$, consider the random matrix $G \in \mathbb{C}^{n \times m}$
whose entries are i.i.d. circular Gaussian variables with zero mean and variance $\frac{1}{n}$.
Then,
\begin{align*}
\mathbb{P}\left( \norm{G^* G - I_m} \geq 3 \sqrt{\frac{2m}{n}} + t \right) {} & \leq 4 \exp\left(-\frac{nt^2}{18}\right).
\end{align*}
for $t > 0$ satisfying
\begin{equation*}
\sqrt{\frac{2m}{n}} + \frac{t}{3} \leq 1.
\end{equation*}
\label{lemma:davsza:complex}
\end{lemma}

\begin{IEEEproof}[Proof of Proposition~\ref{prop:estss_aca}]
Recall that, given the sample covariance matrix $\Gamma_Y \triangleq \frac{YY^*}{N}$ of the snapshots $Y$,
Algorithm~1 computes $\widehat{\Gamma}$ defined by
\begin{equation*}
\widehat{\Gamma} = \Gamma_Y - \lambda_{m}(\Gamma_Y) I_m.
\end{equation*}
Also recall that $\Gamma_S$ is defined by
\begin{equation*}
\Gamma_S \triangleq \frac{A_{J_0} X_0^{J_0} (X_0^{J_0})^* A_{J_0}^*}{N} = \frac{A_{J_0} \Psi\Lambda\Phi\Phi^*\Lambda\Psi^* A_{J_0}^*}{N}.
\end{equation*}

Let $D \in \mathbb{K}^{m \times m}$ denote the distortion matrix defined by
\begin{equation*}
D \triangleq \widehat{\Gamma} - \Gamma_S.
\end{equation*}
Then, $D$ is decomposed as
\begin{equation*}
D = D_\text{noise} + D_\text{cross} + D_\text{bias}
\end{equation*}
where
{\allowdisplaybreaks
\begin{align*}
D_\text{noise} {} & \triangleq \frac{WW^*}{N} - \sigma_w^2 I_m, \\
D_\text{cross} {} & \triangleq \frac{A_{J_0} X_0^{J_0} W^*}{N} + \frac{W (X_0^{J_0})^* A_{J_0}^*}{N}, \\
D_\text{bias} {} & \triangleq \left[\sigma_w^2 - \lambda_m\left(\Gamma_Y\right)\right] I_m.
\end{align*}} %

It follows that the sample covariance matrix $\Gamma_Y$ is decomposed as
$\Gamma_Y = \Gamma_S + \sigma_w^2 I_m + D_\text{cross} + D_\text{noise}$.
Viewing $\Gamma_Y$ as a perturbed version of $\Gamma_S + \sigma_w^2 I_m$ with distortion $D_\text{noise} + D_\text{cross}$,
we bound the perturbation in the $m$-th eigenvalue by Weyl's Theorem as
\begin{align*}
{} & \underbrace{| \lambda_m(\Gamma_Y) - \lambda_m(\Gamma_S + \sigma_w^2 I_m) |}_{(\ast)} \\
{} & \leq \norm{\Gamma_Y - \Gamma_S - \sigma_w^2 I_m} \\
{} & = \norm{D_\text{noise} + D_\text{cross}} \\
{} & \leq \norm{D_\text{noise}} + \norm{D_\text{cross}}.
\end{align*}
Now, since $\rank(\Gamma_S) \leq s$ and $s < m$, $(\ast)$ reduces to
\begin{equation*}
\left|\sigma_w^2 - \lambda_m(\Gamma_Y)\right| = \norm{D_\text{bias}}.
\end{equation*}
Therefore, $\norm{D_\text{bias}}$ satisfies
\begin{equation*}
\norm{D_\text{bias}} \leq \norm{D_\text{noise}} + \norm{D_\text{cross}}.
\end{equation*}

For $k \in \mathbb{N}$ where $k < N$, let $Z_k \in \mathbb{K}^{k \times N}$ be an i.i.d. Gaussian matrix such that $\mathbb{E} Z_k = 0$
and $\mathbb{E} \frac{Z_k (Z_k)^*}{N} = I_k$.
Then, we define for $k = 1,\ldots, N-1$ a family of random variables
\begin{equation*}
\xi_k \triangleq \Big\| \frac{Z_k (Z_k)^*}{N} - I_k \Big\|.
\end{equation*}
Using $\xi_k$, we bound the distortion terms from above by
\begin{align*}
\norm{D_\text{noise}} {} & = \sigma_w^2 \Big\| \frac{WW^*}{\sigma_w^2 N} - I_m \Big\| \sim \xi_m \sigma_w^2,
\intertext{and}
\norm{D_\text{cross}} {} & \leq 2 \sigma_w \norm{A_{J_0} \Psi \Lambda} \Big\| \frac{\Phi W^*}{\sigma_w N} \Big\| \\
{} & \leq 2 \sigma_w \lambda_1^{1/2}(\Gamma) \\
{} & \quad \cdot \Big\| \frac{1}{N} \Big[ \Phi^*,~ W^*/\sigma_w \Big]^* \Big[ \Phi^*,~ W^*/\sigma_w \Big] - I_{m+M} \Big\| \\
{} & \sim 2 \xi_{m+M} \sigma_w \lambda_1^{1/2}(\Gamma).
\end{align*}
where $X_1 \sim X_2$ denotes that random variables $X_1$ and $X_2$ have the same distribution.
Therefore, for any fixed $c > 0$, since $\mathbb{P}(\xi_k \leq c)$ decreases in $k$,
it follows that
\begin{align}
\mathbb{P}\left(\norm{D} \leq c\right)
\geq \mathbb{P}\left( 2 \xi_m \sigma_w^2 + 2 \xi_{m+s} \sigma_w \lambda_1^{1/2}(\Gamma) \leq c\right) \nonumber \\
\geq \mathbb{P}\left( 2 \xi_{m+s} \sigma_w \left[ \sigma_w + 2 \lambda_1^{1/2}(\Gamma)\right] \leq c\right).
\label{eq:proof:prop:estss_aca:eq1}
\end{align}

\noindent\textbf{Rank estimation}
From the sample covariance matrix $\widehat{\Gamma}$,
Algorithm~1 determines the rank $r$ as the maximal number $k$ that satisfies
\begin{equation*}
\lambda_k(\widehat{\Gamma}) - \lambda_{k+1}(\widehat{\Gamma}) \geq \tau \lambda_1(\widehat{\Gamma}).
\end{equation*}
Therefore, to guarantee that Algorithm~\ref{alg:estss} stops at the desired rank $r$,
we need sufficient conditions for
\begin{align}
\lambda_r(\widehat{\Gamma}) - \lambda_{r+1}(\widehat{\Gamma}) {} & \geq \tau \lambda_1(\widehat{\Gamma}), \label{eq:proof:prop:estss_aca:eq2} \\
\lambda_k(\widehat{\Gamma}) - \lambda_{k+1}(\widehat{\Gamma}) {} & < \tau \lambda_1(\widehat{\Gamma}), \quad \forall k > r. \label{eq:proof:prop:estss_aca:eq3}
\end{align}

By Weyl's perturbation theorem \cite[Corollary~III.2.6]{Bha97}, it follows that
\begin{equation*}
\max_{k \in [m]} \left| \lambda_k(\widehat{\Gamma}) - \lambda_k(\Gamma_S) \right| \leq \norm{\widehat{\Gamma} - \Gamma_S}
\end{equation*}
and hence
\begin{equation}
\frac{\lambda_r(\widehat{\Gamma}) - \lambda_{r+1}(\widehat{\Gamma})}{\lambda_1(\widehat{\Gamma})}
\geq \frac{\lambda_r(\Gamma_S) - \lambda_{r+1}(\Gamma_S) - 2 \norm{D}}{\lambda_1(\Gamma_S) + \norm{D}}.
\label{eq:proof:prop:estss_aca:eq4}
\end{equation}

Again, by Weyl's perturbation theorem \cite[Corollary~III.2.6]{Bha97}, it follows that
{\allowdisplaybreaks
\begin{align*}
\left| \lambda_k\left(\frac{\Phi\Phi^*}{N}\right) - 1 \right|
{} & = \left| \lambda_k\left(\frac{\Phi\Phi^*}{N}\right) - \lambda_k(I_M) \right| \\
{} & \leq \Big\| \frac{\Phi\Phi^*}{N} - I_M \Big\| \leq \xi_M
\end{align*}} %
for all $k = 1,\ldots, M$ and hence
\begin{equation}
1 - \xi_M \leq \lambda_M\left(\frac{\Phi\Phi^*}{N}\right) \leq \lambda_1\left(\frac{\Phi\Phi^*}{N}\right) \leq 1 + \xi_M
\label{eq:proof:prop:estss_aca:eq5}
\end{equation}

By (\ref{eq:proof:prop:estss_aca:eq5}), it follows that
\begin{align}
\lambda_k(\Gamma_S) {} & \geq (1-\xi_M) \lambda_k(\Gamma) \label{eq:proof:prop:estss_aca:eq6} \\
\lambda_k(\Gamma_S) {} & \leq (1+\xi_M) \lambda_k(\Gamma) \label{eq:proof:prop:estss_aca:eq7}.
\end{align}

By Lemma~\ref{lemma:davsza:complex}, it follows that
\begin{align}
\mathbb{P}\left(
\xi_k > \frac{6\sqrt{k + \ln(4/\epsilon)}}{\sqrt{N}}
\right) \leq \epsilon.
\label{eq:proof:prop:estss_aca:xi}
\end{align}
Therefore, by (\ref{eq:prop:estss_aca:cond1}), (\ref{eq:prop:estss_aca:cond2}), and (\ref{eq:proof:prop:estss_aca:xi}),
it follows that $\xi_M \leq \theta$ with probability $1 - \epsilon/2$.
Therefore, we assume $\xi_M \leq \theta$ in the remaining steps of the proof.

Since $\xi_M \leq \theta$, by (\ref{eq:proof:prop:estss_aca:eq4}), (\ref{eq:proof:prop:estss_aca:eq6}), and (\ref{eq:proof:prop:estss_aca:eq7})
the following condition is a sufficient condition for (\ref{eq:proof:prop:estss_aca:eq2}):
\begin{align*}
\frac{(1-\theta)\lambda_r(\Gamma) - (1+\theta)\lambda_{r+1}(\Gamma) - 2 \norm{D}}{(1+\theta)\lambda_1(\Gamma) + \norm{D}}
\geq \tau,
\end{align*}
which is equivalent to
\begin{equation}
\norm{D} \leq \frac{(1-\theta)\lambda_r(\Gamma) - (1+\theta)\lambda_{r+1}(\Gamma) - (1+\theta)\tau\lambda_1(\Gamma)}{2+\tau}.
\label{eq:proof:prop:estss_aca:eq8}
\end{equation}

Since $\Gamma$ satisfies (\ref{eq:gapGamma1}),
we obtain a sufficient condition for (\ref{eq:proof:prop:estss_aca:eq8}) given by
\begin{equation}
\frac{\norm{D}}{\lambda_1(\Gamma)} \leq \frac{(1+\theta)\nu\tau}{2+\tau}.
\label{eq:proof:prop:estss_aca:eq9}
\end{equation}
Similarly, since $\Gamma$ satisfies (\ref{eq:gapGamma2}),
we verify that (\ref{eq:proof:prop:estss_aca:eq9}) is also a sufficient condition for (\ref{eq:proof:prop:estss_aca:eq3}).

Recall that Algorithm~1 computes $\widehat{S}$ as the subspace spanned by the $r$ dominant eigenvectors of $\widehat{\Gamma}$.
Next, we derive conditions that guarantees that $\norm{P_{\widehat{S}} - P_{\bar{S}}} \leq \eta$
where $\bar{S}$ is the subspace spanned by the $r$ dominant eigenvectors of $\Gamma_S$.

We apply $\Gamma_1 = \widehat{\Gamma}$ and $\Gamma_2 = \Gamma_S$ to Lemma~\ref{lemma:iss2} and obtain
\begin{equation}
3 \norm{D} \geq \norm{P_{\widehat{S}} - P_{\bar{S}}} \cdot \left[\lambda_r(\Gamma_S) - \lambda_{r+1}(\Gamma_S)\right].
\label{eq:proof:prop:estss_aca:eq10}
\end{equation}
Since $\xi_M \leq \theta$ and $\Gamma$ satisfies (\ref{eq:gapGamma1}),
\begin{align}
\lambda_r(\Gamma_S) - \lambda_{r+1}(\Gamma_S)
{} & \geq (1-\theta)\lambda_r(\Gamma) - (1+\theta)\lambda_{r+1}(\Gamma) \nonumber \\
{} & \geq (1+\theta)(1+\nu)\tau\lambda_1(\Gamma). \label{eq:proof:prop:estss_aca:eq11}
\end{align}

By (\ref{eq:proof:prop:estss_aca:eq10}) and (\ref{eq:proof:prop:estss_aca:eq11}),
we note that $\norm{P_{\widehat{S}} - P_{\bar{S}}} \leq \eta$ is implied by
\begin{equation}
\frac{3\norm{D}}{\lambda_1(\Gamma)} \leq \eta(1+\theta)(1+\nu)\tau\lambda_1(\Gamma).
\label{eq:proof:prop:estss_aca:eq12}
\end{equation}

Recall that $C_{\eta,\nu,\theta,\tau}$ is defined by
\begin{equation*}
C_{\eta,\nu,\theta,\tau} \triangleq (1+\theta)\tau \min\left\{ \frac{(1+\nu)\eta}{3},~ \frac{\nu}{2+\tau} \right\}.
\end{equation*}
Then, by (\ref{eq:proof:prop:estss_aca:eq1})
\begin{equation}
\xi_{m+s} \leq \frac{C_{\eta,\nu,\theta,\tau}}{2[\sigma_w^2/\lambda_1(\Gamma) + 2(\sigma_w^2/\lambda_1(\Gamma))^{1/2}]}
\label{eq:proof:prop:estss_aca:eq13}
\end{equation}
implies (\ref{eq:proof:prop:estss_aca:eq9}) and (\ref{eq:proof:prop:estss_aca:eq12}).

Finally, we note that, by (\ref{eq:prop:estss_aca:cond1}), (\ref{eq:prop:estss_aca:cond2}), and (\ref{eq:proof:prop:estss_aca:xi}),
it follows that (\ref{eq:proof:prop:estss_aca:eq13}) holds with probability $1 - \epsilon/2$.
\end{IEEEproof}

\begin{IEEEproof}[Proof of Lemma~\ref{lemma:davsza:complex}]
We use the following lemma to prove Lemma~\ref{lemma:davsza:complex}
\begin{lemma}[{\cite[Lemma~36]{Ver10}}]
Consider a matrix $B \in \mathbb{K}^{n \times m}$ ($m \leq n$) that satisfies
\begin{equation*}
\norm{B^*B - I} \leq \max(\delta,\delta^2)
\end{equation*}
for some $\delta > 0$. Then,
\begin{equation}
1 - \delta \leq \sigma_m(B) \leq \sigma_1(B) \leq 1 + \delta.
\label{eq:ver10eq2}
\end{equation}
Conversely, if $B$ satisfies (\ref{eq:ver10eq2}) for some $\delta > 0$, then $\norm{B^*B - I} \leq 3 \max(\delta,\delta^2)$.
\label{lemma:ver10}
\end{lemma}

We can write $G$ as
\begin{equation*}
G = \frac{1}{\sqrt{2}} (G_\text{Re} + j G_\text{Im})
\end{equation*}
where $G_\text{Re}, G_\text{Im} \in \mathbb{R}^{n \times m}$ are mutually independent i.i.d. Gaussian matrices
whose entries follow $\mathcal{N}(0,\frac{1}{n})$.
Then,
{\allowdisplaybreaks
\begin{align*}
{} & \norm{G^* G - I_s} \\
{} & = \Bigg\|
\underbrace{
\frac{1}{2}
\left[ \begin{array}{cc} G_\text{Re} & -G_\text{Im} \\ G_\text{Im} & G_\text{Re} \end{array} \right]^*
\left[ \begin{array}{cc} G_\text{Re} & -G_\text{Im} \\ G_\text{Im} & G_\text{Re} \end{array} \right]}_{(\ast)}
- I_{2s}
\Bigg\|
\end{align*}} %
where $(\ast)$ is a block matrix
\begin{equation*}
\left[ \begin{array}{cc} A_{11} & A_{12} \\ A_{21} & A_{22} \end{array} \right]
\end{equation*}
whose block entries are given by
{\allowdisplaybreaks
\begin{align*}
A_{11} {} & = \frac{1}{2} \left[(G_\text{Re}^* G_\text{Re} - I_s) + (G_\text{Im}^* G_\text{Im} - I_s) \right], \\
A_{12} {} & = \frac{1}{2} \left( G_\text{Im}^* G_\text{Re} - G_\text{Re}^* G_\text{Im} \right), \\
A_{21} {} & = A_{12}^*, \\
A_{22} {} & = A_{11}.
\end{align*}} %
Then,
{\allowdisplaybreaks
\begin{align}
{} & \norm{G^* G - I_s} \nonumber \\
{} & \leq
\frac{1}{2}
\left\|
\left[
\begin{array}{cc}
(-G_\text{Re})^* (-G_\text{Re}) - I_s & (-G_\text{Re})^* G_\text{Im} \\
G_\text{Im}^* (-G_\text{Re}) & G_\text{Im}^* G_\text{Im} - I_s
\end{array}
\right]
\right\| \nonumber \\
{} & \quad +
\frac{1}{2}
\left\|
\left[
\begin{array}{cc}
G_\text{Im}^* G_\text{Im} - I_s & G_\text{Im}^* G_\text{Re} \\
G_\text{Re}^* G_\text{Im} & G_\text{Re}^* G_\text{Re} - I_s
\end{array}
\right]
\right\| \nonumber \\
{} & =
\frac{1}{2} \left\|
[-G_\text{Re} ~ G_\text{Im}]^* [-G_\text{Re} ~ G_\text{Im}] - I_{2s}
\right\| \nonumber \\
{} & + \frac{1}{2} \left\|
[G_\text{Re} ~ G_\text{Im}]^* [G_\text{Re} ~ G_\text{Im}] - I_{2s}
\right\|. \label{eq:proof:lemma:davsza:complex:eq1}
\end{align}} %

By Theorem~\ref{thm:davsza} and Lemma~\ref{lemma:ver10}, we have
{\allowdisplaybreaks
\begin{align}
{} & \mathbb{P}\left(\norm{[G_\text{Re} ~ G_\text{Im}]^* [G_\text{Re} ~ G_\text{Im}] - I_{2s}} \nonumber \geq 3 \sqrt{\frac{2m}{n}} + t\right) \nonumber \\
{} & \leq 2 \exp\left(-\frac{nt^2}{18}\right)
\label{eq:proof:lemma:davsza:complex:eq2}
\end{align}} %
for $t > 0$ satisfying
\begin{equation*}
\sqrt{\frac{2m}{n}} + \frac{t}{3} \leq 1.
\end{equation*}

By the symmetry of the Gaussian distribution, we also have
{\allowdisplaybreaks
\begin{align}
{} & \mathbb{P}\left(\norm{[-G_\text{Re} ~ G_\text{Im}]^* [-G_\text{Re} ~ G_\text{Im}] - I_{2s}} \nonumber \geq 3 \sqrt{\frac{2m}{n}} + t\right) \nonumber \\
{} & \leq 2 \exp\left(-\frac{nt^2}{18}\right).
\label{eq:proof:lemma:davsza:complex:eq3}
\end{align}} %

Combining (\ref{eq:proof:lemma:davsza:complex:eq1})--(\ref{eq:proof:lemma:davsza:complex:eq3}) completes the proof.
\end{IEEEproof} % [Proof of Lemma~\ref{lemma:davsza:complex}]

\bibliographystyle{IEEEtran}
%\bibliography{IEEEabrv,mmv,cs,etc}

\end{document}